\pdfoutput=1
\documentclass[journal, twoside]{IEEEtran}
\usepackage[T1]{fontenc}

\usepackage{tikz}
\usepackage{adjustbox}
\usetikzlibrary{intersections, angles, patterns, fit, shadings, decorations.text, matrix, arrows, through, calc, math, backgrounds, external,decorations.pathreplacing}
\usepackage{amsmath,amssymb}
\usepackage{pgfplotstable}
\usepackage{subcaption}
\usepackage{dblfloatfix}
\usepackage{import}
\usepackage{siunitx}
\usepackage{tikz}
\usepackage{import}

\usetikzlibrary{positioning}
\usetikzlibrary{fit}
\usetikzlibrary{calc}
\usetikzlibrary{shapes}
\usetikzlibrary{math}
\usetikzlibrary{fpu}
\usetikzlibrary{decorations.markings}
\usetikzlibrary{decorations.pathreplacing}
\usetikzlibrary{arrows.meta}
\usetikzlibrary{intersections}

\makeatletter

\tikzset{outer sep=0}
\tikzset{inner sep=0}

\def\pgfaddtoshape#1#2{
	\begingroup
	\def\pgf@sm@shape@name{#1}%
	\let\anchor\pgf@sh@anchor
	#2%
	\endgroup
}

\newcommand{\anchorlet}[2]{
	\global\expandafter
	\let\csname pgf@anchor@\pgf@sm@shape@name @#1\expandafter\endcsname
	\csname pgf@anchor@\pgf@sm@shape@name @#2\endcsname
}

\pgfmathsetmacro{\NODESIZE}{42}
\pgfmathsetmacro{\NODETHICKNESS}{1.0}
\pgfmathsetmacro{\ROUNDEDCORNERS}{0.5mm}
\tikzset{node distance=0.5*\NODESIZE pt}

\def\FNODESIZE{\NODESIZE pt}

\def\QNODESIZE{0.25*\NODESIZE pt}

\tikzstyle{textstyle} = [text height=1.5ex, text depth=.5ex]
\tikzset{every label/.style=textstyle}

\tikzstyle{linestyle} = [line width = \NODETHICKNESS, rounded corners = \ROUNDEDCORNERS]
\tikzstyle{arrowstyle} = [>=stealth, linestyle]
\tikzstyle{<--} = [<-, arrowstyle]
\tikzstyle{-->} = [->, arrowstyle]
\tikzstyle{->-} = [linestyle, decoration={markings,	mark=at position 0.5 with {\arrow[arrowstyle]{>}}}, postaction={decorate}] 
\tikzstyle{-<-} = [linestyle, decoration={markings,	mark=at position 0.5 with {\arrow[arrowstyle]{<}}}, postaction={decorate}] 

\tikzset{   
    -A-/.style args={#1}{%
        linestyle, 
        decoration={markings, mark=at position #1 with {\arrow[>=Triangle, scale=.025*\NODESIZE]{>}}},
        postaction={decorate}
    },
    -A-/.default = {0.75}
}

\tikzset{   
    -AA-/.style args={#1, #2, #3}{%
        linestyle,
        decoration={markings, mark=at position #1 with {
            \node[amp={color=#2, width=#3, height=#3}, rotate=\pgfdecoratedangle] at (0,0) (inline_amp) {};
        }},
        postaction={decorate}
    },
    -AA-/.default = {0.5, E, 0.15}
}

\tikzstyle{<-->} = [<->, arrowstyle]
\tikzstyle{---} = [arrowstyle]

\tikzset{
    -PC-/.style args={#1}{
        linestyle, 
        decoration={markings, mark=at position #1 with {
            \draw[---,FO]
                (.05*\NODESIZE*\pgflinewidth,.05*\NODESIZE*\pgflinewidth) circle[radius=.05*\NODESIZE*\pgflinewidth];
            \draw[---,FO]
                (-.05*\NODESIZE*\pgflinewidth,.05*\NODESIZE*\pgflinewidth) circle[radius=.05*\NODESIZE*\pgflinewidth];
            \draw[---,FO]
                (0,-.05*\NODESIZE*\pgflinewidth) circle[radius=.05*\NODESIZE*\pgflinewidth];}},
        postaction={decorate}
        },
        -PC-/.default = {0.5}
}

\definecolor{C0}{RGB}{031, 119, 180} 
\definecolor{C1}{RGB}{031, 119, 180} 
\definecolor{C2}{RGB}{255, 127, 014} 
\definecolor{C3}{RGB}{044, 160, 044} 
\definecolor{C4}{RGB}{215, 039, 040} 
\definecolor{C6}{RGB}{148, 103, 189} 
\definecolor{C100}{RGB}{140, 086, 075} 
\definecolor{C7}{RGB}{227, 119, 194} 
\definecolor{C8}{RGB}{127, 127, 127} 
\definecolor{C9}{RGB}{188, 189, 034} 
\definecolor{C5}{RGB}{023, 190, 207} 

\definecolor{C0l}{RGB}{174, 199, 232} 
\definecolor{C11}{RGB}{174, 199, 232} 
\definecolor{C12}{RGB}{255, 187, 120} 
\definecolor{C13}{RGB}{152, 223, 138} 
\definecolor{C14}{RGB}{255, 152, 150} 
\definecolor{C15}{RGB}{197, 176, 213} 
\definecolor{C16}{RGB}{196, 156, 148} 
\definecolor{C17}{RGB}{247, 182, 210} 
\definecolor{C18}{RGB}{199, 199, 199} 
\definecolor{C19}{RGB}{219, 219, 141} 
\definecolor{C20}{RGB}{158, 218, 229} 

\definecolor{Snow}{HTML}{FBFBFB} 			
\definecolor{TUeRed}{RGB}{200, 25, 25}		
\definecolor{TUeGreen}{RGB}{25, 200, 113}	
\definecolor{TUeBlue}{RGB}{25, 113, 200}	

\definecolor{O}{RGB}{031, 119, 180} 	
\definecolor{Ol}{RGB}{174, 199, 232} 	
\definecolor{E}{RGB}{255, 127, 014} 	
\definecolor{El}{RGB}{255, 187, 120} 	
\definecolor{D}{RGB}{148, 103, 189}     
\definecolor{Dl}{RGB}{197, 176, 213} 	
\definecolor{EO}{RGB}{215, 039, 040} 	
\definecolor{EOl}{RGB}{255, 152, 150}   

\tikzstyle{FW} = [fill=white]			
\tikzstyle{FB} = [fill=white]			
\tikzstyle{FO} = [fill=C0l, draw=C0]	
\tikzstyle{FE} = [fill=C1l, draw=C1]	
\tikzstyle{FD} = [fill=C2l, draw=C2]	

\def\direce{e}
\def\direcw{w}
\def\direcn{n}
\def\direcs{s}

\def\flipfalse{0}
\newcount\portcount
\pgfdeclareshape{basic}{
	\inheritsavedanchors[from=rectangle] 
	
	\inheritanchor[from=rectangle]{center}
	\inheritanchor[from=rectangle]{mid}		
	\inheritanchor[from=rectangle]{base}	
	\inheritanchor[from=rectangle]{north}
	\inheritanchor[from=rectangle]{south}		
	\inheritanchor[from=rectangle]{west}		
	\inheritanchor[from=rectangle]{mid west}				
	\inheritanchor[from=rectangle]{base west}		
	\inheritanchor[from=rectangle]{north west}		
	\inheritanchor[from=rectangle]{south west}		
	\inheritanchor[from=rectangle]{east}
	\inheritanchor[from=rectangle]{mid east}
	\inheritanchor[from=rectangle]{base east}	
	\inheritanchor[from=rectangle]{north east}
	\inheritanchor[from=rectangle]{south east}	
	
	\inheritanchorborder[from=rectangle]	

	\inheritbackgroundpath[from=rectangle]	
}

\pgfaddtoshape{basic}{
	\anchor{west south west}{
		\pgf@process{\northeast}
		\pgf@ya=.5\pgf@y
		\pgf@process{\southwest}
		\pgf@y=1.5\pgf@y
		\advance\pgf@y by \pgf@ya
		\pgf@y=.5\pgf@y
	}
	\anchor{west north west}{
		\pgf@process{\northeast}
		\pgf@ya=1.5\pgf@y
		\pgf@process{\southwest}
		\pgf@y=.5\pgf@y
		\advance\pgf@y by \pgf@ya
		\pgf@y=.5\pgf@y
	}
	\anchor{east north east}{
		\pgf@process{\southwest}
		\pgf@ya=.5\pgf@y
		\pgf@process{\northeast}
		\pgf@y=1.5\pgf@y
		\advance\pgf@y by \pgf@ya
		\pgf@y=.5\pgf@y
	}
	\anchor{east south east}{
		\pgf@process{\southwest}
		\pgf@ya=1.5\pgf@y
		\pgf@process{\northeast}
		\pgf@y=.5\pgf@y
		\advance\pgf@y by \pgf@ya
		\pgf@y=.5\pgf@y
	}
	\anchor{north north west}{
		\pgf@process{\southwest}
		\pgf@xa=1.5\pgf@x
		\pgf@process{\northeast}
		\pgf@x=.5\pgf@x
		\advance\pgf@x by \pgf@xa
		\pgf@x=.5\pgf@x
	}
	\anchor{north north east}{
		\pgf@process{\southwest}
		\pgf@xa=.5\pgf@x
		\pgf@process{\northeast}
		\pgf@x=1.5\pgf@x
		\advance\pgf@x by \pgf@xa
		\pgf@x=.5\pgf@x
	}
	\anchor{south south west}{
		\pgf@process{\northeast}
		\pgf@xa=.5\pgf@x
		\pgf@process{\southwest}
		\pgf@x=1.5\pgf@x
		\advance\pgf@x by \pgf@xa
		\pgf@x=.5\pgf@x
	}
	\anchor{south south east}{
		\pgf@process{\northeast}
		\pgf@xa=1.5\pgf@x
		\pgf@process{\southwest}
		\pgf@x=.5\pgf@x
		\advance\pgf@x by \pgf@xa
		\pgf@x=.5\pgf@x
	}
}

\pgfaddtoshape{basic}{
	\anchorlet{c}{center}		
	\anchorlet{n}{north}
	\anchorlet{e}{east}
	\anchorlet{s}{south}
	\anchorlet{w}{west}			
	\anchorlet{se}{south east}
	\anchorlet{sw}{south west}
	\anchorlet{ne}{north east}
	\anchorlet{nw}{north west}
	\anchorlet{wsw}{west south west}
	\anchorlet{wnw}{west north west}
	\anchorlet{ene}{east north east}
	\anchorlet{ese}{east south east}
	\anchorlet{nnw}{north north west}
	\anchorlet{nne}{north north east}
	\anchorlet{ssw}{south south west}
	\anchorlet{sse}{south south east}
}	

\pgfdeclareshape{ampshape}{
	\savedmacro\direction{
		\edef\direction{\pgfkeysvalueof{/tikz/ampkeys/direction}}%
	}
	\saveddimen\minwidth{
		\pgfmathsetlength\pgf@x{\pgfshapeminwidth}%
	}
	\saveddimen\minheight{
		\pgfmathsetlength\pgf@x{\pgfshapeminheight}%
	}
	\inheritsavedanchors[from=basic] 
	
	\inheritanchor[from=basic]{center}
	\inheritanchor[from=basic]{mid}		
	\inheritanchor[from=basic]{base}	
	\inheritanchor[from=basic]{north}
	\inheritanchor[from=basic]{south}		
	\inheritanchor[from=basic]{west}		
	\inheritanchor[from=basic]{mid west}				
	\inheritanchor[from=basic]{base west}		
	\inheritanchor[from=basic]{north west}		
	\inheritanchor[from=basic]{south west}		
	\inheritanchor[from=basic]{east}
	\inheritanchor[from=basic]{mid east}
	\inheritanchor[from=basic]{base east}	
	\inheritanchor[from=basic]{north east}
	\inheritanchor[from=basic]{south east}
	\inheritanchor[from=basic]{west south west}%
	\inheritanchor[from=basic]{west north west}%
	\inheritanchor[from=basic]{east north east}%
	\inheritanchor[from=basic]{east south east}%
	\inheritanchor[from=basic]{north north west}%
	\inheritanchor[from=basic]{north north east}%
	\inheritanchor[from=basic]{south south west}%
	\inheritanchor[from=basic]{south south east}%
	
	\inheritanchorborder[from=basic]	
	\inheritbackgroundpath[from=basic]

    \pgfutil@g@addto@macro\pgf@sh@s@ampshape{%
        \pgfutil@ifundefined{pgf@anchor@ampshape@in0}{
	        \expandafter\xdef\csname pgf@anchor@ampshape@in0\endcsname{%
	            \noexpand\ampshape@port{0}
	        }%
	    }{}%
        \pgfutil@ifundefined{pgf@anchor@ampshape@in}{
	        \expandafter\xdef\csname pgf@anchor@ampshape@in\endcsname{%
	            \noexpand\ampshape@port{0}
	        }%
	    }{}%
        \pgfutil@ifundefined{pgf@anchor@ampshape@out0}{
	        \expandafter\xdef\csname pgf@anchor@ampshape@out0\endcsname{%
	            \noexpand\ampshape@port{1}
	        }%
	    }{}%
        \pgfutil@ifundefined{pgf@anchor@ampshape@out}{
	        \expandafter\xdef\csname pgf@anchor@ampshape@out\endcsname{%
	            \noexpand\ampshape@port{1}
	        }%
	    }{}%
	}
}

\def\ampshape@port#1{
    \northeast	

    \ifnum#1=0	
	    \if\direction\direce
			\pgf@x=-\pgf@x
		    \pgf@ya= \pgf@y
		    \pgfmathsetlength{\pgf@y}{\pgf@ya-0.5*\minheight}%
		\fi
	    \if\direction\direcw
			\pgf@x=\pgf@x
		    \pgf@ya= \pgf@y
		    \pgfmathsetlength{\pgf@y}{\pgf@ya-0.5*\minheight}%
		\fi
	    \if\direction\direcn
			\pgf@y=-\pgf@y
		    \pgf@xa=\pgf@x
		    \pgfmathsetlength{\pgf@x}{\pgf@xa-0.5*\minwidth}%
		\fi
	    \if\direction\direcs
			\pgf@y=\pgf@y
		    \pgf@xa= \pgf@x
		    \pgfmathsetlength{\pgf@x}{\pgf@xa-0.5*\minwidth}%
		\fi
	\else	
	    \if\direction\direce
			\pgf@x=\pgf@x
		    \pgf@ya= \pgf@y
		    \pgfmathsetlength{\pgf@y}{\pgf@ya-0.5*\minheight}%
		\fi
	    \if\direction\direcw
			\pgf@x=-\pgf@x
		    \pgf@ya= \pgf@y
		    \pgfmathsetlength{\pgf@y}{\pgf@ya-0.5*\minheight}%
		\fi
	    \if\direction\direcn
			\pgf@y=\pgf@y
		    \pgf@xa= \pgf@x
		    \pgfmathsetlength{\pgf@x}{\pgf@xa-0.5*\minwidth}%
		\fi
	    \if\direction\direcs
			\pgf@y=-\pgf@y
		    \pgf@xa= \pgf@x
		    \pgfmathsetlength{\pgf@x}{\pgf@xa-0.5*\minwidth}%
		\fi
	\fi
}

\pgfaddtoshape{ampshape}{
	\anchorlet{c}{center}		
	\anchorlet{n}{north}
	\anchorlet{e}{east}
	\anchorlet{s}{south}
	\anchorlet{w}{west}			
	\anchorlet{se}{south east}
	\anchorlet{sw}{south west}
	\anchorlet{ne}{north east}
	\anchorlet{nw}{north west}
	\anchorlet{wsw}{west south west}
	\anchorlet{wnw}{west north west}
	\anchorlet{ene}{east north east}
	\anchorlet{ese}{east south east}
	\anchorlet{nnw}{north north west}
	\anchorlet{nne}{north north east}
	\anchorlet{ssw}{south south west}
	\anchorlet{sse}{south south east}
}

\tikzset{
	/tikz/ampkeys/.cd,
	height/.initial=0.5,
	width/.initial=0.5,
	color/.initial=O,
	direction/.initial=e,
	linestyle/.initial={linestyle, rounded corners = 0},
	/tikz/amp/.code={
		\pgfqkeys{/tikz/ampkeys}{#1}%
		\tikzset{/tikz/ampkeys/drawer/.expanded=%
			{\pgfkeysvalueof{/tikz/ampkeys/direction}}%
			{\pgfkeysvalueof{/tikz/ampkeys/height}}%
			{\pgfkeysvalueof{/tikz/ampkeys/width}}%
			{\pgfkeysvalueof{/tikz/ampkeys/color}}%
			{\pgfkeysvalueof{/tikz/ampkeys/linestyle}}%
		}
	},
	/tikz/ampkeys/drawer/.code n args={5}{%
		\tikzset{
			ampshape,
			minimum height=#2*\NODESIZE,
			minimum width=#3*\NODESIZE,
			append after command={
				\pgfextra{\let\bdr=\tikzlastnode%
				\if#1e
					\draw[draw=#4, fill=#4l, #5] (\bdr.sw) to (\bdr.nw) to (\bdr.e) to cycle {};
				\fi
				\if#1w
					\draw[draw=#4, fill=#4l, #5] (\bdr.se) to (\bdr.ne) to (\bdr.w) to cycle {};
				\fi
				\if#1n
					\draw[draw=#4, fill=#4l, #5] (\bdr.se) to (\bdr.sw) to (\bdr.n) to cycle {};
				\fi
				\if#1s
					\draw[draw=#4, fill=#4l, #5] (\bdr.ne) to (\bdr.nw) to (\bdr.s) to cycle {};
				\fi
				}
			}
		}
	},
}
\newcount\portcount

\pgfdeclareshape{aomshape}{
	\savedmacro\direction{
		\edef\direction{\pgfkeysvalueof{/tikz/aomkeys/direction}}%
	}
	\saveddimen\minwidth{
		\pgfmathsetlength\pgf@x{\pgfshapeminwidth}%
	}
	\saveddimen\minheight{
		\pgfmathsetlength\pgf@x{\pgfshapeminheight}%
	}
	\inheritsavedanchors[from=basic]

	\inheritanchor[from=basic]{center}
	\inheritanchor[from=basic]{mid}
	\inheritanchor[from=basic]{base}
	\inheritanchor[from=basic]{north}
	\inheritanchor[from=basic]{south}
	\inheritanchor[from=basic]{west}
	\inheritanchor[from=basic]{mid west}
	\inheritanchor[from=basic]{base west}
	\inheritanchor[from=basic]{north west}
	\inheritanchor[from=basic]{south west}
	\inheritanchor[from=basic]{east}
	\inheritanchor[from=basic]{mid east}
	\inheritanchor[from=basic]{base east}
	\inheritanchor[from=basic]{north east}
	\inheritanchor[from=basic]{south east}
	\inheritanchor[from=basic]{west south west}%
	\inheritanchor[from=basic]{west north west}%
	\inheritanchor[from=basic]{east north east}%
	\inheritanchor[from=basic]{east south east}%
	\inheritanchor[from=basic]{north north west}%
	\inheritanchor[from=basic]{north north east}%
	\inheritanchor[from=basic]{south south west}%
	\inheritanchor[from=basic]{south south east}%

	\inheritanchorborder[from=basic]
	\inheritbackgroundpath[from=basic]

	\pgfutil@g@addto@macro\pgf@sh@s@aomshape{%
		\pgfutil@ifundefined{pgf@anchor@aomshape@in0}{
			\expandafter\xdef\csname pgf@anchor@aomshape@in0\endcsname{%
				\noexpand\aomshape@port{0}
			}%
		}{}%
		\pgfutil@ifundefined{pgf@anchor@aomshape@in}{
			\expandafter\xdef\csname pgf@anchor@aomshape@in\endcsname{%
				\noexpand\aomshape@port{0}
			}%
		}{}%
		\pgfutil@ifundefined{pgf@anchor@aomshape@out0}{
			\expandafter\xdef\csname pgf@anchor@aomshape@out0\endcsname{%
				\noexpand\aomshape@port{1}
			}%
		}{}%
		\pgfutil@ifundefined{pgf@anchor@aomshape@out}{
			\expandafter\xdef\csname pgf@anchor@aomshape@out\endcsname{%
				\noexpand\aomshape@port{1}
			}%
		}{}%
	}
}

\def\aomshape@port#1{
	\northeast	

	\ifnum#1=0	
		\if\direction\direce
			\pgf@x=-\pgf@x
			\pgf@ya= \pgf@y
			\pgfmathsetlength{\pgf@y}{\pgf@ya-0.5*\minheight}%
		\fi
		\if\direction\direcw
			\pgf@x=\pgf@x
			\pgf@ya= \pgf@y
			\pgfmathsetlength{\pgf@y}{\pgf@ya-0.5*\minheight}%
		\fi
		\if\direction\direcn
			\pgf@y=-\pgf@y
			\pgf@xa=\pgf@x
			\pgfmathsetlength{\pgf@x}{\pgf@xa-0.5*\minwidth}%
		\fi
		\if\direction\direcs
			\pgf@y=\pgf@y
			\pgf@xa= \pgf@x
			\pgfmathsetlength{\pgf@x}{\pgf@xa-0.5*\minwidth}%
		\fi
	\else	
		\if\direction\direce
			\pgf@x=\pgf@x
			\pgf@ya= \pgf@y
			\pgfmathsetlength{\pgf@y}{\pgf@ya-0.5*\minheight}%
		\fi
		\if\direction\direcw
			\pgf@x=-\pgf@x
			\pgf@ya= \pgf@y
			\pgfmathsetlength{\pgf@y}{\pgf@ya-0.5*\minheight}%
		\fi
		\if\direction\direcn
			\pgf@y=\pgf@y
			\pgf@xa= \pgf@x
			\pgfmathsetlength{\pgf@x}{\pgf@xa-0.5*\minwidth}%
		\fi
		\if\direction\direcs
			\pgf@y=-\pgf@y
			\pgf@xa= \pgf@x
			\pgfmathsetlength{\pgf@x}{\pgf@xa-0.5*\minwidth}%
		\fi
	\fi
}

\pgfaddtoshape{aomshape}{
	\anchorlet{c}{center}
	\anchorlet{n}{north}
	\anchorlet{e}{east}
	\anchorlet{s}{south}
	\anchorlet{w}{west}
	\anchorlet{se}{south east}
	\anchorlet{sw}{south west}
	\anchorlet{ne}{north east}
	\anchorlet{nw}{north west}
	\anchorlet{wsw}{west south west}
	\anchorlet{wnw}{west north west}
	\anchorlet{ene}{east north east}
	\anchorlet{ese}{east south east}
	\anchorlet{nnw}{north north west}
	\anchorlet{nne}{north north east}
	\anchorlet{ssw}{south south west}
	\anchorlet{sse}{south south east}
}

\tikzset{
/tikz/aomkeys/.cd,
size/.initial=1,
circlesize/.initial=1,
color/.initial=O,
direction/.initial=e,
linestyle/.initial={linestyle, inner sep=0.5mm},
fillgradient/.initial=O,
/tikz/aom/.code={
\pgfqkeys{/tikz/aomkeys}{#1}%
\tikzset{/tikz/aomkeys/drawer/.expanded=%
	{\pgfkeysvalueof{/tikz/aomkeys/size}}%
	{\pgfkeysvalueof{/tikz/aomkeys/color}}%
	{\pgfkeysvalueof{/tikz/aomkeys/linestyle}}%
\if\pgfkeysvalueof{/tikz/aomkeys/direction}e
	{0}%
\fi
\if\pgfkeysvalueof{/tikz/aomkeys/direction}w
	{0}%
\fi
\if\pgfkeysvalueof{/tikz/aomkeys/direction}n
	{1}%
\fi
\if\pgfkeysvalueof{/tikz/aomkeys/direction}s
	{1}%
\fi
{\pgfkeysvalueof{/tikz/aomkeys/direction}}%
{\pgfkeysvalueof{/tikz/aomkeys/circlesize}}%
{\pgfkeysvalueof{/tikz/aomkeys/fillgradient}}%
}
},
/tikz/aomkeys/drawer/.code n args={7}{%
		\tikzset{
			aomshape,
			draw,
			minimum height = #1*\NODESIZE,
			minimum width = #1*\NODESIZE,
			#2,
			#3,
			append after command={
					\pgfextra{\let\bdr=\tikzlastnode%
						\node[#7, fit=(\bdr.nw)(\bdr.se)] (boxgradient){};

						\node[coordinate] at ($(\bdr.in)!0.25!(\bdr.out)$) (circlein){};
						\node[coordinate] at ($(\bdr.in)!0.75!(\bdr.out)$) (circleout){};

						\ifnum#4>0
							\node[coordinate] at (circleout -| \bdr.nne) (circleouttop){};
						\else
							\node[coordinate] at (circleout |- \bdr.ene) (circleouttop){};
						\fi

						\draw[---, #2, #3, fill] (\bdr.in) to (circlein) circle (0.05*#6);
						\draw[---, #2, #3, fill] (\bdr.out) to (circleout) circle (0.05*#6);

						\draw[---, #2, #3] (circlein) to (circleouttop){};

					}
				}
		}
	},
}
\newcount\portcount

\pgfdeclareshape{boxshape}{
	\savedmacro\nin{
		\edef\nin{\pgfkeysvalueof{/tikz/boxkeys/nin}}%
	}
	\savedmacro\nout{
		\edef\nout{\pgfkeysvalueof{/tikz/boxkeys/nout}}%
	}
	\savedmacro\direction{
		\edef\direction{\pgfkeysvalueof{/tikz/boxkeys/direction}}%
	}
	\inheritsavedanchors[from=basic]

	\inheritanchor[from=basic]{center}
	\inheritanchor[from=basic]{mid}
	\inheritanchor[from=basic]{base}
	\inheritanchor[from=basic]{north}
	\inheritanchor[from=basic]{south}
	\inheritanchor[from=basic]{west}
	\inheritanchor[from=basic]{mid west}
	\inheritanchor[from=basic]{base west}
	\inheritanchor[from=basic]{north west}
	\inheritanchor[from=basic]{south west}
	\inheritanchor[from=basic]{east}
	\inheritanchor[from=basic]{mid east}
	\inheritanchor[from=basic]{base east}
	\inheritanchor[from=basic]{north east}
	\inheritanchor[from=basic]{south east}
	\inheritanchor[from=basic]{west south west}%
	\inheritanchor[from=basic]{west north west}%
	\inheritanchor[from=basic]{east north east}%
	\inheritanchor[from=basic]{east south east}%
	\inheritanchor[from=basic]{north north west}%
	\inheritanchor[from=basic]{north north east}%
	\inheritanchor[from=basic]{south south west}%
	\inheritanchor[from=basic]{south south east}%

	\inheritanchorborder[from=basic]
	\inheritbackgroundpath[from=basic]

	\pgfutil@g@addto@macro\pgf@sh@s@boxshape{%
		\pgfmathsetcount{\portcount}{0}
		\pgfmathloop%
		\ifnum\the\portcount<\nin
		\pgfutil@ifundefined{pgf@anchor@boxshape@in\the\portcount}{
			\expandafter\xdef\csname pgf@anchor@boxshape@in\the\portcount\endcsname{%
				\noexpand\boxshape@port[\the\portcount]{0}
			}%
		}{}%
		\ifnum\the\portcount=0
			\pgfutil@ifundefined{pgf@anchor@boxshape@in}{%
				\expandafter\xdef\csname pgf@anchor@boxshape@in\endcsname{%
					\noexpand\boxshape@port[\the\portcount]{0}
				}%
			}{}%
		\fi
		\pgfmathaddtocount{\portcount}{1}	
		\repeatpgfmathloop
		\pgfmathsetcount{\portcount}{0}
		\pgfmathloop%
		\ifnum\the\portcount<\nout
		\pgfutil@ifundefined{pgf@anchor@boxshape@out\the\portcount}{%
			\expandafter\xdef\csname pgf@anchor@boxshape@out\the\portcount\endcsname{%
				\noexpand\boxshape@port[\the\portcount]{1}
			}%
		}{}%
		\ifnum\the\portcount=0
			\pgfutil@ifundefined{pgf@anchor@boxshape@out}{%
				\expandafter\xdef\csname pgf@anchor@boxshape@out\endcsname{%
					\noexpand\boxshape@port[\the\portcount]{1}
				}%
			}{}%
		\fi
		\pgfmathaddtocount{\portcount}{1}	
		\repeatpgfmathloop
	}
}

\def\boxshape@port[#1]#2{
	\northeast \pgf@xa=\pgf@x \pgf@ya=\pgf@y
	\southwest \pgf@xb=\pgf@x \pgf@yb=\pgf@y

	\ifnum#2=0	
		\if\direction\direce	
			\pgf@x=\pgf@xb
			\pgf@yc=\pgf@ya \advance\pgf@yc by -\pgf@yb	
			\pgfmathsetlength{\pgf@y}{\pgf@ya-(#1 + 0.5)*(\pgf@yc/\nin)}%
		\fi
		\if\direction\direcw
			\pgf@x=\pgf@xa
			\pgf@yc=\pgf@ya \advance\pgf@yc by -\pgf@yb	
			\pgfmathsetlength{\pgf@y}{\pgf@ya-(#1 + 0.5)*(\pgf@yc/\nin)}%
		\fi
		\if\direction\direcn
			\pgf@y=\pgf@yb
			\pgf@xc=\pgf@xa \advance\pgf@xc by -\pgf@xb	
			\pgfmathsetlength{\pgf@x}{\pgf@xb+(#1 + 0.5)*(\pgf@xc/\nin)}%
		\fi
		\if\direction\direcs
			\pgf@y=\pgf@ya
			\pgf@xc=\pgf@xa \advance\pgf@xc by -\pgf@xb	
			\pgfmathsetlength{\pgf@x}{\pgf@xb+(#1 + 0.5)*(\pgf@xc/\nin)}%
		\fi
	\else	
		\if\direction\direce	
			\pgf@x=\pgf@xa
			\pgf@yc=\pgf@ya \advance\pgf@yc by -\pgf@yb	
			\pgfmathsetlength{\pgf@y}{\pgf@ya-(#1 + 0.5)*(\pgf@yc/\nout)}%
		\fi
		\if\direction\direcw
			\pgf@x=\pgf@xb
			\pgf@yc=\pgf@ya \advance\pgf@yc by -\pgf@yb	
			\pgfmathsetlength{\pgf@y}{\pgf@ya-(#1 + 0.5)*(\pgf@yc/\nout)}%
		\fi
		\if\direction\direcn
			\pgf@y=\pgf@ya
			\pgf@xc=\pgf@xa \advance\pgf@xc by -\pgf@xb	
			\pgfmathsetlength{\pgf@x}{\pgf@xb+(#1 + 0.5)*(\pgf@xc/\nout)}%
		\fi
		\if\direction\direcs
			\pgf@y=\pgf@yb
			\pgf@xc=\pgf@xa \advance\pgf@xc by -\pgf@xb	
			\pgfmathsetlength{\pgf@x}{\pgf@xb+(#1 + 0.5)*(\pgf@xc/\nout)}%
		\fi
	\fi
}

\pgfaddtoshape{boxshape}{
	\anchorlet{c}{center}
	\anchorlet{n}{north}
	\anchorlet{e}{east}
	\anchorlet{s}{south}
	\anchorlet{w}{west}
	\anchorlet{se}{south east}
	\anchorlet{sw}{south west}
	\anchorlet{ne}{north east}
	\anchorlet{nw}{north west}
	\anchorlet{wsw}{west south west}
	\anchorlet{wnw}{west north west}
	\anchorlet{ene}{east north east}
	\anchorlet{ese}{east south east}
	\anchorlet{nnw}{north north west}
	\anchorlet{nne}{north north east}
	\anchorlet{ssw}{south south west}
	\anchorlet{sse}{south south east}
}

\tikzset{
/tikz/boxkeys/.cd,
height/.initial=0.5,
width/.initial=1,
color/.initial=O,
direction/.initial=e,
linestyle/.initial={linestyle, inner sep=0.5mm},
nin/.initial=1,
nout/.initial=1,
draw/.initial=1,
/tikz/box/.code={
\pgfqkeys{/tikz/boxkeys}{#1}%
\tikzset{/tikz/boxkeys/drawer/.expanded=%
\if\pgfkeysvalueof{/tikz/boxkeys/direction}e
	{\pgfkeysvalueof{/tikz/boxkeys/width}}%
	{\pgfkeysvalueof{/tikz/boxkeys/height}}%
	{0}%
	{-90}%
\fi
\if\pgfkeysvalueof{/tikz/boxkeys/direction}w
	{\pgfkeysvalueof{/tikz/boxkeys/width}}%
	{\pgfkeysvalueof{/tikz/boxkeys/height}}%
	{0}%
	{90}%
\fi
\if\pgfkeysvalueof{/tikz/boxkeys/direction}n
	{\pgfkeysvalueof{/tikz/boxkeys/width}}%
	{\pgfkeysvalueof{/tikz/boxkeys/height}}%
	{1}%
	{0}%
\fi
\if\pgfkeysvalueof{/tikz/boxkeys/direction}s
	{\pgfkeysvalueof{/tikz/boxkeys/width}}%
	{\pgfkeysvalueof{/tikz/boxkeys/height}}%
	{1}%
	{180}%
\fi
{\pgfkeysvalueof{/tikz/boxkeys/color}}%
{\pgfkeysvalueof{/tikz/boxkeys/linestyle}}%
\ifnum\pgfkeysvalueof{/tikz/boxkeys/draw}>0%
	{draw}%
\else
	{}
\fi
}
},
/tikz/boxkeys/drawer/.code n args={7}{%
		\tikzset{
			boxshape,
			#7,
			#6,
			#5,
			minimum height=
			\ifnum#3>0	
				#1*\NODESIZE
			\else
				#2*\NODESIZE
			\fi
			,minimum width=
			\ifnum#3>0
				#2*\NODESIZE
			\else
				#1*\NODESIZE
			\fi
		}
	},
}
\pgfdeclareshape{beshape}{ 
    \savedmacro\nports{
        \edef\nports{\pgfkeysvalueof{/tikz/bekeys/nports}}%
    }
    \savedmacro\direction{
        \edef\direction{\pgfkeysvalueof{/tikz/bekeys/direction}}%
    }
    \savedmacro\inverted{
        \edef\inverted{\pgfkeysvalueof{/tikz/bekeys/inverted}}%
    }
    \savedmacro\ninports{
        \ifnum\inverted=0
        \edef\ninports{\nports}
        \else
        \edef\ninports{1}%
        \fi
    }
    \savedmacro\noutports{
        \ifnum\inverted=0
        \edef\noutports{1}%
        \else
        \edef\noutports{\nports}%
        \fi
    }
    \inheritsavedanchors[from=basic]

    \inheritanchor[from=basic]{center}
    \inheritanchor[from=basic]{mid}
    \inheritanchor[from=basic]{base}
    \inheritanchor[from=basic]{north}
    \inheritanchor[from=basic]{south}
    \inheritanchor[from=basic]{west}
    \inheritanchor[from=basic]{mid west}
    \inheritanchor[from=basic]{base west}
    \inheritanchor[from=basic]{north west}
    \inheritanchor[from=basic]{south west}
    \inheritanchor[from=basic]{east}
    \inheritanchor[from=basic]{mid east}
    \inheritanchor[from=basic]{base east}
    \inheritanchor[from=basic]{north east}
    \inheritanchor[from=basic]{south east}
    \inheritanchor[from=basic]{west south west}%
    \inheritanchor[from=basic]{west north west}%
    \inheritanchor[from=basic]{east north east}%
    \inheritanchor[from=basic]{east south east}%
    \inheritanchor[from=basic]{north north west}%
    \inheritanchor[from=basic]{north north east}%
    \inheritanchor[from=basic]{south south west}%
    \inheritanchor[from=basic]{south south east}%

    \inheritanchorborder[from=basic]
    \inheritbackgroundpath[from=basic]

    \pgfutil@g@addto@macro\pgf@sh@s@beshape{%
        \pgfmathsetcount{\portcount}{0}
        \pgfmathloop%
        \ifnum\the\portcount<\nports
        \ifnum\the\portcount<\ninports
        \pgfutil@ifundefined{pgf@anchor@beshape@in\the\portcount}{
            \expandafter\xdef\csname pgf@anchor@beshape@in\the\portcount\endcsname{%
                \noexpand\beshape@port[\the\portcount]{0}
            }%
        }{}%
        \ifnum\the\portcount=0
        \pgfutil@ifundefined{pgf@anchor@beshape@in}{%
            \expandafter\xdef\csname pgf@anchor@beshape@in\endcsname{%
                \noexpand\beshape@port[\the\portcount]{0}
            }%
        }{}%
        \fi
        \fi
        \ifnum\the\portcount<\noutports
        \pgfutil@ifundefined{pgf@anchor@beshape@out\the\portcount}{%
            \expandafter\xdef\csname pgf@anchor@beshape@out\the\portcount\endcsname{%
                \noexpand\beshape@port[\the\portcount]{1}
            }%
        }{}%
        \ifnum\the\portcount=0
        \pgfutil@ifundefined{pgf@anchor@beshape@out}{%
            \expandafter\xdef\csname pgf@anchor@beshape@out\endcsname{%
                \noexpand\beshape@port[\the\portcount]{1}
            }%
        }{}%
        \fi
        \fi
        \pgfmathaddtocount{\portcount}{1}    
        \repeatpgfmathloop
    }
}

\def\beshape@port[#1]#2{
    \northeast \pgf@xa=\pgf@x \pgf@ya=\pgf@y
    \southwest \pgf@xb=\pgf@x \pgf@yb=\pgf@y

    \ifnum#2=0
    \ifnum\inverted=0
    \def\chooseports{0}
    \else
    \def\chooseports{1}
    \fi
    \else
    \ifnum\inverted=0
    \def\chooseports{1}
    \else
    \def\chooseports{0}
    \fi
    \fi

    \ifnum\chooseports=0    
    \if\direction\direce
    \pgf@x=\pgf@xb
    \pgf@yc=\pgf@ya \advance\pgf@yc by -\pgf@yb    
    \pgfmathsetlength{\pgf@y}{\pgf@ya-(#1 + 0.5)*(\pgf@yc/\nports)}%
    \fi
    \if\direction\direcw
    \pgf@x=\pgf@xa
    \pgf@yc=\pgf@ya \advance\pgf@yc by -\pgf@yb    
    \pgfmathsetlength{\pgf@y}{\pgf@ya-(#1 + 0.5)*(\pgf@yc/\nports)}%
    \fi
    \if\direction\direcn
    \pgf@y=\pgf@yb
    \pgf@xc=\pgf@xa \advance\pgf@xc by -\pgf@xb    
    \pgfmathsetlength{\pgf@x}{\pgf@xb+(#1 + 0.5)*(\pgf@xc/\nports)}%
    \fi
    \if\direction\direcs
    \pgf@y=\pgf@ya
    \pgf@xc=\pgf@xa \advance\pgf@xc by -\pgf@xb    
    \pgfmathsetlength{\pgf@x}{\pgf@xb+(#1 + 0.5)*(\pgf@xc/\nports)}%
    \fi
    \else    
    \if\direction\direce
    \pgf@x=\pgf@xa
    \pgf@yc=\pgf@ya \advance\pgf@yc by -\pgf@yb    
    \pgfmathsetlength{\pgf@y}{\pgf@ya-0.5\pgf@yc}%
    \fi
    \if\direction\direcw
    \pgf@x=\pgf@xb
    \pgf@yc=\pgf@ya \advance\pgf@yc by -\pgf@yb    
    \pgfmathsetlength{\pgf@y}{\pgf@ya-0.5\pgf@yc}%
    \fi
    \if\direction\direcn
    \pgf@y=\pgf@ya
    \pgf@xc=\pgf@xa \advance\pgf@xc by -\pgf@xb    
    \pgfmathsetlength{\pgf@x}{\pgf@xa-0.5\pgf@xc}%
    \fi
    \if\direction\direcs
    \pgf@y=\pgf@yb
    \pgf@xc=\pgf@xa \advance\pgf@xc by -\pgf@xb    
    \pgfmathsetlength{\pgf@x}{\pgf@xa-0.5\pgf@xc}%
    \fi
    \fi
}

\pgfaddtoshape{beshape}{
    \anchorlet{c}{center}
    \anchorlet{n}{north}
    \anchorlet{e}{east}
    \anchorlet{s}{south}
    \anchorlet{w}{west}
    \anchorlet{se}{south east}
    \anchorlet{sw}{south west}
    \anchorlet{ne}{north east}
    \anchorlet{nw}{north west}
    \anchorlet{wsw}{west south west}
    \anchorlet{wnw}{west north west}
    \anchorlet{ene}{east north east}
    \anchorlet{ese}{east south east}
    \anchorlet{nnw}{north north west}
    \anchorlet{nne}{north north east}
    \anchorlet{ssw}{south south west}
    \anchorlet{sse}{south south east}
}

\tikzset{
    /tikz/bekeys/.cd,
    height/.initial=1,
    width/.initial=0.5,
    color/.initial=O,
    direction/.initial=e,
    linestyle/.initial={linestyle, rounded corners = 0},
    nports/.initial=3,
    inverted/.initial=0,    
    sbe/.initial=0,
    angle/.initial=60,
    /tikz/be/.code={
        \pgfqkeys{/tikz/bekeys}{#1}%
        \tikzset{/tikz/bekeys/drawer/.expanded=%
            \if\pgfkeysvalueof{/tikz/bekeys/direction}e
                {\pgfkeysvalueof{/tikz/bekeys/width}}%
                {\pgfkeysvalueof{/tikz/bekeys/height}}%
                {0}%
                {-90}%
            \fi
            \if\pgfkeysvalueof{/tikz/bekeys/direction}w
                {\pgfkeysvalueof{/tikz/bekeys/width}}%
                {\pgfkeysvalueof{/tikz/bekeys/height}}%
                {0}%
                {90}%
            \fi
            \if\pgfkeysvalueof{/tikz/bekeys/direction}n
                {\pgfkeysvalueof{/tikz/bekeys/width}}%
                {\pgfkeysvalueof{/tikz/bekeys/height}}%
                {1}%
                {0}%
            \fi
            \if\pgfkeysvalueof{/tikz/bekeys/direction}s
                {\pgfkeysvalueof{/tikz/bekeys/width}}%
                {\pgfkeysvalueof{/tikz/bekeys/height}}%
                {1}%
                {180}%
            \fi
            {\pgfkeysvalueof{/tikz/bekeys/color}}%
            {\pgfkeysvalueof{/tikz/bekeys/linestyle}}%
            {\pgfkeysvalueof{/tikz/bekeys/sbe}}%
            {\pgfkeysvalueof{/tikz/bekeys/angle}}%
        }
    },
    /tikz/bekeys/drawer/.code n args={8}{%
        \tikzset{
            beshape,
            #6,
            #5,
            minimum height=
            \ifnum#3>0    
            #1*\NODESIZE
            \else
            #2*\NODESIZE
            \fi
            ,minimum width=
            \ifnum#3>0
            #2*\NODESIZE
            \else
            #1*\NODESIZE
            \fi
            ,append after command={
                \pgfextra{
                    \let\bdr=\tikzlastnode%
                    \node[trapezium, line width = \NODETHICKNESS, minimum height=#1*\NODESIZE, minimum width=#2*\NODESIZE, trapezium stretches=true, rotate=#4, trapezium angle=70, inner sep=0.001mm, #5, #6] at (\bdr) (trap) {};

                    \node[rectangle, line width = \NODETHICKNESS, minimum height=#1*\NODESIZE, minimum width=#2*\NODESIZE, anchor=north, rotate=#4, #5, #6] at (trap.south) (r1) {};

                    \tikzmath{coordinate \C;
                    \C = (trap.top left corner)-(trap.top right corner);
                    \distAB = sqrt((\Cx)^2+(\Cy)^2);
                    }

                    \node[rectangle, line width = \NODETHICKNESS, minimum height=#1*\NODESIZE, minimum width=\distAB, anchor=south, rotate=#4, #5, #6, red] at (trap.north) (r2) {};

					\draw[#5, #6] (trap.top left corner) to (r2.north west) to (r2.north east) to (trap.top right corner) to (trap.bottom right corner) to (r1.south east) to (r1.south west) to (r1.north west) to cycle;

                }
            }
        }
    },
}
\pgfdeclareshape{couplershape}{	

	\inheritsavedanchors[from=basic] 

	\inheritanchor[from=basic]{center}
	\inheritanchor[from=basic]{mid}		
	\inheritanchor[from=basic]{base}	
	\inheritanchor[from=basic]{north}
	\inheritanchor[from=basic]{south}		
	\inheritanchor[from=basic]{west}		
	\inheritanchor[from=basic]{mid west}				
	\inheritanchor[from=basic]{base west}		
	\inheritanchor[from=basic]{north west}		
	\inheritanchor[from=basic]{south west}		
	\inheritanchor[from=basic]{east}
	\inheritanchor[from=basic]{mid east}
	\inheritanchor[from=basic]{base east}	
	\inheritanchor[from=basic]{north east}
	\inheritanchor[from=basic]{south east}
	\inheritanchor[from=basic]{west south west}%
	\inheritanchor[from=basic]{west north west}%
	\inheritanchor[from=basic]{east north east}%
	\inheritanchor[from=basic]{east south east}%
	\inheritanchor[from=basic]{north north west}%
	\inheritanchor[from=basic]{north north east}%
	\inheritanchor[from=basic]{south south west}%
	\inheritanchor[from=basic]{south south east}%
	
	\inheritanchorborder[from=basic]	
	\inheritbackgroundpath[from=basic]
}

\pgfaddtoshape{couplershape}{
	\anchorlet{in}{center}		
	\anchorlet{in0}{center}
	\anchorlet{out}{center}
	\anchorlet{out0}{center}
	\anchorlet{c}{center}		
	\anchorlet{n}{north}
	\anchorlet{e}{east}
	\anchorlet{s}{south}
	\anchorlet{w}{west}			
	\anchorlet{se}{south east}
	\anchorlet{sw}{south west}
	\anchorlet{ne}{north east}
	\anchorlet{nw}{north west}
	\anchorlet{wsw}{west south west}
	\anchorlet{wnw}{west north west}
	\anchorlet{ene}{east north east}
	\anchorlet{ese}{east south east}
	\anchorlet{nnw}{north north west}
	\anchorlet{nne}{north north east}
	\anchorlet{ssw}{south south west}
	\anchorlet{sse}{south south east}
}

\tikzset{
	/tikz/couplerkeys/.cd,
	size/.initial=0.2,
	color/.initial=O,
	rotation/.initial=0,
	heightwidthratio/.initial=0.5,
	/tikz/coupler/.code={
		\pgfqkeys{/tikz/couplerkeys}{#1}%
		\tikzset{/tikz/couplerkeys/drawer/.expanded=%
			{\pgfkeysvalueof{/tikz/couplerkeys/size}}%
			{\pgfkeysvalueof{/tikz/couplerkeys/color}}%
			{\pgfkeysvalueof{/tikz/couplerkeys/rotation}}%
			{\pgfkeysvalueof{/tikz/couplerkeys/heightwidthratio}}%
		}
	},
	/tikz/couplerkeys/drawer/.code n args={4}{%
		\tikzset{
			couplershape,
			minimum height=#1*\NODESIZE
			\ifnum#3<1
				\ifnum#3>-1
					*#4
				\fi
			\fi
			,minimum width=#1*\NODESIZE
			\ifnum#3<91
				\ifnum#3>89
					*#4
				\fi
			\fi
			\ifnum#3<-89
				\ifnum#3>-91
					*#4
				\fi
			\fi
			,#2,
			append after command={
				\pgfextra{\let\bdr=\tikzlastnode%
				\node[ellipse, fill, #2, rotate=#3, outer sep = 0, minimum width=#1*\NODESIZE, minimum height=#1*#4*\NODESIZE] at (\bdr.center){};
				}
			}
		}
	},
}
\pgfdeclareshape{fibershape}{
	\savedmacro\direction{
		\edef\direction{\pgfkeysvalueof{/tikz/fiberkeys/direction}}%
	}
	\savedmacro\flip{
		\edef\flip{\pgfkeysvalueof{/tikz/fiberkeys/flip}}%
	}
	\saveddimen\minwidth{
		\pgfmathsetlength\pgf@x{\pgfshapeminwidth}%
	}
	\saveddimen\minheight{
		\pgfmathsetlength\pgf@x{\pgfshapeminheight}%
	}
	\inheritsavedanchors[from=basic] 
	
	\inheritanchor[from=basic]{center}
	\inheritanchor[from=basic]{mid}		
	\inheritanchor[from=basic]{base}	
	\inheritanchor[from=basic]{north}
	\inheritanchor[from=basic]{south}		
	\inheritanchor[from=basic]{west}		
	\inheritanchor[from=basic]{mid west}				
	\inheritanchor[from=basic]{base west}		
	\inheritanchor[from=basic]{north west}		
	\inheritanchor[from=basic]{south west}		
	\inheritanchor[from=basic]{east}
	\inheritanchor[from=basic]{mid east}
	\inheritanchor[from=basic]{base east}	
	\inheritanchor[from=basic]{north east}
	\inheritanchor[from=basic]{south east}
	\inheritanchor[from=basic]{west south west}%
	\inheritanchor[from=basic]{west north west}%
	\inheritanchor[from=basic]{east north east}%
	\inheritanchor[from=basic]{east south east}%
	\inheritanchor[from=basic]{north north west}%
	\inheritanchor[from=basic]{north north east}%
	\inheritanchor[from=basic]{south south west}%
	\inheritanchor[from=basic]{south south east}%
	
	\inheritanchorborder[from=basic]	
	\inheritbackgroundpath[from=basic]

    \pgfutil@g@addto@macro\pgf@sh@s@fibershape{%
        \pgfutil@ifundefined{pgf@anchor@fibershape@in0}{
	        \expandafter\xdef\csname pgf@anchor@fibershape@in0\endcsname{%
	            \noexpand\fibershape@port{0}
	        }%
	    }{}%
        \pgfutil@ifundefined{pgf@anchor@fibershape@in}{
	        \expandafter\xdef\csname pgf@anchor@fibershape@in\endcsname{%
	            \noexpand\fibershape@port{0}
	        }%
	    }{}%
        \pgfutil@ifundefined{pgf@anchor@fibershape@out0}{
	        \expandafter\xdef\csname pgf@anchor@fibershape@out0\endcsname{%
	            \noexpand\fibershape@port{1}
	        }%
	    }{}%
        \pgfutil@ifundefined{pgf@anchor@fibershape@out}{
	        \expandafter\xdef\csname pgf@anchor@fibershape@out\endcsname{%
	            \noexpand\fibershape@port{1}
	        }%
	    }{}%
	}
}

\def\fibershape@port#1{
    \northeast	

    \ifnum#1=0	
	    \if\direction\direce
			\pgf@x=-\pgf@x
	    	\if\flip\flipfalse
		    	\pgf@y=-\pgf@y
		    \else
		    	\pgf@y=\pgf@y
		    \fi
		\fi
	    \if\direction\direcw
			\pgf@x=\pgf@x
	    	\if\flip\flipfalse
		    	\pgf@y=-\pgf@y
		    \else
		    	\pgf@y=\pgf@y
		    \fi
		\fi
	    \if\direction\direcn
			\pgf@y=-\pgf@y
	    	\if\flip\flipfalse
		    	\pgf@x=-\pgf@x
		    \else
		    	\pgf@x=\pgf@x
		    \fi
		\fi
	    \if\direction\direcs
			\pgf@y=\pgf@y
	    	\if\flip\flipfalse
		    	\pgf@x=-\pgf@x
		    \else
		    	\pgf@x=\pgf@x
		    \fi
		\fi
	\else	
	    \if\direction\direce
			\pgf@x=\pgf@x
	    	\if\flip\flipfalse
		    	\pgf@y=-\pgf@y
		    \else
		    	\pgf@y=\pgf@y
		    \fi
		\fi
	    \if\direction\direcw
			\pgf@x=-\pgf@x
	    	\if\flip\flipfalse
		    	\pgf@y=-\pgf@y
		    \else
		    	\pgf@y=\pgf@y
		    \fi
		\fi
	    \if\direction\direcn
			\pgf@y=\pgf@y
	    	\if\flip\flipfalse
		    	\pgf@x=-\pgf@x
		    \else
		    	\pgf@x=\pgf@x
		    \fi
		\fi
	    \if\direction\direcs
			\pgf@y=-\pgf@y
	    	\if\flip\flipfalse
		    	\pgf@x=-\pgf@x
		    \else
		    	\pgf@x=\pgf@x
		    \fi
		\fi
	\fi
}

\pgfaddtoshape{fibershape}{
	\anchorlet{c}{center}		
	\anchorlet{n}{north}
	\anchorlet{e}{east}
	\anchorlet{s}{south}
	\anchorlet{w}{west}			
	\anchorlet{se}{south east}
	\anchorlet{sw}{south west}
	\anchorlet{ne}{north east}
	\anchorlet{nw}{north west}
	\anchorlet{wsw}{west south west}
	\anchorlet{wnw}{west north west}
	\anchorlet{ene}{east north east}
	\anchorlet{ese}{east south east}
	\anchorlet{nnw}{north north west}
	\anchorlet{nne}{north north east}
	\anchorlet{ssw}{south south west}
	\anchorlet{sse}{south south east}
}

\tikzset{
	/tikz/fiberkeys/.cd,
	size/.initial=1,
	color/.initial=C0,
	direction/.initial=e,
	linestyle/.initial={linestyle},
	flip/.initial={0},
	drawbase/.initial={1},
	/tikz/fiber/.code={
		\pgfqkeys{/tikz/fiberkeys}{#1}%
		\tikzset{/tikz/fiberkeys/drawer/.expanded=%
			{\pgfkeysvalueof{/tikz/fiberkeys/direction}}%
			{\pgfkeysvalueof{/tikz/fiberkeys/size}}%
			{\pgfkeysvalueof{/tikz/fiberkeys/color}}%
			{\pgfkeysvalueof{/tikz/fiberkeys/linestyle}}%
			\if\pgfkeysvalueof{/tikz/fiberkeys/direction}e
				{a}%
				\if\pgfkeysvalueof{/tikz/fiberkeys/flip}0
					{south}%
				\else
					{north}%
				\fi
				{\pgfkeysvalueof{/tikz/fiberkeys/size}}
				{\pgfkeysvalueof{/tikz/fiberkeys/size} * 0.5}
			\fi
			\if\pgfkeysvalueof{/tikz/fiberkeys/direction}w
				{a}%
				\if\pgfkeysvalueof{/tikz/fiberkeys/flip}0
					{south}%
				\else
					{north}%
				\fi
				{\pgfkeysvalueof{/tikz/fiberkeys/size}}
				{\pgfkeysvalueof{/tikz/fiberkeys/size} * 0.5}
			\fi
			\if\pgfkeysvalueof{/tikz/fiberkeys/direction}n
				{b}%
				\if\pgfkeysvalueof{/tikz/fiberkeys/flip}0
					{west}%
				\else
					{east}%
				\fi
				{\pgfkeysvalueof{/tikz/fiberkeys/size} * 0.5}
				{\pgfkeysvalueof{/tikz/fiberkeys/size}}
			\fi
			\if\pgfkeysvalueof{/tikz/fiberkeys/direction}s
				{b}%
				\if\pgfkeysvalueof{/tikz/fiberkeys/flip}0
					{west}%
				\else
					{east}%
				\fi
				{\pgfkeysvalueof{/tikz/fiberkeys/size} * 0.5}
				{\pgfkeysvalueof{/tikz/fiberkeys/size}}
			\fi
			{\pgfkeysvalueof{/tikz/fiberkeys/drawbase}}%
		}
	},
	/tikz/fiberkeys/drawer/.code n args={9}{%
		\tikzset{
			fibershape,
			minimum width=#7*\NODESIZE,
			minimum height=#8*\NODESIZE,
			append after command={
				\pgfextra{\let\bdr=\tikzlastnode%
				\if#5a	
					\ifnum#9>0
						\draw[#3, #4] (\bdr.#6 west) to (\bdr.#6 east) {};
					\fi
					\node[draw=#3, #4, circle, minimum size=#2*0.5*\NODESIZE, anchor=#6] at ([xshift=-0.1*#2*\NODESIZE]\bdr.#6) () {};
					\node[draw=#3, #4, circle, minimum size=#2*0.5*\NODESIZE, anchor=#6] at (\bdr.#6) () {};
					\node[draw=#3, #4, circle, minimum size=#2*0.5*\NODESIZE, anchor=#6] at ([xshift=0.1*#2*\NODESIZE]\bdr.#6) () {};
				\fi
				\if#5b	
					\ifnum#9>0
						\draw[#3, #4] (\bdr.north #6) to (\bdr.south #6) {};
					\fi
					\node[draw=#3, #4, circle, minimum size=#2*0.5*\NODESIZE, anchor=#6] at ([yshift=0.1*#2*\NODESIZE]\bdr.#6) () {};
					\node[draw=#3, #4, circle, minimum size=#2*0.5*\NODESIZE, anchor=#6] at (\bdr.#6) () {};
					\node[draw=#3, #4, circle, minimum size=#2*0.5*\NODESIZE, anchor=#6] at ([yshift=-0.1*#2*\NODESIZE]\bdr.#6) () {};
				\fi
				}
			}
		}
	},
}
\newcount\portcount

\pgfdeclareshape{fiberswitchshape}{
	\savedmacro\nin{
		\edef\nin{\pgfkeysvalueof{/tikz/fiberswitchkeys/nin}}%
	}
	\savedmacro\nout{
		\edef\nout{\pgfkeysvalueof{/tikz/fiberswitchkeys/nout}}%
	}
	\savedmacro\direction{
		\edef\direction{\pgfkeysvalueof{/tikz/fiberswitchkeys/direction}}%
	}
	\inheritsavedanchors[from=basic] 
	
	\inheritanchor[from=basic]{center}
	\inheritanchor[from=basic]{mid}		
	\inheritanchor[from=basic]{base}	
	\inheritanchor[from=basic]{north}
	\inheritanchor[from=basic]{south}		
	\inheritanchor[from=basic]{west}		
	\inheritanchor[from=basic]{mid west}				
	\inheritanchor[from=basic]{base west}		
	\inheritanchor[from=basic]{north west}		
	\inheritanchor[from=basic]{south west}		
	\inheritanchor[from=basic]{east}
	\inheritanchor[from=basic]{mid east}
	\inheritanchor[from=basic]{base east}	
	\inheritanchor[from=basic]{north east}
	\inheritanchor[from=basic]{south east}
	\inheritanchor[from=basic]{west south west}%
	\inheritanchor[from=basic]{west north west}%
	\inheritanchor[from=basic]{east north east}%
	\inheritanchor[from=basic]{east south east}%
	\inheritanchor[from=basic]{north north west}%
	\inheritanchor[from=basic]{north north east}%
	\inheritanchor[from=basic]{south south west}%
	\inheritanchor[from=basic]{south south east}%
	
	\inheritanchorborder[from=basic]	
	\inheritbackgroundpath[from=basic]

    \pgfutil@g@addto@macro\pgf@sh@s@fiberswitchshape{%
        \pgfmathsetcount{\portcount}{0}
        \pgfmathloop%
        \ifnum\the\portcount<\nin
	        \pgfutil@ifundefined{pgf@anchor@fiberswitchshape@in\the\portcount}{
		        \expandafter\xdef\csname pgf@anchor@fiberswitchshape@in\the\portcount\endcsname{%
		            \noexpand\fiberswitchshape@port[\the\portcount]{0}
		        }%
		    }{}%
	        \ifnum\the\portcount=0
    		    \pgfutil@ifundefined{pgf@anchor@fiberswitchshape@in}{%
		        \expandafter\xdef\csname pgf@anchor@fiberswitchshape@in\endcsname{%
		            \noexpand\fiberswitchshape@port[\the\portcount]{0}
		        }%
		        }{}%
		    \fi
	        \pgfmathaddtocount{\portcount}{1}	
	        \repeatpgfmathloop
        \pgfmathsetcount{\portcount}{0}
        \pgfmathloop%
    	\ifnum\the\portcount<\nout
	        \pgfutil@ifundefined{pgf@anchor@fiberswitchshape@out\the\portcount}{%
		        \expandafter\xdef\csname pgf@anchor@fiberswitchshape@out\the\portcount\endcsname{%
		            \noexpand\fiberswitchshape@port[\the\portcount]{1}
		        }%
		    }{}%
	        \ifnum\the\portcount=0
    		    \pgfutil@ifundefined{pgf@anchor@fiberswitchshape@out}{%
		        \expandafter\xdef\csname pgf@anchor@fiberswitchshape@out\endcsname{%
		            \noexpand\fiberswitchshape@port[\the\portcount]{1}
		        }%
		        }{}%
		    \fi
	        \pgfmathaddtocount{\portcount}{1}	
	        \repeatpgfmathloop
	}
}

\def\fiberswitchshape@port[#1]#2{
    \northeast \pgf@xa=\pgf@x \pgf@ya=\pgf@y
    \southwest \pgf@xb=\pgf@x \pgf@yb=\pgf@y
    
    \ifnum#2=0	
	    \if\direction\direce	
	    	\pgf@x=\pgf@xb
		    \pgf@yc=\pgf@ya \advance\pgf@yc by -\pgf@yb	
		    \pgfmathsetlength{\pgf@y}{\pgf@ya-(#1 + 0.5)*(\pgf@yc/\nin)}%
	    \fi
	    \if\direction\direcw
	    	\pgf@x=\pgf@xa
		    \pgf@yc=\pgf@ya \advance\pgf@yc by -\pgf@yb	
		    \pgfmathsetlength{\pgf@y}{\pgf@ya-(#1 + 0.5)*(\pgf@yc/\nin)}%
	    \fi
	    \if\direction\direcn
	    	\pgf@y=\pgf@yb
		    \pgf@xc=\pgf@xa \advance\pgf@xc by -\pgf@xb	
		    \pgfmathsetlength{\pgf@x}{\pgf@xb+(#1 + 0.5)*(\pgf@xc/\nin)}%
	    \fi
	    \if\direction\direcs
	    	\pgf@y=\pgf@ya
		    \pgf@xc=\pgf@xa \advance\pgf@xc by -\pgf@xb	
		    \pgfmathsetlength{\pgf@x}{\pgf@xb+(#1 + 0.5)*(\pgf@xc/\nin)}%
	    \fi
	\else	
	    \if\direction\direce	
	    	\pgf@x=\pgf@xa
		    \pgf@yc=\pgf@ya \advance\pgf@yc by -\pgf@yb	
		    \pgfmathsetlength{\pgf@y}{\pgf@ya-(#1 + 0.5)*(\pgf@yc/\nout)}%
	    \fi
	    \if\direction\direcw
	    	\pgf@x=\pgf@xb
		    \pgf@yc=\pgf@ya \advance\pgf@yc by -\pgf@yb	
		    \pgfmathsetlength{\pgf@y}{\pgf@ya-(#1 + 0.5)*(\pgf@yc/\nout)}%
	    \fi
	    \if\direction\direcn
	    	\pgf@y=\pgf@ya
		    \pgf@xc=\pgf@xa \advance\pgf@xc by -\pgf@xb	
		    \pgfmathsetlength{\pgf@x}{\pgf@xb+(#1 + 0.5)*(\pgf@xc/\nout)}%
	    \fi
	    \if\direction\direcs
	    	\pgf@y=\pgf@yb
		    \pgf@xc=\pgf@xa \advance\pgf@xc by -\pgf@xb	
		    \pgfmathsetlength{\pgf@x}{\pgf@xb+(#1 + 0.5)*(\pgf@xc/\nout)}%
	    \fi
	\fi
}

\pgfaddtoshape{fiberswitchshape}{
	\anchorlet{c}{center}		
	\anchorlet{n}{north}
	\anchorlet{e}{east}
	\anchorlet{s}{south}
	\anchorlet{w}{west}			
	\anchorlet{se}{south east}
	\anchorlet{sw}{south west}
	\anchorlet{ne}{north east}
	\anchorlet{nw}{north west}
	\anchorlet{wsw}{west south west}
	\anchorlet{wnw}{west north west}
	\anchorlet{ene}{east north east}
	\anchorlet{ese}{east south east}
	\anchorlet{nnw}{north north west}
	\anchorlet{nne}{north north east}
	\anchorlet{ssw}{south south west}
	\anchorlet{sse}{south south east}
}

\tikzset{
	/tikz/fiberswitchkeys/.cd,
	size/.initial=1,
	color/.initial=O,
	direction/.initial=e,
	linestyle/.initial={linestyle, inner sep=0.5mm},
	nin/.initial=1,	
	nout/.initial=3,
	/tikz/fiberswitch/.code={
		\pgfqkeys{/tikz/fiberswitchkeys}{#1}%
		\tikzset{/tikz/fiberswitchkeys/drawer/.expanded=%
			{\pgfkeysvalueof{/tikz/fiberswitchkeys/size}}%
			{\pgfkeysvalueof{/tikz/fiberswitchkeys/color}}%
			{\pgfkeysvalueof{/tikz/fiberswitchkeys/linestyle}}%
			{\pgfkeysvalueof{/tikz/fiberswitchkeys/nout}}%
			\if\pgfkeysvalueof{/tikz/fiberswitchkeys/direction}e
				{0}%
			\fi
			\if\pgfkeysvalueof{/tikz/fiberswitchkeys/direction}w
				{0}%
			\fi
			\if\pgfkeysvalueof{/tikz/fiberswitchkeys/direction}n
				{1}%
			\fi
			\if\pgfkeysvalueof{/tikz/fiberswitchkeys/direction}s
				{1}%
			\fi
			{\pgfkeysvalueof{/tikz/fiberswitchkeys/direction}}%
		}
	},
	/tikz/fiberswitchkeys/drawer/.code n args={6}{%
		\tikzset{
			fiberswitchshape,
			draw,
			minimum height = #1*\NODESIZE,
			minimum width = #1*\NODESIZE,
			#2,
			#3,
			append after command={
				\pgfextra{\let\bdr=\tikzlastnode%
						\ifnum#5>0
							\node[coordinate] at ($(\bdr.in)!0.25!(\bdr.out0 -| \bdr.in)$) (circlein){};
							\foreach \n [evaluate=\n as \nport using int(\n-1)] in {1,...,#4}{
								\node[coordinate] at ($(\bdr.out\nport)!0.25!(\bdr.in -| \bdr.out\nport)$) (circleout\nport){};
							}
						\else
							\node[coordinate] at ($(\bdr.in)!0.25!(\bdr.out0 |- \bdr.in)$) (circlein){};
							\foreach \n [evaluate=\n as \nport using int(\n-1)] in {1,...,#4}{
								\node[coordinate] at ($(\bdr.out\nport)!0.25!(\bdr.in |- \bdr.out\nport)$) (circleout\nport){};
							}
						\fi

						\draw[---, #2, #3, fill] (\bdr.in) to (circlein) circle (0.05);
						\foreach \n [evaluate=\n as \nport using int(\n-1)] in {1,...,#4}{
							\draw[---, #2, #3, fill] (\bdr.out\nport) to (circleout\nport) circle (0.05);
						}

						\tikzmath{
							int \nportmax;
							\nportmax = int(#4-1);
						}
						\draw[---, #2, #3] (circlein) to (circleout0){};
						\if#6e
							\draw[-->, #2, #3, looseness=0.8] ($(circlein)!0.6!(circleout0)$) to [out=-60, in=60]($(circlein)!0.6!(circleout\nportmax)$) {};
						\fi
						\if#6w
							\draw[-->, #2, #3, looseness=0.8] ($(circlein)!0.6!(circleout0)$) to [out=-120, in=120]($(circlein)!0.6!(circleout\nportmax)$) {};
						\fi
						\if#6s
							\draw[-->, #2, #3, looseness=0.8] ($(circlein)!0.6!(circleout0)$) to [out=-30, in=-150]($(circlein)!0.6!(circleout\nportmax)$) {};
						\fi
						\if#6n
							\draw[-->, #2, #3, looseness=0.8] ($(circlein)!0.6!(circleout0)$) to [out=30, in=150]($(circlein)!0.6!(circleout\nportmax)$) {};
						\fi

				}
			}
		}
	},
}
\pgfdeclareshape{filtershape}{
	\savedmacro\direction{
		\edef\direction{\pgfkeysvalueof{/tikz/filterkeys/direction}}%
	}
	\saveddimen\minwidth{
		\pgfmathsetlength\pgf@x{\pgfshapeminwidth}%
	}
	\saveddimen\minheight{
		\pgfmathsetlength\pgf@x{\pgfshapeminheight}%
	}
	\inheritsavedanchors[from=basic]

	\inheritanchor[from=basic]{center}
	\inheritanchor[from=basic]{mid}
	\inheritanchor[from=basic]{base}
	\inheritanchor[from=basic]{north}
	\inheritanchor[from=basic]{south}
	\inheritanchor[from=basic]{west}
	\inheritanchor[from=basic]{mid west}
	\inheritanchor[from=basic]{base west}
	\inheritanchor[from=basic]{north west}
	\inheritanchor[from=basic]{south west}
	\inheritanchor[from=basic]{east}
	\inheritanchor[from=basic]{mid east}
	\inheritanchor[from=basic]{base east}
	\inheritanchor[from=basic]{north east}
	\inheritanchor[from=basic]{south east}
	\inheritanchor[from=basic]{west south west}%
	\inheritanchor[from=basic]{west north west}%
	\inheritanchor[from=basic]{east north east}%
	\inheritanchor[from=basic]{east south east}%
	\inheritanchor[from=basic]{north north west}%
	\inheritanchor[from=basic]{north north east}%
	\inheritanchor[from=basic]{south south west}%
	\inheritanchor[from=basic]{south south east}%

	\inheritanchorborder[from=basic]
	\inheritbackgroundpath[from=basic]

	\pgfutil@g@addto@macro\pgf@sh@s@filtershape{%
		\pgfutil@ifundefined{pgf@anchor@filtershape@in0}{
			\expandafter\xdef\csname pgf@anchor@filtershape@in0\endcsname{%
				\noexpand\filtershape@port{0}
			}%
		}{}%
		\pgfutil@ifundefined{pgf@anchor@filtershape@in}{
			\expandafter\xdef\csname pgf@anchor@filtershape@in\endcsname{%
				\noexpand\filtershape@port{0}
			}%
		}{}%
		\pgfutil@ifundefined{pgf@anchor@filtershape@out0}{
			\expandafter\xdef\csname pgf@anchor@filtershape@out0\endcsname{%
				\noexpand\filtershape@port{1}
			}%
		}{}%
		\pgfutil@ifundefined{pgf@anchor@filtershape@out}{
			\expandafter\xdef\csname pgf@anchor@filtershape@out\endcsname{%
				\noexpand\filtershape@port{1}
			}%
		}{}%
	}
}

\def\filtershape@port#1{
	\northeast	

	\ifnum#1=0	
		\if\direction\direce
			\pgf@x=-\pgf@x
			\pgf@ya= \pgf@y
			\pgfmathsetlength{\pgf@y}{\pgf@ya-0.5*\minheight}%
		\fi
		\if\direction\direcw
			\pgf@x=\pgf@x
			\pgf@ya= \pgf@y
			\pgfmathsetlength{\pgf@y}{\pgf@ya-0.5*\minheight}%
		\fi
		\if\direction\direcn
			\pgf@y=-\pgf@y
			\pgf@xa=\pgf@x
			\pgfmathsetlength{\pgf@x}{\pgf@xa-0.5*\minwidth}%
		\fi
		\if\direction\direcs
			\pgf@y=\pgf@y
			\pgf@xa= \pgf@x
			\pgfmathsetlength{\pgf@x}{\pgf@xa-0.5*\minwidth}%
		\fi
	\else	
		\if\direction\direce
			\pgf@x=\pgf@x
			\pgf@ya= \pgf@y
			\pgfmathsetlength{\pgf@y}{\pgf@ya-0.5*\minheight}%
		\fi
		\if\direction\direcw
			\pgf@x=-\pgf@x
			\pgf@ya= \pgf@y
			\pgfmathsetlength{\pgf@y}{\pgf@ya-0.5*\minheight}%
		\fi
		\if\direction\direcn
			\pgf@y=\pgf@y
			\pgf@xa= \pgf@x
			\pgfmathsetlength{\pgf@x}{\pgf@xa-0.5*\minwidth}%
		\fi
		\if\direction\direcs
			\pgf@y=-\pgf@y
			\pgf@xa= \pgf@x
			\pgfmathsetlength{\pgf@x}{\pgf@xa-0.5*\minwidth}%
		\fi
	\fi
}

\pgfaddtoshape{filtershape}{
	\anchorlet{c}{center}
	\anchorlet{n}{north}
	\anchorlet{e}{east}
	\anchorlet{s}{south}
	\anchorlet{w}{west}
	\anchorlet{se}{south east}
	\anchorlet{sw}{south west}
	\anchorlet{ne}{north east}
	\anchorlet{nw}{north west}
	\anchorlet{wsw}{west south west}
	\anchorlet{wnw}{west north west}
	\anchorlet{ene}{east north east}
	\anchorlet{ese}{east south east}
	\anchorlet{nnw}{north north west}
	\anchorlet{nne}{north north east}
	\anchorlet{ssw}{south south west}
	\anchorlet{sse}{south south east}
}

\pgfmathsetmacro{\WSSSINEHEIGHT}{0.06}

\tikzset{
	/tikz/filterkeys/.cd,
	size/.initial=0.5,
	color/.initial=O,
	direction/.initial=e,
	linestyle/.initial={linestyle},
	fillgradient/.initial=O,
	/tikz/filter/.code={
			\pgfqkeys{/tikz/filterkeys}{#1}%
			\tikzset{/tikz/filterkeys/drawer/.expanded=%
					{\pgfkeysvalueof{/tikz/filterkeys/direction}}%
					{\pgfkeysvalueof{/tikz/filterkeys/size}}%
					{\pgfkeysvalueof{/tikz/filterkeys/color}}%
					{\pgfkeysvalueof{/tikz/filterkeys/linestyle}}%
					{\pgfkeysvalueof{/tikz/filterkeys/fillgradient}}%
			}
		},
	/tikz/filterkeys/drawer/.code n args={5}{%
			\tikzset{
				filtershape,
				minimum height=#2*\NODESIZE,
				minimum width=#2*\NODESIZE,
				#3,
				#4,
				draw,
				append after command={
						\pgfextra{\let\bdr=\tikzlastnode%
							\node[#5, fit=(\bdr.nw)(\bdr.se)] (boxgradient){};

							\node[coordinate] at (\bdr.wnw -| \bdr.nnw) (hl){};
							\node[coordinate] at (\bdr.ene -| \bdr.nne) (hr){};
							\draw[#3,---, rounded corners = 0] (hl) sin ($(hl)!0.25!(hr) + (0,0.002*#2*\NODESIZE)$) cos ($(hl)!0.5!(hr)$) sin ($(hl)!0.75!(hr) + (0,-0.002*#2*\NODESIZE)$) cos (hr);

							\node[coordinate] at (\bdr.w -| \bdr.nnw) (ml){};
							\node[coordinate] at (\bdr.e -| \bdr.nne) (mr){};
							\draw[#3,---, rounded corners = 0] (ml) sin ($(ml)!0.25!(mr) + (0,0.002*#2*\NODESIZE)$) cos ($(ml)!0.5!(mr)$) sin ($(ml)!0.75!(mr) + (0,-0.002*#2*\NODESIZE)$) cos (mr);

							\node[coordinate] at (\bdr.wsw -| \bdr.nnw) (ll){};
							\node[coordinate] at (\bdr.ese -| \bdr.nne) (lr){};
							\draw[#3,---, rounded corners = 0] (ll) sin ($(ll)!0.25!(lr) + (0,0.002*#2*\NODESIZE)$) cos ($(ll)!0.5!(lr)$) sin ($(ll)!0.75!(lr) + (0,-0.002*#2*\NODESIZE)$) cos (lr);

							\draw[#3, ---] ($(hl)!0.25!(hr) - (0,0.002*#2*\NODESIZE)$) -- ($(hl)!0.75!(hr) + (0,0.002*#2*\NODESIZE)$);
							\draw[#3, ---] ($(ll)!0.25!(lr) - (0,0.002*#2*\NODESIZE)$) -- ($(ll)!0.75!(lr) + (0,0.002*#2*\NODESIZE)$);
						}
					}
			}
		},
}
\newcount\portcount

\pgfdeclareshape{lensshape}{
	\savedmacro\nport{
		\edef\nport{\pgfkeysvalueof{/tikz/lenskeys/nport}}%
	}
	\inheritsavedanchors[from=basic] 
	
	\inheritanchor[from=basic]{center}
	\inheritanchor[from=basic]{mid}		
	\inheritanchor[from=basic]{base}	
	\inheritanchor[from=basic]{north}
	\inheritanchor[from=basic]{south}		
	\inheritanchor[from=basic]{west}		
	\inheritanchor[from=basic]{mid west}				
	\inheritanchor[from=basic]{base west}		
	\inheritanchor[from=basic]{north west}		
	\inheritanchor[from=basic]{south west}		
	\inheritanchor[from=basic]{east}
	\inheritanchor[from=basic]{mid east}
	\inheritanchor[from=basic]{base east}	
	\inheritanchor[from=basic]{north east}
	\inheritanchor[from=basic]{south east}
	\inheritanchor[from=basic]{west south west}%
	\inheritanchor[from=basic]{west north west}%
	\inheritanchor[from=basic]{east north east}%
	\inheritanchor[from=basic]{east south east}%
	\inheritanchor[from=basic]{north north west}%
	\inheritanchor[from=basic]{north north east}%
	\inheritanchor[from=basic]{south south west}%
	\inheritanchor[from=basic]{south south east}%
	
	\inheritanchorborder[from=basic]	
	\inheritbackgroundpath[from=basic]

    \pgfutil@g@addto@macro\pgf@sh@s@lensshape{%
        \pgfmathsetcount{\portcount}{0}
        \pgfmathloop%
        \ifnum\the\portcount<\nport
	        \pgfutil@ifundefined{pgf@anchor@lensshape@in\the\portcount}{
		        \expandafter\xdef\csname pgf@anchor@lensshape@in\the\portcount\endcsname{%
		            \noexpand\lensshape@port[\the\portcount]
		        }%
		    }{}%
	        \ifnum\the\portcount=0
    		    \pgfutil@ifundefined{pgf@anchor@lensshape@in}{%
		        \expandafter\xdef\csname pgf@anchor@lensshape@in\endcsname{%
		            \noexpand\lensshape@port[\the\portcount]
		        }%
		        }{}%
		    \fi
		    \pgfutil@ifundefined{pgf@anchor@lensshape@out\the\portcount}{%
		        \expandafter\xdef\csname pgf@anchor@lensshape@out\the\portcount\endcsname{%
		            \noexpand\lensshape@port[\the\portcount]
		        }%
		    }{}%
	        \ifnum\the\portcount=0
    		    \pgfutil@ifundefined{pgf@anchor@lensshape@out}{%
		        \expandafter\xdef\csname pgf@anchor@lensshape@out\endcsname{%
		            \noexpand\lensshape@port[\the\portcount]
		        }%
		        }{}%
		    \fi
	        \pgfmathaddtocount{\portcount}{1}	
	        \repeatpgfmathloop
	}
}

\def\lensshape@port[#1]{
    \northeast \pgf@xa=\pgf@x \pgf@ya=\pgf@y
    \southwest \pgf@xb=\pgf@x \pgf@yb=\pgf@y
    
	\pgfmathsetlength{\pgf@x}{0.5*\pgf@xa + 0.5*\pgf@xb}%
    \pgf@yc=\pgf@ya \advance\pgf@yc by -\pgf@yb	
    \pgfmathsetlength{\pgf@y}{\pgf@ya-(#1 + 0.5)*(\pgf@yc/\nport)}%
}

\pgfaddtoshape{lensshape}{
	\anchorlet{c}{center}		
	\anchorlet{n}{north}
	\anchorlet{e}{east}
	\anchorlet{s}{south}
	\anchorlet{w}{west}			
	\anchorlet{se}{south east}
	\anchorlet{sw}{south west}
	\anchorlet{ne}{north east}
	\anchorlet{nw}{north west}
	\anchorlet{wsw}{west south west}
	\anchorlet{wnw}{west north west}
	\anchorlet{ene}{east north east}
	\anchorlet{ese}{east south east}
	\anchorlet{nnw}{north north west}
	\anchorlet{nne}{north north east}
	\anchorlet{ssw}{south south west}
	\anchorlet{sse}{south south east}
}

\tikzset{
	/tikz/lenskeys/.cd,
	height/.initial=1,
	width/.initial=0.3,	
	color/.initial=O,
	rotation/.initial=0,
	linestyle/.initial={linestyle, inner sep=0.5mm},
	nport/.initial=1,
	/tikz/lens/.code={
		\pgfqkeys{/tikz/lenskeys}{#1}%
		\tikzset{/tikz/lenskeys/drawer/.expanded=%
			{\pgfkeysvalueof{/tikz/lenskeys/height}}%
			{\pgfkeysvalueof{/tikz/lenskeys/width}}%
			{\pgfkeysvalueof{/tikz/lenskeys/color}}%
			{\pgfkeysvalueof{/tikz/lenskeys/linestyle}}%
			{\pgfkeysvalueof{/tikz/lenskeys/rotation}}%
		}
	},
	/tikz/lenskeys/drawer/.code n args={5}{%
		\tikzset{
			lensshape,
			rotate=#5,
			minimum height=#1*\NODESIZE,
			minimum width=#2*\NODESIZE,
			append after command={
				\pgfextra{\let\bdr=\tikzlastnode%
					\draw[---, #3, #4, rounded corners = 0] (\bdr.n) to [in=120+#5, out=240+#5] (\bdr.s) to [in=-60+#5, out=60+#5] (\bdr.n) -- cycle;
				}
			}
		}
	},
}

\newcount\portcount

\pgfdeclareshape{mirrorshape}{
	\savedmacro\nport{
		\edef\nport{\pgfkeysvalueof{/tikz/mirrorkeys/nport}}%
	}
	\inheritsavedanchors[from=basic] 
	
	\inheritanchor[from=basic]{center}
	\inheritanchor[from=basic]{mid}		
	\inheritanchor[from=basic]{base}	
	\inheritanchor[from=basic]{north}
	\inheritanchor[from=basic]{south}		
	\inheritanchor[from=basic]{west}		
	\inheritanchor[from=basic]{mid west}				
	\inheritanchor[from=basic]{base west}		
	\inheritanchor[from=basic]{north west}		
	\inheritanchor[from=basic]{south west}		
	\inheritanchor[from=basic]{east}
	\inheritanchor[from=basic]{mid east}
	\inheritanchor[from=basic]{base east}	
	\inheritanchor[from=basic]{north east}
	\inheritanchor[from=basic]{south east}
	\inheritanchor[from=basic]{west south west}%
	\inheritanchor[from=basic]{west north west}%
	\inheritanchor[from=basic]{east north east}%
	\inheritanchor[from=basic]{east south east}%
	\inheritanchor[from=basic]{north north west}%
	\inheritanchor[from=basic]{north north east}%
	\inheritanchor[from=basic]{south south west}%
	\inheritanchor[from=basic]{south south east}%
	
	\inheritanchorborder[from=basic]	
	\inheritbackgroundpath[from=basic]

    \pgfutil@g@addto@macro\pgf@sh@s@mirrorshape{%
        \pgfmathsetcount{\portcount}{0}
        \pgfmathloop%
        \ifnum\the\portcount<\nport
	        \pgfutil@ifundefined{pgf@anchor@mirrorshape@in\the\portcount}{
		        \expandafter\xdef\csname pgf@anchor@mirrorshape@in\the\portcount\endcsname{%
		            \noexpand\mirrorshape@port[\the\portcount]
		        }%
		    }{}%
	        \ifnum\the\portcount=0
    		    \pgfutil@ifundefined{pgf@anchor@mirrorshape@in}{%
		        \expandafter\xdef\csname pgf@anchor@mirrorshape@in\endcsname{%
		            \noexpand\mirrorshape@port[\the\portcount]
		        }%
		        }{}%
		    \fi
		    \pgfutil@ifundefined{pgf@anchor@mirrorshape@out\the\portcount}{%
		        \expandafter\xdef\csname pgf@anchor@mirrorshape@out\the\portcount\endcsname{%
		            \noexpand\mirrorshape@port[\the\portcount]
		        }%
		    }{}%
	        \ifnum\the\portcount=0
    		    \pgfutil@ifundefined{pgf@anchor@mirrorshape@out}{%
		        \expandafter\xdef\csname pgf@anchor@mirrorshape@out\endcsname{%
		            \noexpand\mirrorshape@port[\the\portcount]
		        }%
		        }{}%
		    \fi
	        \pgfmathaddtocount{\portcount}{1}	
	        \repeatpgfmathloop
	}
}

\def\mirrorshape@port[#1]{
    \northeast \pgf@xa=\pgf@x \pgf@ya=\pgf@y
    \southwest \pgf@xb=\pgf@x \pgf@yb=\pgf@y
    
	\pgf@x=\pgf@xb
    \pgf@yc=\pgf@ya \advance\pgf@yc by -\pgf@yb	
    \pgfmathsetlength{\pgf@y}{\pgf@ya-(#1 + 0.5)*(\pgf@yc/\nport)}%
}

\pgfaddtoshape{mirrorshape}{
	\anchorlet{c}{center}		
	\anchorlet{n}{north}
	\anchorlet{e}{east}
	\anchorlet{s}{south}
	\anchorlet{w}{west}			
	\anchorlet{se}{south east}
	\anchorlet{sw}{south west}
	\anchorlet{ne}{north east}
	\anchorlet{nw}{north west}
	\anchorlet{wsw}{west south west}
	\anchorlet{wnw}{west north west}
	\anchorlet{ene}{east north east}
	\anchorlet{ese}{east south east}
	\anchorlet{nnw}{north north west}
	\anchorlet{nne}{north north east}
	\anchorlet{ssw}{south south west}
	\anchorlet{sse}{south south east}
}

\tikzset{
	/tikz/mirrorkeys/.cd,
	height/.initial=1,
	width/.initial=0.15,	
	color/.initial=O,
	rotation/.initial=0,
	linestyle/.initial={linestyle, inner sep=0.5mm},
	nport/.initial=1,
	nlines/.initial=5,
	/tikz/mirror/.code={
		\pgfqkeys{/tikz/mirrorkeys}{#1}%
		\tikzset{/tikz/mirrorkeys/drawer/.expanded=%
			{\pgfkeysvalueof{/tikz/mirrorkeys/height}}%
			{\pgfkeysvalueof{/tikz/mirrorkeys/width}}%
			{\pgfkeysvalueof{/tikz/mirrorkeys/color}}%
			{\pgfkeysvalueof{/tikz/mirrorkeys/linestyle}}%
			{\pgfkeysvalueof{/tikz/mirrorkeys/rotation}}%
			{\pgfkeysvalueof{/tikz/mirrorkeys/nlines}}%
		}
	},
	/tikz/mirrorkeys/drawer/.code n args={6}{%
		\tikzset{
			mirrorshape,
			rotate=#5,
			minimum height=#1*\NODESIZE,
			minimum width=#2*\NODESIZE,
			append after command={
				\pgfextra{\let\bdr=\tikzlastnode%
					\draw[---, #3, #4] (\bdr.nw) to (\bdr.sw){};
					\foreach \nline [evaluate=\nline as \linepos using (\nline-0.8)/(#6-0.6)] in {1,...,#6}{
						\draw[---, #3, #4] ($(\bdr.nw)!\linepos!(\bdr.sw)$) to +(#5-45:#2*1.41421356237*\NODESIZE pt){};
					}
				}
			}
		}
	},
}
\pgfdeclareshape{muxshape}{
	\savedmacro\nports{
		\edef\nports{\pgfkeysvalueof{/tikz/muxkeys/nports}}%
	}
	\savedmacro\direction{
		\edef\direction{\pgfkeysvalueof{/tikz/muxkeys/direction}}%
	}
	\savedmacro\inverted{
		\edef\inverted{\pgfkeysvalueof{/tikz/muxkeys/inverted}}%
	}
	\savedmacro\ninports{
		\ifnum\inverted=0
			\edef\ninports{\nports}
		\else
			\edef\ninports{1}%
		\fi
	}
	\savedmacro\noutports{
		\ifnum\inverted=0
			\edef\noutports{1}%
		\else
			\edef\noutports{\nports}%
		\fi
	}
	\inheritsavedanchors[from=basic]

	\inheritanchor[from=basic]{center}
	\inheritanchor[from=basic]{mid}
	\inheritanchor[from=basic]{base}
	\inheritanchor[from=basic]{north}
	\inheritanchor[from=basic]{south}
	\inheritanchor[from=basic]{west}
	\inheritanchor[from=basic]{mid west}
	\inheritanchor[from=basic]{base west}
	\inheritanchor[from=basic]{north west}
	\inheritanchor[from=basic]{south west}
	\inheritanchor[from=basic]{east}
	\inheritanchor[from=basic]{mid east}
	\inheritanchor[from=basic]{base east}
	\inheritanchor[from=basic]{north east}
	\inheritanchor[from=basic]{south east}
	\inheritanchor[from=basic]{west south west}%
	\inheritanchor[from=basic]{west north west}%
	\inheritanchor[from=basic]{east north east}%
	\inheritanchor[from=basic]{east south east}%
	\inheritanchor[from=basic]{north north west}%
	\inheritanchor[from=basic]{north north east}%
	\inheritanchor[from=basic]{south south west}%
	\inheritanchor[from=basic]{south south east}%

	\inheritanchorborder[from=basic]
	\inheritbackgroundpath[from=basic]

	\pgfutil@g@addto@macro\pgf@sh@s@muxshape{%
		\pgfmathsetcount{\portcount}{0}
		\pgfmathloop%
		\ifnum\the\portcount<\nports
		\ifnum\the\portcount<\ninports
			\pgfutil@ifundefined{pgf@anchor@muxshape@in\the\portcount}{
				\expandafter\xdef\csname pgf@anchor@muxshape@in\the\portcount\endcsname{%
					\noexpand\muxshape@port[\the\portcount]{0}
				}%
			}{}%
			\ifnum\the\portcount=0
				\pgfutil@ifundefined{pgf@anchor@muxshape@in}{%
					\expandafter\xdef\csname pgf@anchor@muxshape@in\endcsname{%
						\noexpand\muxshape@port[\the\portcount]{0}
					}%
				}{}%
			\fi
		\fi
		\ifnum\the\portcount<\noutports
			\pgfutil@ifundefined{pgf@anchor@muxshape@out\the\portcount}{%
				\expandafter\xdef\csname pgf@anchor@muxshape@out\the\portcount\endcsname{%
					\noexpand\muxshape@port[\the\portcount]{1}
				}%
			}{}%
			\ifnum\the\portcount=0
				\pgfutil@ifundefined{pgf@anchor@muxshape@out}{%
					\expandafter\xdef\csname pgf@anchor@muxshape@out\endcsname{%
						\noexpand\muxshape@port[\the\portcount]{1}
					}%
				}{}%
			\fi
		\fi
		\pgfmathaddtocount{\portcount}{1}	
		\repeatpgfmathloop
	}
}

\def\muxshape@port[#1]#2{
	\northeast \pgf@xa=\pgf@x \pgf@ya=\pgf@y
	\southwest \pgf@xb=\pgf@x \pgf@yb=\pgf@y

	\ifnum#2=0
		\ifnum\inverted=0
			\def\chooseports{0}
		\else
			\def\chooseports{1}
		\fi
	\else
		\ifnum\inverted=0
			\def\chooseports{1}
		\else
			\def\chooseports{0}
		\fi
	\fi

	\ifnum\chooseports=0	
		\if\direction\direce
			\pgf@x=\pgf@xb
			\pgf@yc=\pgf@ya \advance\pgf@yc by -\pgf@yb	
			\pgfmathsetlength{\pgf@y}{\pgf@ya-(#1 + 0.5)*(\pgf@yc/\nports)}%
		\fi
		\if\direction\direcw
			\pgf@x=\pgf@xa
			\pgf@yc=\pgf@ya \advance\pgf@yc by -\pgf@yb	
			\pgfmathsetlength{\pgf@y}{\pgf@ya-(#1 + 0.5)*(\pgf@yc/\nports)}%
		\fi
		\if\direction\direcn
			\pgf@y=\pgf@yb
			\pgf@xc=\pgf@xa \advance\pgf@xc by -\pgf@xb	
			\pgfmathsetlength{\pgf@x}{\pgf@xb+(#1 + 0.5)*(\pgf@xc/\nports)}%
		\fi
		\if\direction\direcs
			\pgf@y=\pgf@ya
			\pgf@xc=\pgf@xa \advance\pgf@xc by -\pgf@xb	
			\pgfmathsetlength{\pgf@x}{\pgf@xb+(#1 + 0.5)*(\pgf@xc/\nports)}%
		\fi
	\else	
		\if\direction\direce
			\pgf@x=\pgf@xa
			\pgf@yc=\pgf@ya \advance\pgf@yc by -\pgf@yb	
			\pgfmathsetlength{\pgf@y}{\pgf@ya-0.5\pgf@yc}%
		\fi
		\if\direction\direcw
			\pgf@x=\pgf@xb
			\pgf@yc=\pgf@ya \advance\pgf@yc by -\pgf@yb	
			\pgfmathsetlength{\pgf@y}{\pgf@ya-0.5\pgf@yc}%
		\fi
		\if\direction\direcn
			\pgf@y=\pgf@ya
			\pgf@xc=\pgf@xa \advance\pgf@xc by -\pgf@xb	
			\pgfmathsetlength{\pgf@x}{\pgf@xa-0.5\pgf@xc}%
		\fi
		\if\direction\direcs
			\pgf@y=\pgf@yb
			\pgf@xc=\pgf@xa \advance\pgf@xc by -\pgf@xb	
			\pgfmathsetlength{\pgf@x}{\pgf@xa-0.5\pgf@xc}%
		\fi
	\fi
}

\pgfaddtoshape{muxshape}{
	\anchorlet{c}{center}
	\anchorlet{n}{north}
	\anchorlet{e}{east}
	\anchorlet{s}{south}
	\anchorlet{w}{west}
	\anchorlet{se}{south east}
	\anchorlet{sw}{south west}
	\anchorlet{ne}{north east}
	\anchorlet{nw}{north west}
	\anchorlet{wsw}{west south west}
	\anchorlet{wnw}{west north west}
	\anchorlet{ene}{east north east}
	\anchorlet{ese}{east south east}
	\anchorlet{nnw}{north north west}
	\anchorlet{nne}{north north east}
	\anchorlet{ssw}{south south west}
	\anchorlet{sse}{south south east}
}

\tikzset{
/tikz/muxkeys/.cd,
height/.initial=1,
width/.initial=0.5,
color/.initial=O,
direction/.initial=e,
linestyle/.initial={linestyle, rounded corners = 0},
nports/.initial=3,
inverted/.initial=0,	
smux/.initial=0,
angle/.initial=60,
fillgradient/.initial=O,
/tikz/mux/.code={
\pgfqkeys{/tikz/muxkeys}{#1}%
\tikzset{/tikz/muxkeys/drawer/.expanded=%
\if\pgfkeysvalueof{/tikz/muxkeys/direction}e
	{\pgfkeysvalueof{/tikz/muxkeys/width}}%
	{\pgfkeysvalueof{/tikz/muxkeys/height}}%
	{0}%
	{-90}%
\fi
\if\pgfkeysvalueof{/tikz/muxkeys/direction}w
	{\pgfkeysvalueof{/tikz/muxkeys/width}}%
	{\pgfkeysvalueof{/tikz/muxkeys/height}}%
	{0}%
	{90}%
\fi
\if\pgfkeysvalueof{/tikz/muxkeys/direction}n
	{\pgfkeysvalueof{/tikz/muxkeys/width}}%
	{\pgfkeysvalueof{/tikz/muxkeys/height}}%
	{1}%
	{0}%
\fi
\if\pgfkeysvalueof{/tikz/muxkeys/direction}s
	{\pgfkeysvalueof{/tikz/muxkeys/width}}%
	{\pgfkeysvalueof{/tikz/muxkeys/height}}%
	{1}%
	{180}%
\fi
{\pgfkeysvalueof{/tikz/muxkeys/color}}%
{\pgfkeysvalueof{/tikz/muxkeys/linestyle}}%
{\pgfkeysvalueof{/tikz/muxkeys/smux}}%
{\pgfkeysvalueof{/tikz/muxkeys/angle}}%
{\pgfkeysvalueof{/tikz/muxkeys/fillgradient}}%
}
},
/tikz/muxkeys/drawer/.code n args={9}{%
		\tikzset{
			muxshape,
			#6,
			#5,
			minimum height=
			\ifnum#3>0	
				#1*\NODESIZE
			\else
				#2*\NODESIZE
			\fi
			,minimum width=
			\ifnum#3>0
				#2*\NODESIZE
			\else
				#1*\NODESIZE
			\fi
			,append after command={
					\pgfextra{\let\bdr=\tikzlastnode%
						\node[trapezium, line width = \NODETHICKNESS, minimum height=#1*\NODESIZE, minimum width=#2*\NODESIZE, trapezium stretches=true, rotate=#4, trapezium angle=#8, inner sep=0.001mm] at (\bdr) (trap) {};	
						\ifnum#7=0
							\draw[#9, #5, #6] (trap.bottom left corner) to (trap.top left corner) to (trap.top right corner) to (trap.bottom right corner) to cycle;
						\else
							\draw[#9, #5, #6] (trap.bottom left corner) to[in=#4-90, out=#4+90] (trap.top left corner) to (trap.top right corner) to[in=#4+90,out=#4-90] (trap.bottom right corner) to cycle;
						\fi
					}
				}
		}
	},
}
\newcount\portcount

\pgfdeclareshape{polswitchshape}{
	\savedmacro\nin{
		\edef\nin{\pgfkeysvalueof{/tikz/polswitchkeys/nin}}%
	}
	\savedmacro\nout{
		\edef\nout{\pgfkeysvalueof{/tikz/polswitchkeys/nout}}%
	}
	\savedmacro\direction{
		\edef\direction{\pgfkeysvalueof{/tikz/polswitchkeys/direction}}%
	}
	\inheritsavedanchors[from=basic] 
	
	\inheritanchor[from=basic]{center}
	\inheritanchor[from=basic]{mid}		
	\inheritanchor[from=basic]{base}	
	\inheritanchor[from=basic]{north}
	\inheritanchor[from=basic]{south}		
	\inheritanchor[from=basic]{west}		
	\inheritanchor[from=basic]{mid west}				
	\inheritanchor[from=basic]{base west}		
	\inheritanchor[from=basic]{north west}		
	\inheritanchor[from=basic]{south west}		
	\inheritanchor[from=basic]{east}
	\inheritanchor[from=basic]{mid east}
	\inheritanchor[from=basic]{base east}	
	\inheritanchor[from=basic]{north east}
	\inheritanchor[from=basic]{south east}
	\inheritanchor[from=basic]{west south west}%
	\inheritanchor[from=basic]{west north west}%
	\inheritanchor[from=basic]{east north east}%
	\inheritanchor[from=basic]{east south east}%
	\inheritanchor[from=basic]{north north west}%
	\inheritanchor[from=basic]{north north east}%
	\inheritanchor[from=basic]{south south west}%
	\inheritanchor[from=basic]{south south east}%
	
	\inheritanchorborder[from=basic]	
	\inheritbackgroundpath[from=basic]

    \pgfutil@g@addto@macro\pgf@sh@s@polswitchshape{%
        \pgfmathsetcount{\portcount}{0}
        \pgfmathloop%
        \ifnum\the\portcount<\nin
	        \pgfutil@ifundefined{pgf@anchor@polswitchshape@in\the\portcount}{
		        \expandafter\xdef\csname pgf@anchor@polswitchshape@in\the\portcount\endcsname{%
		            \noexpand\polswitchshape@port[\the\portcount]{0}
		        }%
		    }{}%
	        \ifnum\the\portcount=0
    		    \pgfutil@ifundefined{pgf@anchor@polswitchshape@in}{%
		        \expandafter\xdef\csname pgf@anchor@polswitchshape@in\endcsname{%
		            \noexpand\polswitchshape@port[\the\portcount]{0}
		        }%
		        }{}%
		    \fi
	        \pgfmathaddtocount{\portcount}{1}	
	        \repeatpgfmathloop
        \pgfmathsetcount{\portcount}{0}
        \pgfmathloop%
    	\ifnum\the\portcount<\nout
	        \pgfutil@ifundefined{pgf@anchor@polswitchshape@out\the\portcount}{%
		        \expandafter\xdef\csname pgf@anchor@polswitchshape@out\the\portcount\endcsname{%
		            \noexpand\polswitchshape@port[\the\portcount]{1}
		        }%
		    }{}%
	        \ifnum\the\portcount=0
    		    \pgfutil@ifundefined{pgf@anchor@polswitchshape@out}{%
		        \expandafter\xdef\csname pgf@anchor@polswitchshape@out\endcsname{%
		            \noexpand\polswitchshape@port[\the\portcount]{1}
		        }%
		        }{}%
		    \fi
	        \pgfmathaddtocount{\portcount}{1}	
	        \repeatpgfmathloop
	}
}

\def\polswitchshape@port[#1]#2{
    \northeast \pgf@xa=\pgf@x \pgf@ya=\pgf@y
    \southwest \pgf@xb=\pgf@x \pgf@yb=\pgf@y
    
    \ifnum#2=0	
	    \if\direction\direce	
	    	\pgf@x=\pgf@xb
		    \pgf@yc=\pgf@ya \advance\pgf@yc by -\pgf@yb	
		    \pgfmathsetlength{\pgf@y}{\pgf@ya-(#1 + 0.5)*(\pgf@yc/\nin)}%
	    \fi
	    \if\direction\direcw
	    	\pgf@x=\pgf@xa
		    \pgf@yc=\pgf@ya \advance\pgf@yc by -\pgf@yb	
		    \pgfmathsetlength{\pgf@y}{\pgf@ya-(#1 + 0.5)*(\pgf@yc/\nin)}%
	    \fi
	    \if\direction\direcn
	    	\pgf@y=\pgf@yb
		    \pgf@xc=\pgf@xa \advance\pgf@xc by -\pgf@xb	
		    \pgfmathsetlength{\pgf@x}{\pgf@xb+(#1 + 0.5)*(\pgf@xc/\nin)}%
	    \fi
	    \if\direction\direcs
	    	\pgf@y=\pgf@ya
		    \pgf@xc=\pgf@xa \advance\pgf@xc by -\pgf@xb	
		    \pgfmathsetlength{\pgf@x}{\pgf@xb+(#1 + 0.5)*(\pgf@xc/\nin)}%
	    \fi
	\else	
	    \if\direction\direce	
	    	\pgf@x=\pgf@xa
		    \pgf@yc=\pgf@ya \advance\pgf@yc by -\pgf@yb	
		    \pgfmathsetlength{\pgf@y}{\pgf@ya-(#1 + 0.5)*(\pgf@yc/\nout)}%
	    \fi
	    \if\direction\direcw
	    	\pgf@x=\pgf@xb
		    \pgf@yc=\pgf@ya \advance\pgf@yc by -\pgf@yb	
		    \pgfmathsetlength{\pgf@y}{\pgf@ya-(#1 + 0.5)*(\pgf@yc/\nout)}%
	    \fi
	    \if\direction\direcn
	    	\pgf@y=\pgf@ya
		    \pgf@xc=\pgf@xa \advance\pgf@xc by -\pgf@xb	
		    \pgfmathsetlength{\pgf@x}{\pgf@xb+(#1 + 0.5)*(\pgf@xc/\nout)}%
	    \fi
	    \if\direction\direcs
	    	\pgf@y=\pgf@yb
		    \pgf@xc=\pgf@xa \advance\pgf@xc by -\pgf@xb	
		    \pgfmathsetlength{\pgf@x}{\pgf@xb+(#1 + 0.5)*(\pgf@xc/\nout)}%
	    \fi
	\fi
}

\pgfaddtoshape{polswitchshape}{
	\anchorlet{c}{center}		
	\anchorlet{n}{north}
	\anchorlet{e}{east}
	\anchorlet{s}{south}
	\anchorlet{w}{west}			
	\anchorlet{se}{south east}
	\anchorlet{sw}{south west}
	\anchorlet{ne}{north east}
	\anchorlet{nw}{north west}
	\anchorlet{wsw}{west south west}
	\anchorlet{wnw}{west north west}
	\anchorlet{ene}{east north east}
	\anchorlet{ese}{east south east}
	\anchorlet{nnw}{north north west}
	\anchorlet{nne}{north north east}
	\anchorlet{ssw}{south south west}
	\anchorlet{sse}{south south east}
}

\tikzset{
	/tikz/polswitchkeys/.cd,
	size/.initial=1,
	color/.initial=O,
	direction/.initial=e,
	linestyle/.initial={linestyle, inner sep=0.5mm},
	nin/.initial=1,	
	nout/.initial=1, 
	/tikz/polswitch/.code={
		\pgfqkeys{/tikz/polswitchkeys}{#1}%
		\tikzset{/tikz/polswitchkeys/drawer/.expanded=%
			{\pgfkeysvalueof{/tikz/polswitchkeys/size}}%
			{\pgfkeysvalueof{/tikz/polswitchkeys/color}}%
			{\pgfkeysvalueof{/tikz/polswitchkeys/linestyle}}%
			{\pgfkeysvalueof{/tikz/polswitchkeys/nout}}%
			\if\pgfkeysvalueof{/tikz/polswitchkeys/direction}e
				{0}%
			\fi
			\if\pgfkeysvalueof{/tikz/polswitchkeys/direction}w
				{0}%
			\fi
			\if\pgfkeysvalueof{/tikz/polswitchkeys/direction}n
				{1}%
			\fi
			\if\pgfkeysvalueof{/tikz/polswitchkeys/direction}s
				{1}%
			\fi
			{\pgfkeysvalueof{/tikz/polswitchkeys/direction}}%
		}
	},
	/tikz/polswitchkeys/drawer/.code n args={6}{%
		\tikzset{
			polswitchshape,
			draw,
			minimum height = #1*\NODESIZE,
			minimum width = #1*\NODESIZE,
			#2,
			#3,
			append after command={
				\pgfextra{\let\bdr=\tikzlastnode%
						\node[coordinate] at ($(\bdr.in)!0.25!(\bdr.out)$) (circlein){};
						\node[coordinate] at ($(\bdr.in)!0.75!(\bdr.out)$) (circleout){};

						\draw[---, #2, #3, fill] (\bdr.in) to (circlein) circle (0.05);
						\draw[---, #2, #3, fill] (\bdr.out) to (circleout) circle (0.05);

						\node[coordinate] at ($(\bdr.in)!0.6!(\bdr.out)$) (circlemiddle){};
						\ifnum#5>0
							\node[coordinate] at ($(circlemiddle)!0.5!(circlemiddle -| \bdr.e)$) (circletopcor){};
							\node[coordinate] at ($(circlemiddle)!0.5!(circlemiddle -| \bdr.w)$) (circlebotcor){};
						\else
							\node[coordinate] at ($(circlemiddle)!0.5!(circlemiddle |- \bdr.n)$) (circletopcor){};
							\node[coordinate] at ($(circlemiddle)!0.5!(circlemiddle |- \bdr.s)$) (circlebotcor){};
						\fi

						\node[draw, circle, #2, #3, minimum size=0.3*\FNODESIZE] at (circletopcor) (circletop){};
						\node[draw, circle, #2, #3, minimum size=0.3*\FNODESIZE] at (circlebotcor) (circlebot){};
						\draw[-->, #2, #3] (circletop.south) to (circletop.north){};
						\draw[-->, #2, #3] (circlebot.west) to (circlebot.east){};

						\if#6e
							\draw[-->, #2, #3] ([xshift=-0.05*\FNODESIZE]circletop.south west) to [out=-120, in=120] ([xshift=-0.05*\FNODESIZE]circlebot.north west){};
						\fi
						\if#6w
							\draw[-->, #2, #3] ([xshift=0.05*\FNODESIZE]circletop.south east) to [out=-60, in=60] ([xshift=0.05*\FNODESIZE]circlebot.north east){};
						\fi
						\if#6s
							\draw[-->, #2, #3] ([yshift=0.05*\FNODESIZE]circletop.north west) to [out=150, in=30] ([yshift=0.05*\FNODESIZE]circlebot.north east){};
						\fi
						\if#6n
							\draw[-->, #2, #3] ([yshift=-0.05*\FNODESIZE]circletop.south west) to [out=-150, in=-30] ([yshift=-0.05*\FNODESIZE]circlebot.south east){};
						\fi

				}
			}
		}
	},
}
\pgfdeclareshape{pdshape}{
	\savedmacro\direction{
		\edef\direction{\pgfkeysvalueof{/tikz/pdkeys/direction}}%
	}
	\saveddimen\minwidth{
		\pgfmathsetlength\pgf@x{\pgfshapeminwidth}%
	}
	\saveddimen\minheight{
		\pgfmathsetlength\pgf@x{\pgfshapeminheight}%
	}
	\inheritsavedanchors[from=basic]

	\inheritanchor[from=basic]{center}
	\inheritanchor[from=basic]{mid}
	\inheritanchor[from=basic]{base}
	\inheritanchor[from=basic]{north}
	\inheritanchor[from=basic]{south}
	\inheritanchor[from=basic]{west}
	\inheritanchor[from=basic]{mid west}
	\inheritanchor[from=basic]{base west}
	\inheritanchor[from=basic]{north west}
	\inheritanchor[from=basic]{south west}
	\inheritanchor[from=basic]{east}
	\inheritanchor[from=basic]{mid east}
	\inheritanchor[from=basic]{base east}
	\inheritanchor[from=basic]{north east}
	\inheritanchor[from=basic]{south east}
	\inheritanchor[from=basic]{west south west}%
	\inheritanchor[from=basic]{west north west}%
	\inheritanchor[from=basic]{east north east}%
	\inheritanchor[from=basic]{east south east}%
	\inheritanchor[from=basic]{north north west}%
	\inheritanchor[from=basic]{north north east}%
	\inheritanchor[from=basic]{south south west}%
	\inheritanchor[from=basic]{south south east}%

	\inheritanchorborder[from=basic]
	\inheritbackgroundpath[from=basic]

	\pgfutil@g@addto@macro\pgf@sh@s@pdshape{%
		\pgfutil@ifundefined{pgf@anchor@pdshape@in0}{
			\expandafter\xdef\csname pgf@anchor@pdshape@in0\endcsname{%
				\noexpand\pdshape@port{0}
			}%
		}{}%
		\pgfutil@ifundefined{pgf@anchor@pdshape@in}{
			\expandafter\xdef\csname pgf@anchor@pdshape@in\endcsname{%
				\noexpand\pdshape@port{0}
			}%
		}{}%
		\pgfutil@ifundefined{pgf@anchor@pdshape@out0}{
			\expandafter\xdef\csname pgf@anchor@pdshape@out0\endcsname{%
				\noexpand\pdshape@port{1}
			}%
		}{}%
		\pgfutil@ifundefined{pgf@anchor@pdshape@out}{
			\expandafter\xdef\csname pgf@anchor@pdshape@out\endcsname{%
				\noexpand\pdshape@port{1}
			}%
		}{}%
	}
}

\def\pdshape@port#1{
	\northeast	

	\ifnum#1=0	
		\if\direction\direce
			\pgf@x=-\pgf@x
			\pgf@ya= \pgf@y
			\pgfmathsetlength{\pgf@y}{\pgf@ya-0.5*\minheight}%
		\fi
		\if\direction\direcw
			\pgf@x=\pgf@x
			\pgf@ya= \pgf@y
			\pgfmathsetlength{\pgf@y}{\pgf@ya-0.5*\minheight}%
		\fi
		\if\direction\direcn
			\pgf@y=-\pgf@y
			\pgf@xa=\pgf@x
			\pgfmathsetlength{\pgf@x}{\pgf@xa-0.5*\minwidth}%
		\fi
		\if\direction\direcs
			\pgf@y=\pgf@y
			\pgf@xa= \pgf@x
			\pgfmathsetlength{\pgf@x}{\pgf@xa-0.5*\minwidth}%
		\fi
	\else	
		\if\direction\direce
			\pgf@x=\pgf@x
			\pgf@ya= \pgf@y
			\pgfmathsetlength{\pgf@y}{\pgf@ya-0.5*\minheight}%
		\fi
		\if\direction\direcw
			\pgf@x=-\pgf@x
			\pgf@ya= \pgf@y
			\pgfmathsetlength{\pgf@y}{\pgf@ya-0.5*\minheight}%
		\fi
		\if\direction\direcn
			\pgf@y=\pgf@y
			\pgf@xa= \pgf@x
			\pgfmathsetlength{\pgf@x}{\pgf@xa-0.5*\minwidth}%
		\fi
		\if\direction\direcs
			\pgf@y=-\pgf@y
			\pgf@xa= \pgf@x
			\pgfmathsetlength{\pgf@x}{\pgf@xa-0.5*\minwidth}%
		\fi
	\fi
}

\pgfaddtoshape{pdshape}{
	\anchorlet{c}{center}
	\anchorlet{n}{north}
	\anchorlet{e}{east}
	\anchorlet{s}{south}
	\anchorlet{w}{west}
	\anchorlet{se}{south east}
	\anchorlet{sw}{south west}
	\anchorlet{ne}{north east}
	\anchorlet{nw}{north west}
	\anchorlet{wsw}{west south west}
	\anchorlet{wnw}{west north west}
	\anchorlet{ene}{east north east}
	\anchorlet{ese}{east south east}
	\anchorlet{nnw}{north north west}
	\anchorlet{nne}{north north east}
	\anchorlet{ssw}{south south west}
	\anchorlet{sse}{south south east}
}

\tikzset{
/tikz/pdkeys/.cd,
size/.initial=0.5,
color/.initial=EO,
direction/.initial=e,
linestyle/.initial={linestyle},
fillgradient/.initial=O,
/tikz/pd/.code={
		\pgfqkeys{/tikz/pdkeys}{#1}%
		\tikzset{/tikz/pdkeys/drawer/.expanded=%
				{\pgfkeysvalueof{/tikz/pdkeys/direction}}%
				{\pgfkeysvalueof{/tikz/pdkeys/size}}%
				{\pgfkeysvalueof{/tikz/pdkeys/color}}%
				{\pgfkeysvalueof{/tikz/pdkeys/linestyle}}%
				{\pgfkeysvalueof{/tikz/pdkeys/fillgradient}}%
		}
	},
/tikz/pdkeys/drawer/.code n args={5}{%
\tikzset{
pdshape,
minimum height=#2*\NODESIZE,
minimum width=#2*\NODESIZE,
#3,
#4,
draw,
append after command={
\pgfextra{\let\bdr=\tikzlastnode%
\node[#5, fit=(\bdr.nw)(\bdr.se)] (boxgradient){};
\draw[---,#3] ($(\bdr.s)!.1!(\bdr.n)$) to ($(\bdr.s)!.9!(\bdr.n)$);
\fill[#3] ({$(\bdr.s)!.3!(\bdr.n)$} -| {$(\bdr.w)!.3!(\bdr.e)$}) to ($(\bdr.s)!.7!(\bdr.n)$) to ({$(\bdr.s)!.3!(\bdr.n)$} -| {$(\bdr.w)!.7!(\bdr.e)$}) to cycle;
\draw[---,#3] ({$(\bdr.s)!.7!(\bdr.n)$} -| {$(\bdr.w)!.35!(\bdr.e)$}) to ({$(\bdr.s)!.7!(\bdr.n)$} -| {$(\bdr.w)!.65!(\bdr.e)$});
}
}
}
},
}
\pgfdeclareshape{pbsshape}{
	\savedmacro\direction{
		\edef\direction{\pgfkeysvalueof{/tikz/pbskeys/direction}}%
	}
	\saveddimen\minwidth{
		\pgfmathsetlength\pgf@x{\pgfshapeminwidth}%
	}
	\saveddimen\minheight{
		\pgfmathsetlength\pgf@x{\pgfshapeminheight}%
	}
	\savedmacro\nport{
		\edef\nport{\pgfkeysvalueof{/tikz/pbskeys/nport}}%
	}
	\inheritsavedanchors[from=basic]

	\inheritanchor[from=basic]{center}
	\inheritanchor[from=basic]{mid}
	\inheritanchor[from=basic]{base}
	\inheritanchor[from=basic]{north}
	\inheritanchor[from=basic]{south}
	\inheritanchor[from=basic]{west}
	\inheritanchor[from=basic]{mid west}
	\inheritanchor[from=basic]{base west}
	\inheritanchor[from=basic]{north west}
	\inheritanchor[from=basic]{south west}
	\inheritanchor[from=basic]{east}
	\inheritanchor[from=basic]{mid east}
	\inheritanchor[from=basic]{base east}
	\inheritanchor[from=basic]{north east}
	\inheritanchor[from=basic]{south east}
	\inheritanchor[from=basic]{west south west}%
	\inheritanchor[from=basic]{west north west}%
	\inheritanchor[from=basic]{east north east}%
	\inheritanchor[from=basic]{east south east}%
	\inheritanchor[from=basic]{north north west}%
	\inheritanchor[from=basic]{north north east}%
	\inheritanchor[from=basic]{south south west}%
	\inheritanchor[from=basic]{south south east}%

	\inheritanchorborder[from=basic]
	\inheritbackgroundpath[from=basic]

	\pgfutil@g@addto@macro\pgf@sh@s@pbsshape{%
		\pgfutil@ifundefined{pgf@anchor@pbsshape@in0}{
			\expandafter\xdef\csname pgf@anchor@pbsshape@in0\endcsname{%
				\noexpand\pbsshape@port[0]{0}
			}%
		}{}%
		\pgfutil@ifundefined{pgf@anchor@pbsshape@in1}{
			\expandafter\xdef\csname pgf@anchor@pbsshape@in1\endcsname{%
				\noexpand\pbsshape@port[1]{0}
			}%
		}{}%
		\pgfutil@ifundefined{pgf@anchor@pbsshape@in}{
			\expandafter\xdef\csname pgf@anchor@pbsshape@in\endcsname{%
				\noexpand\pbsshape@port[0]{0}
			}%
		}{}%
		\pgfutil@ifundefined{pgf@anchor@pbsshape@out0}{
			\expandafter\xdef\csname pgf@anchor@pbsshape@out0\endcsname{%
				\noexpand\pbsshape@port[0]{1}
			}%
		}{}%
		\pgfutil@ifundefined{pgf@anchor@pbsshape@out1}{
			\expandafter\xdef\csname pgf@anchor@pbsshape@out1\endcsname{%
				\noexpand\pbsshape@port[1]{1}
			}%
		}{}%
		\pgfutil@ifundefined{pgf@anchor@pbsshape@out}{
			\expandafter\xdef\csname pgf@anchor@pbsshape@out\endcsname{%
				\noexpand\pbsshape@port[0]{1}
			}%
		}{}%
		\pgfmathsetcount{\portcount}{0}
		\pgfmathloop%
		\ifnum\the\portcount<\nport
		\pgfutil@ifundefined{pgf@anchor@pbsshape@p\the\portcount}{
			\expandafter\xdef\csname pgf@anchor@pbsshape@p\the\portcount\endcsname{%
				\noexpand\pbsshape@port[\the\portcount]{2}
			}%
		}{}%
		\pgfmathaddtocount{\portcount}{1}	
		\repeatpgfmathloop
	}
}

\def\pbsshape@port[#1]#2{
	\northeast	

	\ifnum#2=0	
		\if\direction\direce
			\ifnum#1=0
				\pgf@x=-\pgf@x
				\pgf@ya= \pgf@y
				\pgfmathsetlength{\pgf@y}{\pgf@ya-0.5*\minheight}%
			\else
				\pgf@x=0\pgf@x
				\pgf@ya= \pgf@y
				\pgfmathsetlength{\pgf@y}{\pgf@ya-\minheight}%
			\fi
		\fi
		\if\direction\direcw
			\ifnum#1=0
				\pgf@x=\pgf@x
				\pgf@ya= \pgf@y
				\pgfmathsetlength{\pgf@y}{\pgf@ya-0.5*\minheight}%
			\else
				\pgf@x=0\pgf@x
				\pgf@ya= \pgf@y
				\pgfmathsetlength{\pgf@y}{\pgf@ya}%
			\fi
		\fi
		\if\direction\direcn
			\ifnum#1=0
				\pgf@y=-\pgf@y
				\pgf@xa=\pgf@x
				\pgfmathsetlength{\pgf@x}{\pgf@xa-0.5*\minwidth}%
			\else
				\pgf@y=0\pgf@y
				\pgf@xa=\pgf@x
				\pgfmathsetlength{\pgf@x}{\pgf@xa-0*\minwidth}%
			\fi
		\fi
		\if\direction\direcs
			\ifnum#1=0
				\pgf@y=\pgf@y
				\pgf@xa= \pgf@x
				\pgfmathsetlength{\pgf@x}{\pgf@xa-0.5*\minwidth}%
			\else
				\pgf@y=0\pgf@y
				\pgf@xa= \pgf@x
				\pgfmathsetlength{\pgf@x}{\pgf@xa-1*\minwidth}%
			\fi
		\fi
	\else	
		\if\direction\direce
			\ifnum#1=0
				\pgf@x=\pgf@x
				\pgf@ya= \pgf@y
				\pgfmathsetlength{\pgf@y}{\pgf@ya-0.5*\minheight}%
			\else
				\pgf@x=0\pgf@x
				\pgf@ya= \pgf@y
				\pgfmathsetlength{\pgf@y}{\pgf@ya}%
			\fi
		\fi
		\if\direction\direcw
			\ifnum#1=0
				\pgf@x=-\pgf@x
				\pgf@ya= \pgf@y
				\pgfmathsetlength{\pgf@y}{\pgf@ya-0.5*\minheight}%
			\else
				\pgf@x=0\pgf@x
				\pgf@ya= \pgf@y
				\pgfmathsetlength{\pgf@y}{\pgf@ya-\minheight}%
			\fi
		\fi
		\if\direction\direcn
			\ifnum#1=0
				\pgf@y=\pgf@y
				\pgf@xa= \pgf@x
				\pgfmathsetlength{\pgf@x}{\pgf@xa-0.5*\minwidth}%
			\else
				\pgf@y=0\pgf@y
				\pgf@xa= \pgf@x
				\pgfmathsetlength{\pgf@x}{\pgf@xa-1*\minwidth}%
			\fi
		\fi
		\if\direction\direcs
			\ifnum#1=0
				\pgf@y=-\pgf@y
				\pgf@xa= \pgf@x
				\pgfmathsetlength{\pgf@x}{\pgf@xa-0.5*\minwidth}%
			\else
				\pgf@y=0\pgf@y
				\pgf@xa= \pgf@x
				\pgfmathsetlength{\pgf@x}{\pgf@xa-0*\minwidth}%
			\fi
		\fi
	\fi

	\ifnum#2=2	%
		\northeast \pgf@xa=\pgf@x \pgf@ya=\pgf@y
		\southwest \pgf@xb=\pgf@x \pgf@yb=\pgf@y

		\pgf@xc=\pgf@xa \advance\pgf@xc by -\pgf@xb	
		\pgfmathsetlength{\pgf@x}{\pgf@xa-(#1 + 0.5)*(\pgf@xc/\nport)}%
		\pgf@yc=\pgf@ya \advance\pgf@yc by -\pgf@yb	
		\pgfmathsetlength{\pgf@y}{\pgf@ya-(#1 + 0.5)*(\pgf@yc/\nport)}%

		\if\direction\direce
			\pgf@x=-\pgf@x
		\fi
		\if\direction\direcw
			\pgf@x=-\pgf@x
		\fi
	\fi
}

\pgfaddtoshape{pbsshape}{
	\anchorlet{c}{center}
	\anchorlet{n}{north}
	\anchorlet{e}{east}
	\anchorlet{s}{south}
	\anchorlet{w}{west}
	\anchorlet{se}{south east}
	\anchorlet{sw}{south west}
	\anchorlet{ne}{north east}
	\anchorlet{nw}{north west}
	\anchorlet{wsw}{west south west}
	\anchorlet{wnw}{west north west}
	\anchorlet{ene}{east north east}
	\anchorlet{ese}{east south east}
	\anchorlet{nnw}{north north west}
	\anchorlet{nne}{north north east}
	\anchorlet{ssw}{south south west}
	\anchorlet{sse}{south south east}
}

\pgfmathsetmacro{\WSSSINEHEIGHT}{0.06}

\tikzset{
	/tikz/pbskeys/.cd,
	size/.initial=0.5,
	color/.initial=O,
	direction/.initial=e,
	linestyle/.initial={linestyle},
	nport/.initial=1,
	fillgradient/.initial=O,
	/tikz/pbs/.code={
			\pgfqkeys{/tikz/pbskeys}{#1}%
			\tikzset{/tikz/pbskeys/drawer/.expanded=%
					{\pgfkeysvalueof{/tikz/pbskeys/direction}}%
					{\pgfkeysvalueof{/tikz/pbskeys/size}}%
					{\pgfkeysvalueof{/tikz/pbskeys/color}}%
					{\pgfkeysvalueof{/tikz/pbskeys/linestyle}}%
					{\pgfkeysvalueof{/tikz/pbskeys/fillgradient}}%
			}
		},
	/tikz/pbskeys/drawer/.code n args={5}{%
			\tikzset{
				pbsshape,
				minimum height=#2*\NODESIZE,
				minimum width=#2*\NODESIZE,
				#3,
				#4,
				draw,
				append after command={
						\pgfextra{\let\bdr=\tikzlastnode%
							\node[#5, fit=(\bdr.nw)(\bdr.se)] (boxgradient){};

							\if#1e
								\draw[#3, ---] ($(\bdr.nw)!.01!(\bdr.se)$) to ($(\bdr.se)!.01!(\bdr.nw)$);
							\fi
							\if#1w
								\draw[#3, ---] ($(\bdr.nw)!.01!(\bdr.se)$) to ($(\bdr.se)!.01!(\bdr.nw)$);
							\fi
							\if#1n
								\draw[#3, ---] ($(\bdr.ne)!.01!(\bdr.sw)$) to ($(\bdr.sw)!.01!(\bdr.ne)$);
							\fi
							\if#1s
								\draw[#3, ---] ($(\bdr.ne)!.01!(\bdr.sw)$) to ($(\bdr.sw)!.01!(\bdr.ne)$);
							\fi

						}
					}
			}
		},
}

\ifdefined\fontchoice
\else
	\def\fontchoice{times} 
\fi

\usepackage{pdftexcmds}

\ifnum\pdf@strcmp{\fontchoice}{firasans}=0 %
	\usepackage[book]{FiraSans} 
	\usepackage[T1]{fontenc}
	
	\tikzset{every picture/.style={/utils/exec={\sffamily}}}
\fi

\ifnum\pdf@strcmp{\fontchoice}{times}=0 %
	\usepackage{times}
\fi

\ifnum\pdf@strcmp{\fontchoice}{timesnewroman}=0 %
	\usepackage{mathptmx}
	\usepackage[T1]{fontenc}
\fi

\ifnum\pdf@strcmp{\fontchoice}{helvetica}=0 %
	\usepackage[scaled]{helvet}
	
	\usepackage[T1]{fontenc}
\fi

\makeatother

\usepackage[ backend=biber,style=ieee,sorting=none, maxnames = 10]{biblatex}
\bibliography{References} 

\DeclareBibliographyCategory{ignore}

\makeatletter
\def\blfootnote{\gdef\@thefnmark{}\@footnotetext}
\makeatother

\pgfplotscreateplotcyclelist{foo}{
        {thick,C1,mark=*},
        {thick,C2,mark=square*},
        {thick,C3, mark = triangle*},
        {thick,C4,mark=+},
        {thick,C5, mark = diamond*},
        {thick,C6, mark = x},
        {thick,C7, mark = pentagon*},
        {thick, C9, mark = 10-pointed star},
        {thick, C8, mark = Mercedes star},
        {thick, C100, mark = halfcircle*}
      }
\pgfplotsset{compat=1.18}
\begin{document}

\title{Rate-adaptive Reconciliation for Experimental Continuous-variable Quantum Key Distribution with Discrete Modulation over a Free-space Optical Link}

\author{
    Kadir G\" um\" u\c s, \textit{Student Member, IEEE},
    João dos Reis Frazão, \textit{Student Member, IEEE},
    Vincent van Vliet, \textit{Student Member, IEEE},
    Sjoerd van der Heide, \textit{Member, IEEE},
    Menno van den Hout, \textit{Student Member, IEEE},	
    Gabriele~Liga, \textit{Member, IEEE},
    Yunus Can G\" ultekin, \textit{Member, IEEE},
    Aaron Albores-Mejia, \textit{Member, IEEE},
    Thomas Bradley, \textit{Member, IEEE},
    Alex Alvarado, \textit{Senior Member, IEEE}, and Chigo Okonkwo, \textit{Senior Member, IEEE}
 \thanks{This work was supported by the Dutch Ministry of Economic Affairs and Climate Policy (EZK), as part of the PhotonDelta National GrowthFunds Programme on Photonics and the QuantumDeltaNL National Growthfunds on Quantum Technology. This article was presented in part at the Optical Fiber Communications (OFC) Conference, San Diego, CA, USA, March 2024~ \cite{gumucs2023adaptive}. \emph{(Corresponding author: Kadir G\" um\" u\c s})}
 \thanks{Kadir G\" um\" u\c s, João dos Reis Frazão, Vincent van Vliet, Sjoerd van der Heide, Menno van den Hout, Aaron Albores-Mejia, Thomas Bradley, and Chigo Okonkwo are with the High Capacity Optical Transmission Laboratory, Electro-Optical Communications Group, Eindhoven University of Technology, 5600 MB, Eindhoven, The Netherlands. (e-mails: k.gumus@tue.nl, j.c.dos.reis.frazao@tue.nl, v.v.vliet@tue.nl, s.p.v.d.heide@tue.nl, m.v.d.hout@tue.nl, a.albores.mejia@tue.nl, t.d.bradley@tue.nl, cokonkwo@tue.nl)}
 \thanks{Aaron Albores-Mejia, Alex Alvarado, and Chigo Okonkwo are with CubiQ Technologies, De Groene Loper 5, Eindhoven, The Netherlands (e-mails: aaron@cubiq-technologies.com, alex@cubiq-technologies.com, chigo@cubiq-technologies.com)}
 \thanks{Gabriele Liga, Yunus Can G\" ultekin, and Alex Alvarado are with the Information and Communication Theory Lab, Signal Processing Group, Eindhoven University of Technology, 5600MB, The Netherlands (e-mails: g.liga@tue.nl, y.c.g.gultekin@tue.nl, a.alvarado@tue.nl)}
 }%

\markboth{Journal of Lightwave Technologies, vol. xx, no. xx, Month Data, Year}%
{Journal of Lightwave Technologies, vol. xx, no. xx, Month Data, Year}

\maketitle


\begin{abstract}
Continuous-variable quantum key distribution (CV-QKD) has been proposed as a method for securely exchanging keys to protect against the security concerns caused by potential advancements in quantum computing. In addition to optical fibre transmission, the free-space optical (FSO) channel is an interesting channel for CV-QKD, as it is possible to share keys over this channel wirelessly. The instability of the FSO channel caused by turbulence-induced fading, however, can cause a degradation in the system's performance.
One of the most important aspects of CV-QKD is the reconciliation step, which significantly impacts the performance of the CV-QKD system. Hence, rate-adaptive reconciliation is necessary for CV-QKD over FSO to combat the fluctuations in the channel and improve secret key rates (SKRs). Therefore, in this paper, we simulate the impact of discrete modulation on the reconciliation efficiency and consider the use of $d$-dimensional reconciliation with $d > 8$ to mitigate this impact, improving reconciliation efficiencies by up to 3.4\%. We validate our results by experimentally demonstrating CV-QKD over a turbulent FSO link and demonstrate SKR gains by up to 165\%. Furthermore, we optimise the reconciliation efficiency for FSO links, achieving additional SKR gains of up to 7.6\%. 

\end{abstract}


\vspace{10mm}
\section{Introduction}
\IEEEPARstart{Q}{uantum} key distribution (QKD), first proposed in \cite{Bennett_2014}, has attracted considerable attention in recent years as concerns for information security grow. Using Shor's algorithm \cite{shor1999polynomial}, it would be possible to break current public-key cryptography protocols, assuming that sufficiently capable quantum computers can be developed. With continuous advances in quantum computing \cite{gyongyosi2019survey}, these concerns are expected to become a reality in the future. A potential solution to these issues is QKD, as it allows for the sharing of secret keys without a potential eavesdropper (Eve) with infinite computational power being able to recover the keys.

\begin{figure}[t!]
    \centering
    \tikzstyle{gradientc1} = [bottom color=C1!5, top color=C1!15]
\tikzstyle{gradientc4} = [bottom color=C4!5, top color=C4!15]    
\tikzstyle{gradientc6} = [bottom color=C6!5, top color=C6!15]
\tikzstyle{gradientc7} = [bottom color=C7!5, top color=C7!15]
\tikzstyle{gradientc8} = [bottom color=C8!5, top color=C8!15]
\begin{tikzpicture}[>=latex]
\definecolor{pastelblue}{RGB}{128, 206, 225}
\definecolor{pastelpurple}{RGB}{150, 111, 214}
\definecolor{pastelred}{RGB}{254, 216, 177}
\draw[rounded corners, C1, very thick, dashed] (0,0) rectangle(3,7.5) {};
\node[C1](Alice) at (1.5, 7){\large{Alice}};

\draw[rounded corners, C4, very thick, dashed] (5.5,0) rectangle(8.5,7.5) {};

\node[C4](Bob) at (7, 7){\large{Bob}};

\draw[rounded corners, C3, very thick, dashed] (3.25,6) rectangle(5.25,7.5) {};

\node[C3](Eve) at (4.25, 6.75){\large{Eve}};

\draw[rounded corners, gradientc7, C7, dashed] (3.1,0.05) rectangle(5.4,1.45) {};
\node[align = center, C7](Quantum channel) at (4.25, 0.2){\footnotesize Quantum};

\draw[rounded corners, gradientc8, C8, dashed] (3.1,1.55) rectangle(5.4,5.9) {};
\node[align = center, C8] at (4.25, 1.75){\footnotesize Classical};

\filldraw[rounded corners,gradientc1, C1, very thick] (0.25, 0.25) rectangle(2.75,1){};
\node[align = center, C1 ,font=\scriptsize\linespread{0.8}\selectfont](QST Alice) at (1.5,0.625){Transmission \\of quantum states};

\filldraw[rounded corners,gradientc4, C4, very thick] (5.75, 0.25) rectangle(8.25,1){};
\node[align = center, C4,font=\scriptsize\linespread{0.8}\selectfont](QST Bob) at (7,0.625){Measurement \\of quantum states};

\draw[->,thick, C1] (2.75,0.625) -- (5.75,0.625){};

\draw[<-,thick, C3] (3.35,0.625) -- (3.35,2.775){};
\draw[-,thick, C3] (3.35,2.975) -- (3.35,5.025){};
\draw[->,thick, C3] (3.35,5.225) -- (3.35,6){};
\draw[thick, C3](3.35,2.975) arc(90:270:0.1);
\draw[thick, C3](3.35,5.225) arc(90:270:0.1);


\begin{scope}[shift = {(0,-0.75)}]
\filldraw[rounded corners, gradientc1, C1, very thick] (0.25, 3.25) rectangle(2.75,4){};

\draw[->,thick, C1] (1.5,1.75) -- (1.5,3.25) node[midway,right,xshift = 3mm]{\footnotesize $\mathbf{x}$};

\node[align = center, C1](Reconciliation Alice) at (1.5,3.625){\scriptsize Reconciliation};

\filldraw[rounded corners, gradientc4, C4, very thick] (5.75, 3.25) rectangle(8.25,4){};

\node[align = center, C4](Reconciliation Bob) at (7,3.625){\scriptsize Reconciliation};

\draw[->,thick, C4] (7,1.75) -- (7,3.25) node[midway,right,xshift = 3mm]{\footnotesize $\mathbf{y}$};

\draw[<->,thick, C6] (2.75,3.625) -- (5.75,3.625){};

\draw[-,thick, C3] (4.25,3.625) -- (4.25,5.775){};
\draw[->,thick, C3] (4.25,5.975) -- (4.25,6.75){};
\draw[thick, C3](4.25,5.975) arc(90:270:0.1);
\end{scope}


\filldraw[rounded corners, gradientc1, C1, very thick] (0.25, 4.75) rectangle(2.75,5.5){};

\draw[->,thick, C1] (1.5,3.25) -- (1.5,4.75)
node[midway,right,xshift = 3mm]{\footnotesize $\mathbf{s}$};

\node[align = center,C1](PA Alice) at (1.5,5.125){\scriptsize Privacy amplification};

\filldraw[rounded corners, gradientc4, C4, very thick] (5.75, 4.75) rectangle(8.25,5.5){};

\node[align = center,C4](PA Bob) at (7,5.125){\scriptsize Privacy amplification};

\draw[->,thick, C4] (7,3.25) -- (7,4.75)node[midway,right,xshift = 3mm]{\footnotesize $\mathbf{\hat{s}}$};

\draw[->, thick,C1](1.5,5.5) -- (1.5, 6.75);
\node[align = center,C1,font=\scriptsize\linespread{0.8}\selectfont](Sequence Bob) at (2,6.125){ Secret \\ key};
\draw[->, thick,C4](7,5.5) -- (7, 6.75);
\node[align = center,C4,font=\scriptsize\linespread{0.8}\selectfont](Sequence Bob) at (7.5,6.125){Secret \\ key};

\draw[<->,thick, C6] (2.75,5.125) -- (5.75,5.125){};
\draw[->,thick, C3] (5.15,5.125) -- (5.15,6){};
\end{tikzpicture}
    \caption{An overview of a CV-QKD protocol.}
    \label{fig:QKD}
\end{figure}
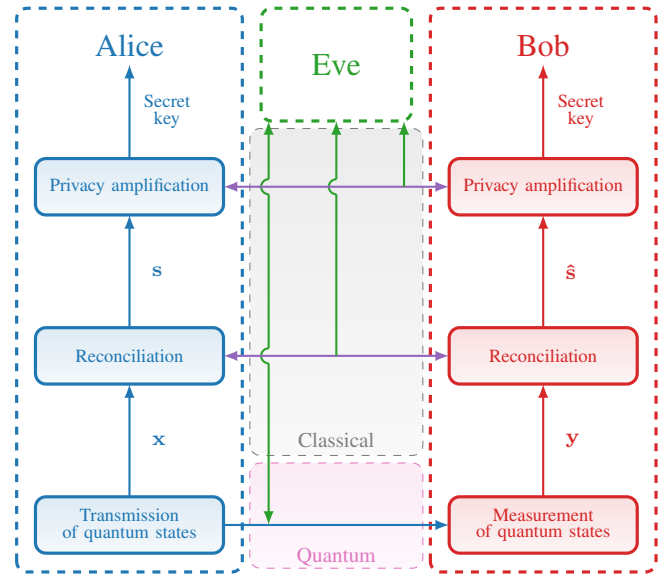

\indent QKD can be broadly categorised into two different variants: discrete-variable (DV) \cite{Bennett_2014} and continuous-variable (CV) \cite{GG02}. DV-QKD involves the transmission of single photons for the distillation of secret keys. Conversely, in CV-QKD, as shown in Fig. \ref{fig:QKD}, quantum random numbers are modulated on the in-phase and quadrature components of coherent light. This light undergoes substantial attenuation to generate weak coherent quantum states. Standard fibre optical telecommunication components can be used for generating and measuring the quantum signal in CV-QKD, which allows for a more cost-effective implementation, as opposed to DV-QKD, which requires single photon detectors~\cite{Laudenbach_2018}. The downside, however, is that post-processing is more challenging for CV-QKD. An important part of the post-processing is the reconciliation, where Alice and Bob try to share bits using error correction. The error correction codes used in CV-QKD are significantly more complex than those used for DV-QKD, and therefore, are the current bottleneck for implementing CV-QKD systems \cite{yang2023information}.

During reconciliation, the two involved parties, Alice and Bob, use their transmitted and measured quantum states, respectively, to exchange bits for secret key distillation. Alice generates the quantum states using either Gaussian modulation \cite{grosshans2003quantum}, in which she randomly generates a symbol based on a Gaussian distribution or discrete modulation \cite{ghorai2019asymptotic}, where she uses a finite-size constellation and randomly picks one of the constellation points. Although Gaussian modulation allows for longer distance QKD, practical implementation is difficult, as it is impossible to obtain a perfect Gaussian distribution using electro-optical modulators and digital-to-analog converters with finite resolution \cite{Kaur2021}. A system using discrete modulation formats mitigates this issue by approximating Gaussian modulation using probabilistically shaped quadrature amplitude modulation (PS-QAM) \cite{cho2019probabilistic}. Analysis of different discrete modulation formats for CV-QKD has been performed in, e.g., \cite{Denys_2021,Notarnicola_2024}.

A commonly used protocol for reconciliation is multi-dimensional reconciliation, first introduced in \cite{Leverrier_2008}. This protocol is especially well-suited for longer-distance CV-QKD systems, as it outperforms slice reconciliation \cite{Assche2004}, which is mostly used for links of very short distance, in this regime \cite{yang2023information}. Low-rate, low-density parity-check (LDPC) codes were designed and simulated for multi-dimensional reconciliation in \cite{Jouguet_2011}. In \cite{Milicevic_2018}, quasi-cyclic LDPC codes were implemented in a graphics processing unit for multi-dimensional reconciliation. Furthermore, in a recent work, real-time error correction for reconciliation was implemented on field programmable gate arrays \cite{Zhou2023}. An open-source library for the simulation of reconciliation has recently become available in \cite{Cil24QCrypt}.

Another important aspect of reconciliation is the rate-adaptivity, i.e., adapting the rate of the error correction code to match the channel signal-to-noise ratio (SNR). A rate-adaptive version of the multi-dimensional reconciliation protocol was proposed in \cite{wang2017efficient}, allowing the CV-QKD system to adapt to the quantum channel fluctuations. This is especially relevant for free-space optical (FSO) channels, as atmospheric turbulence causes changes to the channel over time, resulting in a channel with an unstable SNR \cite{Kaur2021}. To guarantee good error correction performance, the rate of the code needs to be adapted to match the changes in such a fluctuating channel. The FSO channel can be particularly important for CV-QKD, as it allows for wireless key sharing in cases where a direct fibre link is not possible. An additional review of works on reconciliation can be found in \cite{yang2023information}. 

Experimental demonstrations of CV-QKD, such as over fibre in \cite{eriksson2019wavelength, hajomer2022modulation, Zhang_2020} and over free-space in \cite{Joe2024,Wang:21,Shen2019} have been performed. However, reconciliation is rarely discussed in detail, and reconciliation efficiencies and frame error rates (FERs) are often assumed. The performance of the reconciliation significantly influences the achievable distances and key rates of these experimental demonstrations. Therefore, we argue that there is a need for more work focusing on reconciliation in experimental demonstrations, as the current literature on reconciliation is mostly theoretical.

 Multi-dimensional reconciliation involves the use of multiplications and divisions of ($d = \{1,2,4,8\}$-) dimensional numbers constructed using the Cayley-Dickson construction \cite{dickson1919quaternions} in an attempt to construct a virtual channel. The constructed virtual channel is similar to a binary-input additive white Gaussian noise (BI-AWGN) channel. This BI-AWGN channel allows for the design of high-performance error correction codes, which is important as the performance of the error correction greatly affects the achievable secret key rates (SKRs) of CV-QKD \cite{Laudenbach_2018}. As shown in~\cite{zhou2018practical}, The higher the dimensionality $d$ of the reconciliation, the more closely the virtual channel resembles a BI-AWGN channel, hence, higher SKRs can be achieved. The dimensionality of reconciliation is often limited to $d = 8$ in literature. This is because the multiplications and divisions used in multi-dimensional reconciliation are only defined for $d = \{1,2,4,8\}$ when constructing such numbers according to the Cayley-Dickson construction \cite{dickson1919quaternions}.

In \cite{Leverrier_2008, Jouguet_2011} multi-dimensional reconciliation with $d > 8$, which we will refer to as high-dimensional reconciliation from this point forward, has been proposed and analysed. Instead of using multiplications and divisions, a mapping matrix is used. This high-dimensional reconciliation shows significant gains compared to when $d \leq 8$. This method involves the continuous generation of orthogonal random matrices, which will be used for constructing the mapping matrix. A new mapping matrix needs to be constructed for each set of $d$ symbols for security reasons \cite{Leverrier_2008}. 
However, recent works have not investigated further methods for high-dimensional reconciliation, and we conjecture that this is because of the higher complexity of the protocol caused by the generation of the orthogonal matrices. Additionally, to the best of our knowledge, no analysis on high-dimensional reconciliation for discrete modulation formats for different code rates has been performed. 

In this work, we investigate how the use of discrete modulation formats impacts the performance of error correction over a wide range of code rates. We consider the use of high-dimensional reconciliation to improve the reconciliation efficiency, showing improvements in reconciliation efficiencies, especially for short- and mid-range CV-QKD systems, by up to 3.4\%. Furthermore, we experimentally demonstrate high-dimensional reconciliation with CV-QKD transmission over a turbulent FSO channel with varying turbulence strengths and show that we can increase SKRs by up to 165\% compared to conventional multi-dimensional reconciliation. Finally, we show how optimising the reconciliation efficiency to take into account the fluctuations in optical power induced by fading in the FSO channel allows for an additional 7.6\% gain in SKR.

The remainder of the paper is organised as follows. In Section \ref{Multi-dimensional Reconciliation}, we discuss the multi-dimensional reconciliation conventionally used in CV-QKD systems. Section \ref{High-dimensional Reconciliation} expands on this by explaining the high-dimensional reconciliation and shows how the dimensionality of the reconciliation affects the error correction performance for different discrete modulation formats. Section \ref{Rate Adaptivity} explains the idea of rate-adaptivity and how it influences the error correction performance. Our results are experimentally validated in Section \ref{Experimental Results}. Finally, in Section \ref{Conclusion}, we conclude our work and give suggestions for potential future research. 

\section{Multi-dimensional Reconciliation}
\label{Multi-dimensional Reconciliation}

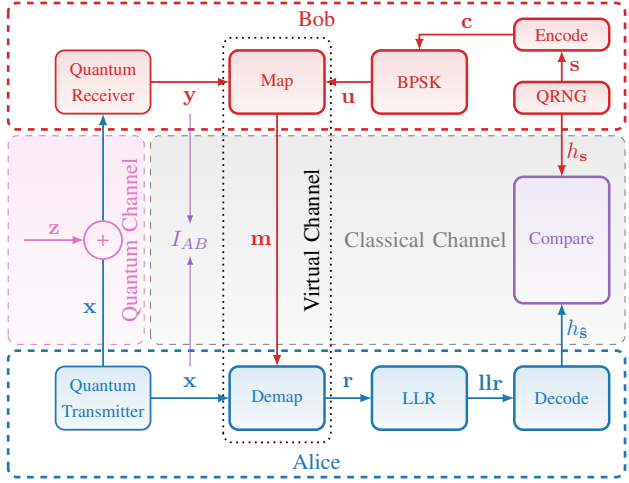
\begin{figure}[t!]
    \centering
    \resizebox{\linewidth}{!}{\tikzstyle{gradientc1} = [bottom color=C1!5, top color=C1!15]
\tikzstyle{gradientc4} = [bottom color=C4!5, top color=C4!15]    
\tikzstyle{gradientc6} = [bottom color=C6!5, top color=C6!15]
\tikzstyle{gradientc8} = [bottom color=C8!5, top color=C8!15]
\tikzstyle{gradientc7} = [bottom color=C7!5, top color=C7!15]
\begin{tikzpicture}[>=latex]

\draw[rounded corners,dashed,gradientc7, C7](-2,2.35) rectangle (0.15,5.65);
\draw[rounded corners,C1,gradientc1,thick] (-1.25,1) rectangle node[midway, align = center]{\footnotesize Quantum \\ \footnotesize Transmitter} (0.25,2); 

\draw[rounded corners,C4,gradientc4,thick] (-1.25,6) rectangle node[midway, align = center]{\footnotesize Quantum \\ \footnotesize Receiver} (0.25,7); 
\draw[C1,->,thick](-0.5,2) -- node[left, yshift = -10.75mm, xshift = -1mm]{$\mathbf{x}$}(-0.5,6);
\draw[thick,C7,gradientc7] (-0.5,4) circle (0.3cm) node {+};
\draw[thick,C7,->](-1.75,4) -- node[midway,above, yshift  = 1mm]{$\mathbf{z}$}(-0.8,4);
\node[rotate = 90,C7] at (-0.05,4){Quantum Channel};

\draw[rounded corners, gradientc8, C8, dashed] (0.25,2.35) rectangle(7.75,5.65) {};
\node[C8] at (4.6,4) {Classical Channel};

\draw[very thick, rounded corners, color = C1,dashed] (-2,0.25) rectangle (7.75,2.25);
\node at (2.875,0.5){\textcolor{C1}{Alice}};

\draw[rounded corners, draw = C1, line width = 0.4mm,gradientc1] (1.5,1) rectangle (3,2);

\draw[rounded corners, draw = C1, line width = 0.4mm,gradientc1] (3.75,1) rectangle (5.25,2);

\draw[rounded corners, draw = C1, line width = 0.4mm,gradientc1] (6,1) rectangle (7.5,2);

\node at (6.75,1.5){\footnotesize \textcolor{C1}{Decode}};
\node at (4.5,1.5){\footnotesize \textcolor{C1}{LLR}};
\node at (2.25,1.5){\footnotesize \textcolor{C1}{Demap}};

\draw[->,thick,draw = C1] (0.25,1.5) -- (1.5,1.5);
\node at (0.875,1.75) {\textcolor{C1}{$\mathbf{x}$}};

\draw[->,thick,draw  =C1] (3,1.5) -- (3.75,1.5);
\node at (3.375,1.75){\textcolor{C1}{$\mathbf{r}$}};

\draw[->,thick, draw = C1] (5.25,1.5) -- (6,1.5);
\node at (5.625,1.75){\textcolor{C1}{$\mathbf{llr}$}};


\draw[very thick, rounded corners, color = C4,dashed] (-2,7.75) rectangle (7.75,5.75);
\node at (2.875,7.5){\textcolor{C4}{Bob}};

\draw[rounded corners,  draw = C4, line width = 0.4mm,gradientc4] (1.5,7) rectangle (3,6);

\draw[rounded corners, draw = C4,line width = 0.4mm,gradientc4] (6,7) rectangle (7.5,7.5);

\draw[rounded corners, draw = C4,line width = 0.4mm,gradientc4] (3.75,6) rectangle (5.25,7);

\draw[rounded corners, draw = C4, line width = 0.4mm,gradientc4] (6,6) rectangle (7.5,6.5);

\node at (6.75,6.25){\footnotesize \textcolor{C4}{QRNG}};
\node at (6.75,7.25){\footnotesize \textcolor{C4}{Encode}};
\node at (4.5,6.5){\footnotesize \textcolor{C4}{BPSK}};
\node at (2.25,6.5){\footnotesize \textcolor{C4}{Map}};

\draw[->,thick, draw  =C4] (0.25,6.5) -- (1.5,6.5);
\node at (0.875,6.25) {\textcolor{C4}{$\mathbf{y}$}};

\draw[->,thick, draw = C4] (6.75,6.5) -- node[midway,right, xshift = 1.25mm]{\textcolor{C4}{$\mathbf{s}$}} (6.75,7);

\draw[->,thick,draw = C4](6,7.25) -- node[midway,above,C4,yshift = 1mm]{$\mathbf{c}$}(4.5,7.25) -- (4.5,7){};

\draw[<-,thick,draw = C4] (3,6.5) -- (3.75,6.5);
\node at (3.375,6.25){\textcolor{C4}{$\mathbf{u}$}};

\draw[->,thick, draw = C4](2.25,6) -- (2.25,2);
\node at (2,4){\textcolor{C4}{$\mathbf{m}$}};
\draw[->,draw = C6](0.875,6) -- (0.875,4.25);
\draw[->,draw = C6](0.875,2) -- (0.875,3.75);
\node at (0.875,4){\textcolor{C6}{$I_{AB}$}};

\draw[rounded corners, draw = C6, line width = 0.4mm,gradientc6] (6,5) rectangle (7.5,3);
\node[C6] at (6.75,4){\footnotesize Compare};
\draw[->,thick, draw = C4](6.75,6) -- (6.75,5);
\draw[->,thick, draw = C1](6.75,2) -- (6.75,3);
\node at (7,2.6){\textcolor{C1}{$h_{\mathbf{\hat{s}}}$}};
\node at (7,5.4){\textcolor{C4}{$h_{\mathbf{s}}$}};



\draw[thick, dotted, rounded corners] (1.4,0.8) rectangle (3.1,7.2);
\node[rotate = 90] at (2.8,4){Virtual Channel};
\end{tikzpicture}}
    \caption{ An overview of the reverse multi-dimensional reconciliation protocol for CV-QKD.}
    \label{fig:Reconciliation}
\end{figure}

In a conventional CV-QKD protocol, reconciliation takes place after the transmission and measurement of the quantum states. During reconciliation, Alice and Bob try to securely exchange a string of bits, which will be used for the secret key distillation in the privacy amplification step. The most common form is reverse reconciliation where the algorithm terminates at the transmitter side Alice.  This achieves longer distances when compared to direct reconciliation, which is affected by the so-called 3~dB limit \cite{Laudenbach_2018}. An overview of reverse reconciliation is given in Fig. \ref{fig:Reconciliation}, and described in what follows. 

The multi-dimensional reconciliation we will describe in this section involves the use of high dimensional numbers and is only defined for  $d \in \{1 \ \text{(real)} ,2 \ \text{(complex)} ,4 \ \text{(quaternion)} ,8 \ \text{(octonion)}\}$ \cite{Leverrier_2008}. All of the multiplications and divisions between vectors of length $d$ in the rest of this section correspond to those between two $d$-dimensional numbers.

At the start of reconciliation, Alice and Bob have their transmitted and measured coherent quantum states, $\mathbf{x}$ and $\mathbf{y}$, respectively. Here, $\mathbf{x}~=~[x_1^I, x_1^Q,x_2^I,\cdots,x_{N/2}^Q]$, is a sequence of real symbols of length $N$, randomly chosen from a constellation $\mathcal{X}$ according to the probabilities assigned during probabilistic shaping, where $I$ and $Q$ refer to the in-phase and quadrature components of the constellation, and
\begin{equation}
    \mathbf{y} = \mathbf{x} + \mathbf{z},
    \label{eq. channel}
\end{equation} where $\mathbf{z} = [z_1^I, z_1^Q,z_2^I,\cdots,z_{N/2}^Q]$ are independent and identically distributed Gaussian noise samples with zero mean and variance $\sigma_z^2/2$.
In this work, we assume that $\mathbb{E}[||x_i^I||^2 + ||x_i^ Q||^2] = 1$ with $\mathbb{E}[||x_i^I||^2] = \mathbb{E}[||x_i^Q||^2] = \frac{1}{2}$, i.e., the variance of a quantum state is 1.
The signal-to-noise ratio (SNR) of the quantum channel is thus defined to be $1/\sigma_z^2$. 
 For the rest of the paper, we will use the notation $\mathbf{x} = [x_1,x_2,\cdots,x_N]$ where $[x_{2i-1}, x_{2i}] = [x_{i}^I, x_{i}^Q] \ \ \forall i \in 1,2,\cdots N/2$. The same notation will be used as well for $\mathbf{y}$ and $\mathbf{z}$. 
\\\indent Using a quantum random number generator (QRNG), Bob generates a string of bits $\mathbf{s} = [s_1,s_2,\cdots,s_K]$ of length $K$, which are the information bits that will be used for key distillation. Bob encodes these bits into a codeword $\mathbf{c} = [c_1,c_2,\cdots c_N]$ using a rate $R = K/N$ error correction code. The bits in $\mathbf{c}$ are then converted to binary phase-shift keying (BPSK) symbols $\mathbf{u}$ using ${u_i} = (-1)^{c_i}$. 

We split $\mathbf{u}$, $\mathbf{y}$, and $\mathbf{x}$ into $N/d$ vectors of length $d$, $\mathbf{u}'_j = [u_{d\cdot j-d+1},u_{d\cdot j-d+2}, \cdots, u_{d\cdot j}]$, $\mathbf{y}'_j = [y_{d\cdot j-d+1},y_{d\cdot j-d+2}, \cdots, y_{d\cdot j}]$, and $\mathbf{x}'_j = [x_{d\cdot j-d+1},x_{d\cdot j-d+2}, \cdots, x_{d\cdot j}]$ $\forall j = 1,2,\cdots N/d$.
Here, $N$ is assumed to be divisible by $d$.
Each new vector of length $d$ corresponds to a $d$-dimensional number. Bob calculates $\mathbf{m}'_j$ using
\begin{equation}
    \mathbf{m}'_j = \mathbf{u}'_j\mathbf{y}'_j, \quad \forall j \in {1,2,\cdots,N/d},
    \label{eq:mod}
\end{equation}
and communicates $\mathbf{m} = [\mathbf{m}'_1, \mathbf{m}'_2, \cdots, \mathbf{m}'_{N/d}]$ to Alice over the classical authentication channel, which is assumed to be error-free. Eve has full access to the classical authentication channel and can obtain any information that is transmitted over that channel.

To recover $\mathbf{u}$ from $\mathbf{m}$, Alice divides $\mathbf{m}$ by her transmitted states $\mathbf{x}$
\begin{equation}
    \mathbf{r}'_j = \frac{\mathbf{m}'_j}{\mathbf{x}'_j} \ \ \ \ \ \ \ \forall j \in {1,2,\cdots,N/d}.
    \label{eq:Demod}
\end{equation}
Doing the multiplication in \eqref{eq:mod} and division in \eqref{eq:Demod} creates a virtual channel closely resembling a BI-AWGN channel. The BI-AWGN channel is a frequently-used model for the design of capacity-approaching error correction code design, i.e., with a high reconciliation efficiency $\beta$. From \eqref{eq. channel} and \eqref{eq:Demod}, we write that
\begin{IEEEeqnarray}{rCl} 
\mathbf{r}'_j = \frac{\mathbf{m}'_j}{\mathbf{x}'_j} &=&  \frac{\mathbf{u}'_j\mathbf{y}'_j}{\mathbf{x}'_j} \nonumber\\
&=& \frac{\mathbf{u}'_j(\mathbf{x}'_j + \mathbf{z}'_j)}{\mathbf{x}'_j} \nonumber\\
&=& \mathbf{u}'_j + \frac{\mathbf{u}'_j\mathbf{z}'_j}{\mathbf{x}'_j}. 
\label{eq:VChannel}
\end{IEEEeqnarray}
Because $\mathbf{y}$ is a noisy version of $\mathbf{x}$, Alice is left with the original BPSK symbols $\mathbf{u}'_j$, but with additive noise $\mathbf{n}'_j = \frac{\mathbf{u}'_j\mathbf{z}'_j}{\mathbf{x}'_j}$. We also define $\mathbf{n} = [\mathbf{n}'_1, \mathbf{n}'_2, \cdots \mathbf{n}'_{N/d}]$. 

We show the noise distribution of the virtual channel in \eqref{eq:VChannel} in the following. We will consider the case where $d = 1$, in which case $n'_j = \frac{u'_jz'_j}{x'_j}$. The random variables are all i.i.d. and therefore do not depend on $j$, hence for the probability distributions we will remove the subscript $j$. As mentioned before, the noise distribution of $z'$ is:
\begin{equation}
    f(z') = \mathcal{N}\Bigg(0,\frac{\sigma_z^2}{2}\Bigg),
\end{equation}
as the source of the noise is the quantum channel, which is assumed to be AWGN with noise variance $\sigma_z^2/2$ per quadrature. Because Alice knows the value of $x'$, the probability distribution of $n'$, conditioned on $x'$, is:
\begin{equation}
    f(n'|x') = \mathcal{N}\Bigg(0,\frac{\sigma_z^2}{2x'^2}\Bigg),
\end{equation}
as the probability distribution does not depend on $u'$, i.e., $f(n'|x',u' = 1) = f(n'|x',u' = -1)$. 
Alice, therefore, observes a channel where each channel use $j$ has a different SNR, which is dependent on $x'^2$. 
In \cite{Milicevic_2018} it is shown that this result generalises for $d = \{2,4,8\}$ to
\begin{equation}
    f(\mathbf{n}'|\mathbf{x}') = \mathcal{N}\Bigg(\mathbf{0},\mathbf{I}_d\frac{d\sigma_z^2}{2||\mathbf{x}'||^2}\Bigg),
\end{equation}
where $\mathbf{I}_d$ is a $d \times d$ identity matrix.
Hence, the noise variance $\sigma_j^2$ for each set of $d$ received symbols is different for each block of $j$ symbols and is dependent on $\mathbf{x}'_j$ via
\begin{equation}
    \sigma_j^2 = \frac{d\sigma_z^2}{2||\textbf{x}'_j||^2}.\label{eq:effectivenoisevariance}
\end{equation}
Alice calculates the log-likelihood ratios (LLRs) of her message $\mathbf{r}'_j$ \cite{Milicevic_2018}
\begin{equation}
    \mathbf{llr}'_j = \frac{2\mathbf{r}'_j}{\sigma_j^2}.
\end{equation}
Using $\mathbf{llr} = [\mathbf{llr}'_1,\mathbf{llr}'_2,\cdots,\mathbf{llr}'_{N/d}]$, Alice attempts to decode the codeword to estimate the information bits $\hat{\mathbf{s}}$. 

\begin{figure}[b!]
    \resizebox{\linewidth}{!}{\begin{tikzpicture}
    \node at (0,0){\begin{tikzpicture}
\begin{axis}[
    width=4cm,
    height=4.5cm,
    xmin=-10, xmax=10,
    ymin=0, ymax=0.3,
    ylabel = Probability Density,
    yticklabel shift = 1mm,
    xticklabel shift = 1mm,
    ylabel shift = 3mm,
    ytick = {0,0.1,0.2,0.3},
    yticklabels = {0,0.1,0.2,0.3},
   scaled y ticks = false, 
]
\pgfplotstableread{Figures/pdf_d1.txt}
\datatable
\addplot+[ybar interval,mark=no, color = C1, fill = C1, fill opacity = 0.2] table [x, y, col sep=space] {\datatable};
\pgfplotstableread{Figures/pdf_Gaussian.txt}
\datatable
\addplot[color = C4, no marks, thick]
         table
         [
          x expr=\thisrowno{0}, 
          y expr=\thisrowno{1} 
         ] {\datatable};
\end{axis}
\end{tikzpicture}};
    \node at (3.55,-0.25){\begin{tikzpicture}
\begin{axis}[
    width=4cm,
    height=4.5cm,
    xmin=-10, xmax=10,
    ymin=0, ymax=0.3,
    xlabel = $n$,
    yticklabel shift = 1mm,
    xticklabel shift = 1mm,
    xlabel shift = 2mm,
    scaled y ticks = false, 
    ytick = {0,0.1,0.2,0.3},
    yticklabels = {0,0.1,0.2,0.3},
]
\pgfplotstableread{Figures/pdf_d8.txt}
\datatable
\addplot+[ybar interval,mark=no, color = C1, fill = C1, fill opacity = 0.2] table [x, y, col sep=space] {\datatable};
\pgfplotstableread{Figures/pdf_Gaussian.txt}
\datatable
\addplot[color = C4, no marks, thick]
         table
         [
          x expr=\thisrowno{0}, 
          y expr=\thisrowno{1} 
         ] {\datatable};
\end{axis}
\end{tikzpicture}};
    \node at (7,0){\begin{tikzpicture}
\begin{axis}[
    width=4cm,
    height=4.5cm,
    xmin=-10, xmax=10,
    ymin=0, ymax=0.3,
    yticklabel shift = 1mm,
    xticklabel shift = 1mm,
    scaled y ticks = false, 
    ytick = {0,0.1,0.2,0.3},
    yticklabels = {0,0.1,0.2,0.3},
]
\pgfplotstableread{Figures/pdf_d128.txt}
\datatable
\addplot+[ybar interval,mark=no, color = C1, fill = C1, fill opacity = 0.2] table [x, y, col sep=space] {\datatable};
\pgfplotstableread{Figures/pdf_Gaussian.txt}
\datatable
\addplot[color = C4, no marks, thick]
         table
         [
          x expr=\thisrowno{0}, 
          y expr=\thisrowno{1} 
         ] {\datatable};
\end{axis}
\end{tikzpicture}};

    \node at (0.5,2.3){\Large $d = 1$};
    \node at (3.85,2.3){\Large $d = 8$};
    \node at (7.2,2.3){\Large $d = 128$};
\end{tikzpicture}}
    \caption{The histograms of $n$ for different $d$ assuming Gaussian modulation for $\mathbf{x}$ with $10^6$ samples at an SNR of -4.7~dB. The red line corresponds to the probability density function of $n$ for a BI-AWGN channel with the same SNR.}
    \label{fig:pdf}
\end{figure}
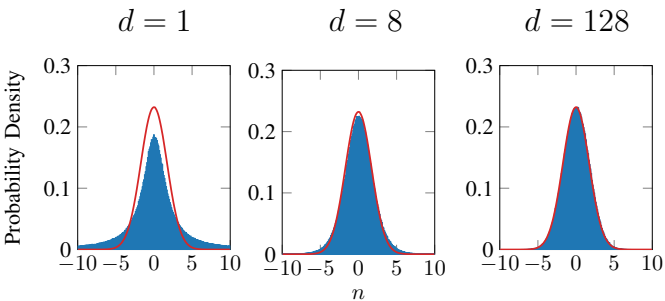

\begin{figure*}[t!]
\begin{subfigure}{0.33\textwidth}
      \begin{tikzpicture}
\begin{axis}[
every axis/.append style={font=\small},
tick label style={font=\footnotesize},
xlabel = $\beta$,
ylabel = FER,
xmin = 0.9, xmax =1,
ymin = 0.001, ymax = 1,
ymode = log,
x tick label style={yshift= -1mm},
y tick label style={xshift= -1mm},
ylabel shift = 1mm,
xlabel shift = 1mm,
width=\textwidth,
height=7cm,
set layers, mark layer=axis tick labels,
grid = major,
 xlabel near ticks,  
 ylabel near ticks,  
 xticklabel style={/pgf/number format/fixed},
every axis plot/.append style={thick},legend style={at={(0.55,0.3)},anchor=west, font = \scriptsize,row sep=-0.75ex,inner sep=0.2ex},
legend cell align={left},
cycle list name = foo
]

\pgfplotstableread{Figures/FER_R5.txt}
\datatable
\foreach \x in {1,2,3,4,5,6,7,8,9,10}{
\addplot+
         table
         [
          x expr=\thisrowno{0}, 
          y expr=\thisrowno{\x} 
         ] {\datatable};
}
\addlegendentry{BI-AWGN}
\addlegendentry{$d = 1$}
\addlegendentry{$d = 2$}
\addlegendentry{$d = 4$}
\addlegendentry{$d = 8$}
\addlegendentry{$d = 16$}
\addlegendentry{$d = 32$}
\addlegendentry{$d = 64$}
\addlegendentry{$d = 128$}
\addlegendentry{$d = 256$}
\end{axis}
\end{tikzpicture}
\end{subfigure}
\begin{subfigure}{0.33\textwidth}
      \begin{tikzpicture}
\begin{axis}[
every axis/.append style={font=\small},
tick label style={font=\footnotesize},
xlabel = $\beta$,
ylabel = FER,
xmin = 0.9, xmax =1,
ymin = 0.001, ymax = 1,
ymode = log,
x tick label style={yshift= -1mm},
y tick label style={xshift= -1mm},
ylabel shift = 1mm,
xlabel shift = 1mm,
width=\textwidth,
set layers, mark layer=axis tick labels,
grid = major,
height=7cm,
 xlabel near ticks,  
 ylabel near ticks,  
 xticklabel style={/pgf/number format/fixed},
every axis plot/.append style={thick},legend style={at={(0.55,0.3)},anchor=west, font = \scriptsize,row sep=-0.75ex,inner sep=0.2ex},
legend cell align={left},
cycle list name = foo
]

\pgfplotstableread{Figures/FER_R10.txt}
\datatable
\foreach \x in {1,2,3,4,5,6,7,8,9,10}{
\addplot+
         table
         [
          x expr=\thisrowno{0}, 
          y expr=\thisrowno{\x} 
         ] {\datatable};
}
\addlegendentry{BI-AWGN}
\addlegendentry{$d = 1$}
\addlegendentry{$d = 2$}
\addlegendentry{$d = 4$}
\addlegendentry{$d = 8$}
\addlegendentry{$d = 16$}
\addlegendentry{$d = 32$}
\addlegendentry{$d = 64$}
\addlegendentry{$d = 128$}
\addlegendentry{$d = 256$}
\end{axis}
\end{tikzpicture}
\end{subfigure}
\begin{subfigure}{0.33\textwidth}
      \begin{tikzpicture}
\begin{axis}[
every axis/.append style={font=\small},
tick label style={font=\footnotesize},
xlabel = $\beta$,
ylabel = FER,
xmin = 0.9, xmax =1,
ymin = 0.001, ymax = 1,
ymode = log,
x tick label style={yshift= -1mm},
y tick label style={xshift= -1mm},
ylabel shift = 1mm,
xlabel shift = 1mm,
width=\textwidth,
height=7cm,
set layers, mark layer=axis tick labels,
grid = major,
 xlabel near ticks,  
 ylabel near ticks,  
 xticklabel style={/pgf/number format/fixed},
every axis plot/.append style={thick},legend style={at={(0.55,0.3)},anchor=west, font = \scriptsize,row sep=-0.75ex,inner sep=0.2ex},
legend cell align={left},
cycle list name = foo
]

\pgfplotstableread{Figures/FER_R50.txt}
\datatable
\foreach \x in {1,2,3,4,5,6,7,8,9,10}{
\addplot+
         table
         [
          x expr=\thisrowno{0}, 
          y expr=\thisrowno{\x} 
         ] {\datatable};
}
\addlegendentry{BI-AWGN}
\addlegendentry{$d = 1$}
\addlegendentry{$d = 2$}
\addlegendentry{$d = 4$}
\addlegendentry{$d = 8$}
\addlegendentry{$d = 16$}
\addlegendentry{$d = 32$}
\addlegendentry{$d = 64$}
\addlegendentry{$d = 128$}
\addlegendentry{$d = 256$}
\end{axis}
\end{tikzpicture}
\end{subfigure}
    \caption{Simulated FER vs. $\beta$ for different rate (Left: $R = \frac{1}{5}$, Middle: $R = \frac{1}{10}$, Right: $R = \frac{1}{50}$) TPB-LDPC codes for different $d$.}
    \label{FER_MD}
\end{figure*}
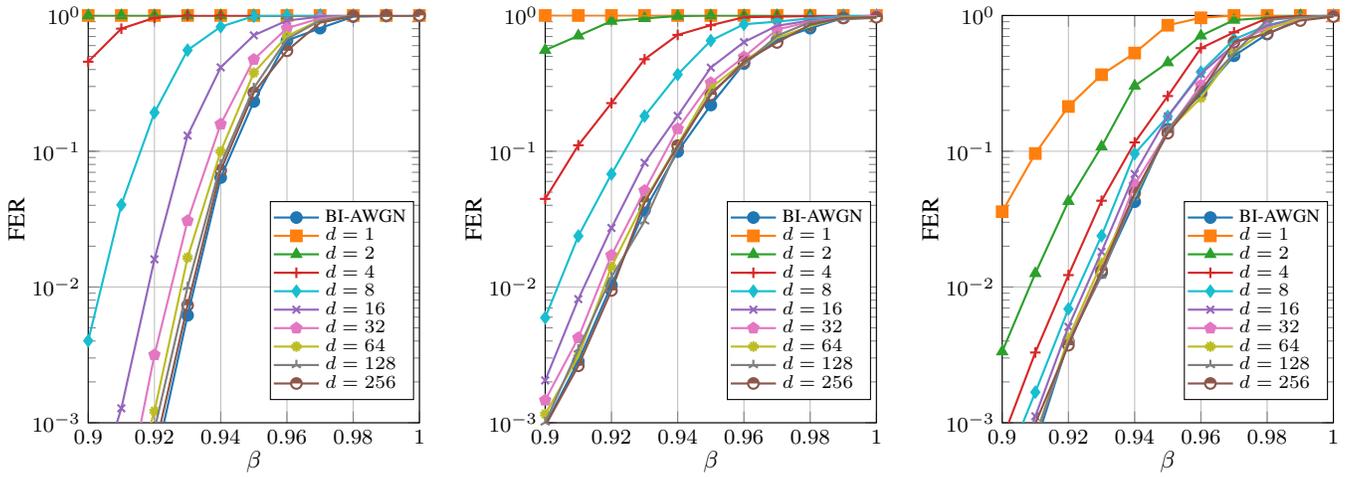

As mentioned before, the virtual channel has a different noise variance for each block of $d$ symbols. The distribution of the noise, therefore, does not correspond to that of a BI-AWGN channel where each channel use has the same noise variance. The distribution of $n$, were $n$ is a value randomly sampled from $\mathbf{n}$, is not Gaussian, as shown in Fig. \ref{fig:pdf} for $d = 1$ and $d = 8$. For $d = 1$, the distribution of $n$ is wider and less Gaussian compared to $d = 8$. The noise distribution more closely resembles that of a fading channel \cite{Li2005}, which is because of fluctuations in $\sigma_j^2$. The capacity of the fading channel is lower compared to an AWGN channel with the same average SNR, hence the capacity of the virtual channel is lower than the quantum channel. As a result, the reconciliation efficiency suffers as the error correction codes are applied to a channel with lower capacity, i.e., the code rate decreases.

For the virtual channel to be BI-AWGN, we can use a modulation format where each symbol has equal power (such as quadrature phase-shift keying (QPSK)), such that $\sigma_j^2$ is constant as $||\mathbf{x}'_j||^2$ is always the same for QPSK.
For other modulation formats, we can increase $d$ because of the following: let $||\mathbf{x}'_j||^2 = \sum_{i = 1}^d ||x_{\cdot j - d + i}||^2$, where $||x_{\cdot j - d + i}||^2$ is i.i.d. with mean $\mu_{||x||^2}$ and variance $\sigma^2_{||x||^2}$, which are both finite. The law of large numbers \cite{revesz2014laws} states that if we sample and sum $d$ independent random samples from an i.i.d. process, the sum will converge to $d\mu$ as $d \to \infty$, and the variance of the sum will be equal to $\frac{\sigma^2}{d}$. Therefore, because $\sigma^2_{||x||^2}$ is finite, $\text{Var}(||\mathbf{x}'||^2) \to 0$ , and consequently $\text{Var}(\sigma_j^2) \to 0$, when $d \to \infty$. This means that $\sigma_j^2$ will have the same value for each $j$ when $d$ is infinite. By increasing $d$, we are effectively reducing the amount of fading in the channel as we are reducing the variance of $||\mathbf{x}'||^2$. Note that QPSK modulation heavily limits the transmission distance for CV-QKD systems when compared to PS-QAM or Gaussian modulation \cite{Kaur2021}. Therefore, using higher dimensional reconciliation is necessary, which will be discussed in the next section.  

Because of the relatively high FERs used in CV-QKD systems, the probability of a miscorrection on Alice's side is quite high.
Therefore, after decoding, Alice and Bob confirm that $\mathbf{s}$ and $\hat{\mathbf{s}}$ are the same using a cyclic redundancy check (CRC). Although it is still possible that $\mathbf{s} \neq \mathbf{\hat{s}}$ after the CRC, it can be shown that the probability of this happening is close to 0 \cite{Milicevic_2018}. If the decoding is successful, the bits will be used during privacy amplification to distil the secret key, otherwise, a frame error has occurred and the frame will be discarded.   

The SKR for a CV-QKD protocol using multi-dimensional reconciliation depends on both the rate of the code and the frame error rate (FER) via \cite{Leverrier_2008}
\begin{equation}
    \text{SKR} = (1-\text{FER})(\beta I_{AB} - \chi_{BE} -\Delta),
\end{equation}
where $\beta = \frac{R}{I_{AB}}$ is the reconciliation efficiency, $I_{AB}$ is the mutual information between Alice's transmitted and Bob's measured quantum states, $\chi_{BE}$ is the Holevo information between Bob's and Eve's measurements, and $\Delta$ is the penalty caused by the finite size effects of the privacy amplification \cite{Milicevic_2018}. We calculate $\Delta$ using:
\begin{equation}
    \Delta = 7\sqrt{\frac{\log_2(2/\epsilon)}{N_{privacy}}},
\end{equation} where $\epsilon$ is the security parameter, which was chosen to be $10^{-10}$ \cite{leverrier2010finite}, and $N_{privacy}$ is the privacy amplification block size. There is a trade-off between $\beta$ and FER, as the FER increases when $\beta$ increases, as we are operating with a code rate closer to capacity \cite{Eriksson:20}. The optimum SKR is often achieved at a FER around 10-30\% \cite{yang2023information}.

\section{High-dimensional Reconciliation}
\label{High-dimensional Reconciliation}
Instead of using multiplication and division during the construction of the virtual channel, it is possible to go beyond $d = 8$ for the multi-dimensional reconciliation using a different construction as proposed in \cite{Leverrier_2008}. We will refer to multi-dimensional reconciliation with $d > 8$ as high-dimensional reconciliation in the rest of this section. We will use a $d\times d$ mapping matrix $\mathbf{R}_j$ which maps $\mathbf{y}'_j$ to $\mathbf{u}'_j$:
\begin{equation}
    \mathbf{R}_j\mathbf{y}'_j = \mathbf{u}'_j.
\end{equation}
This mapping matrix can be transmitted over the channel without leaking any information as $\mathbf{R}_j$ and $\mathbf{u}'_j$ are statistically independent if $\mathbf{R}_j$ is constructed in the following manner \cite{Leverrier_2008}. 

 First, we generate a random orthogonal $d\times d$ matrix $\mathbf{Q}_j$ according to the Haar measure \cite{collins2006integration}. This can be done by taking the orthogonal part of the QR decomposition of a normalised random Gaussian matrix. The generation of $\mathbf{Q}_j$ in this way has complexity $\mathcal{O}(d^3)$, but a simplified method exists with complexity $\mathcal{O}(d^2)$ as described in \cite{Jouguet_2011}. The complexity of high-dimensional reconciliation is dominated by the generation of $\mathbf{Q}_j$. $\mathbf{Q}_j$ must be randomly generated for each $j$ to prevent information leakage \cite{Jouguet_2011}. In this paper, for simulation purposes, for each codeword, only one $\mathbf{Q}$ gets generated and used for each $\mathbf{Q}_j$ to simplify the simulation.
 
We then create a Householder matrix \cite{householder1958unitary} $\mathbf{S}_j$ between $\mathbf{u}'_j$ and $\mathbf{Q}_j\mathbf{y}'_j$
\begin{equation}
    \mathbf{S}_j = \mathbf{I}_d - 2 \frac{\mathbf{a}_j\mathbf{a}_j^T}{||\mathbf{a}_j||^2},
\end{equation}
where
\begin{equation}
    \mathbf{a}_j = \mathbf{Q}_j\mathbf{y}'_j - \mathbf{u}'_j\sqrt{\frac{||\mathbf{y}'_j||^2}{d}}.
\end{equation}
We scale $\mathbf{u}'_j$ by $\sqrt{\frac{||\mathbf{y}'_j||^2}{d}}$ so that the hyperplane between this vector and $\mathbf{Q}_j\mathbf{y}'_j$ goes through the origin, as this is a requirement for creating a Householder matrix \cite{householder1958unitary}. By doing this, we now map $\mathbf{y}'_j$ to $\mathbf{u}'_j\sqrt{\frac{||\mathbf{y}'_j||^2}{d}}$  instead of $\mathbf{u}'_j$. This scaling factor will be taken into account during the LLR calculations. 
We can then calculate $\mathbf{R}_j$ as 
\begin{equation}
    \mathbf{R}_j = \mathbf{S}_j\mathbf{Q}_j.
\end{equation}
The resulting mapping matrix $\mathbf{R}_j$ is orthogonal, as it is a multiplication of two orthogonal matrices. Bob then transmits $\mathbf{R}_j$ to Alice over the classical channel along with $||\mathbf{y}'_j||^2$.
 
 Alice attempts to reconstruct $\mathbf{u}'_j$ by applying the mapping matrix $\mathbf{R}_j$ to $\mathbf{x}'_j$
\begin{IEEEeqnarray}{rCl} 
    \mathbf{R}_j\mathbf{x}'_j &=& \mathbf{R}_j(\mathbf{y}'_j - \mathbf{z}'_j)\nonumber\\ &=& \mathbf{R}_j\mathbf{y}'_j - \mathbf{R}_j\mathbf{z}'_j\nonumber\\ &=& \mathbf{u}'_j\sqrt{\frac{||\mathbf{y}'_j||^2}{d}} - \mathbf{R}_j\mathbf{z}'_j.
\end{IEEEeqnarray}
Because $\mathbf{y}'_j$ is a noisy version of $\mathbf{x}'_j$, $\mathbf{R}_j$ will map $\mathbf{x}'_j$ to a point close to $\mathbf{u}'_j\sqrt{\frac{||\mathbf{y}'_j||^2}{d}}$. Hence, after applying $\mathbf{R}_j$ to $\mathbf{x}'_j$, Alice will have $\mathbf{u}'_j\sqrt{\frac{||\mathbf{y}'_j||^2}{d}}$ with additive noise $\mathbf{n}'_j = -\mathbf{R}_j\mathbf{z}'_j$. The probability distribution of $\mathbf{n}'_j$ given $\mathbf{R}_j$ is known to Alice, can be determined as follows. Consider the probability distribution of $\mathbf{z}'$:
\begin{equation}
    f(\mathbf{z}') = \mathcal{N}\Bigg(\mathbf{0},\mathbf{I}_d\frac{\sigma_z^2}{2}\Bigg).
\end{equation}
Multiplying $\mathbf{z}'$ by $\mathbf{R}$, where $\mathbf{R}$ is an instance of $\mathbf{R}_j$, does not change the mean value of the distribution, as 
\begin{equation}
    \mathbb{E}[-\mathbf{R}\mathbf{z}'] = -\mathbf{R}\mathbb{E}[\mathbf{z}'] = -\mathbf{R}\mathbf{\mu}_{\mathbf{z}'} = \mathbf{0}.
\end{equation}

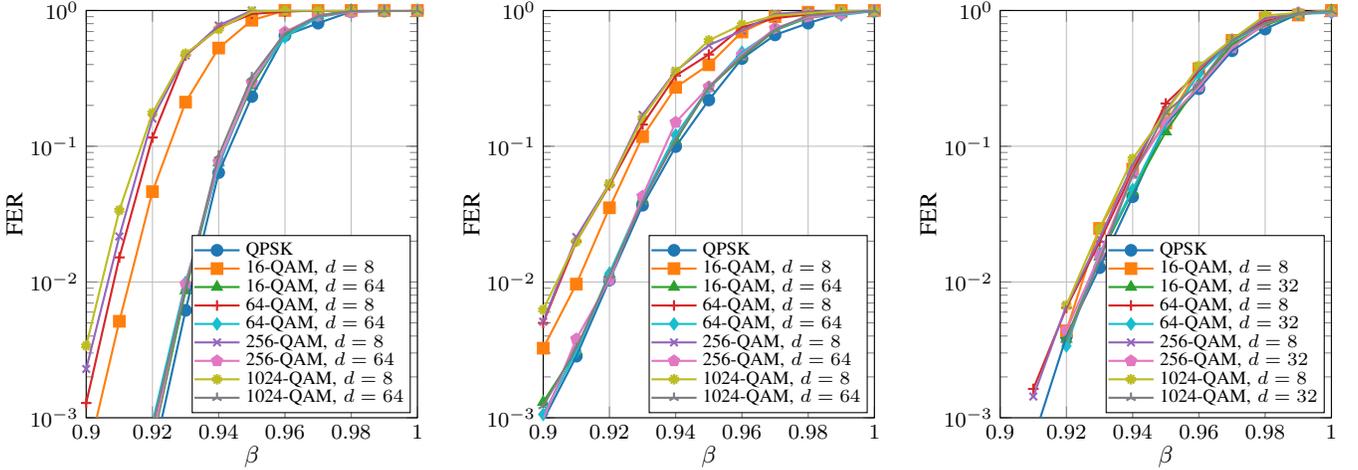
\begin{figure*}[t!]
\begin{subfigure}{0.33\textwidth}
      \begin{tikzpicture}
\begin{axis}[
every axis/.append style={font=\small},
tick label style={font=\footnotesize},
xlabel = $\beta$,
ylabel = FER,
xmin = 0.9, xmax =1,
ymin = 0.001, ymax = 1,
ymode = log,
x tick label style={yshift= -1mm},
y tick label style={xshift= -1mm},
ylabel shift = 1mm,
xlabel shift = 1mm,
width=\textwidth,
height=7cm,
set layers, mark layer=axis tick labels,
grid = major,
 xlabel near ticks,  
 ylabel near ticks,  
 xticklabel style={/pgf/number format/fixed},
every axis plot/.append style={thick},legend style={at={(0.32,0.23)},anchor=west, font = \scriptsize,row sep=-0.75ex,inner sep=0.1ex},
legend cell align={left},
cycle list name = foo
]

\pgfplotstableread{Figures/FER_R5_mod.txt}
\datatable
\foreach \x in {1,2,3,4,5,6,7,8,9}{
\addplot+
         table
         [
          x expr=\thisrowno{0}, 
          y expr=\thisrowno{\x} 
         ] {\datatable};
}
\addlegendentry{QPSK}
\addlegendentry{16-QAM, $d = 8$}
\addlegendentry{16-QAM, $d = 64$}
\addlegendentry{64-QAM, $d = 8$}
\addlegendentry{64-QAM, $d = 64$}
\addlegendentry{256-QAM, $d = 8$}
\addlegendentry{256-QAM, $d = 64$}
\addlegendentry{1024-QAM, $d = 8$}
\addlegendentry{1024-QAM, $d = 64$}
\end{axis}
\end{tikzpicture}
\end{subfigure}
\begin{subfigure}{0.33\textwidth}
      \begin{tikzpicture}
\begin{axis}[
every axis/.append style={font=\small},
tick label style={font=\footnotesize},
xlabel = $\beta$,
ylabel = FER,
xmin = 0.9, xmax =1,
ymin = 0.001, ymax = 1,
ymode = log,
x tick label style={yshift= -1mm},
y tick label style={xshift= -1mm},
ylabel shift = 1mm,
xlabel shift = 1mm,
width=\textwidth,
height=7cm,
set layers, mark layer=axis tick labels,
grid = major,
 xlabel near ticks,  
 ylabel near ticks,  
 xticklabel style={/pgf/number format/fixed},
every axis plot/.append style={thick},legend style={at={(0.32,0.23)},anchor=west, font = \scriptsize,row sep=-0.75ex,inner sep=0.1ex},
legend cell align={left},
cycle list name = foo
]

\pgfplotstableread{Figures/FER_R10_mod.txt}
\datatable
\foreach \x in {1,2,3,4,5,6,7,8,9}{
\addplot+
         table
         [
          x expr=\thisrowno{0}, 
          y expr=\thisrowno{\x} 
         ] {\datatable};
}
\addlegendentry{QPSK}
\addlegendentry{16-QAM, $d = 8$}
\addlegendentry{16-QAM, $d = 64$}
\addlegendentry{64-QAM, $d = 8$}
\addlegendentry{64-QAM, $d = 64$}
\addlegendentry{256-QAM, $d = 8$}
\addlegendentry{256-QAM, $d = 64$}
\addlegendentry{1024-QAM, $d = 8$}
\addlegendentry{1024-QAM, $d = 64$}
\end{axis}
\end{tikzpicture}
\end{subfigure}
\begin{subfigure}{0.33\textwidth}
      \begin{tikzpicture}
\begin{axis}[
every axis/.append style={font=\small},
tick label style={font=\footnotesize},
xlabel = $\beta$,
ylabel = FER,
xmin = 0.9, xmax =1,
ymin = 0.001, ymax = 1,
ymode = log,
x tick label style={yshift= -1mm},
y tick label style={xshift= -1mm},
ylabel shift = 1mm,
xlabel shift = 1mm,
width=\textwidth,
set layers, mark layer=axis tick labels,
grid = major,
height=7cm,
 xlabel near ticks,  
 ylabel near ticks,  
 xticklabel style={/pgf/number format/fixed},
every axis plot/.append style={thick},legend style={at={(0.32,0.23)},anchor=west, font = \scriptsize,row sep=-0.75ex,inner sep=0.1ex},
legend cell align={left},
cycle list name = foo
]

\pgfplotstableread{Figures/FER_R50_mod.txt}
\datatable
\foreach \x in {1,2,3,4,5,6,7,8,9}{
\addplot+
         table
         [
          x expr=\thisrowno{0}, 
          y expr=\thisrowno{\x} 
         ] {\datatable};
}
\addlegendentry{QPSK}
\addlegendentry{16-QAM, $d = 8$}
\addlegendentry{16-QAM, $d = 32$}
\addlegendentry{64-QAM, $d = 8$}
\addlegendentry{64-QAM, $d = 32$}
\addlegendentry{256-QAM, $d = 8$}
\addlegendentry{256-QAM, $d = 32$}
\addlegendentry{1024-QAM, $d = 8$}
\addlegendentry{1024-QAM, $d = 32$}
\end{axis}
\end{tikzpicture}
\end{subfigure}
    \caption{Simulated FER vs. $\beta$ for different rate (Left: $R = \frac{1}{5}$, Middle: $R = \frac{1}{10}$, Right: $R = \frac{1}{50}$)  TPB-LDPC codes for different $d$ and modulation formats.}
    \label{FER_Mod}
\end{figure*}

The covariance matrix of $\mathbf{R}\mathbf{z}'$ is then:
\begin{IEEEeqnarray}{rCl}
    \text{cov}(-\mathbf{R}\mathbf{z}') &=& \mathbb{E}[(-\mathbf{R}\mathbf{z}')(-\mathbf{R}\mathbf{z}')^T] \nonumber\\ &=& \mathbf{R}\mathbb{E}[\mathbf{z}'\mathbf{z}'^T]\mathbf{R}
^T \nonumber\\ &=& \frac{\sigma_z^2}{2}\mathbf{R}\mathbf{I}_d\mathbf{R}^T \nonumber\\ &=& \frac{\sigma_z^2}{2}\mathbf{I}_d,
\end{IEEEeqnarray}
because $\mathbf{R}$ is an orthogonal matrix, hence $\mathbf{R}\mathbf{R}^T = \mathbf{I}_d$. Therefore, the probability distribution of $\mathbf{n}'$ is:
\begin{equation}
    f(\mathbf{n}'|\mathbf{R}) = \mathcal{N}\Bigg(\mathbf{0},\mathbf{I}_d\frac{\sigma_z^2}{2}\Bigg).
\end{equation}

Once again, this distribution is i.i.d., and therefore independent of $j$.
For each block of $j$ symbols, we have created a BI-AWGN channel with signal power $\frac{||\mathbf{y}'_j||^2}{d}$ and noise variance $\sigma_z^2/2$.
Alice can then calculate $\mathbf{llr}'_j$
\begin{equation}
    \mathbf{llr}'_j = 2\frac{\mathbf{R}_j\mathbf{x}'_j}{\sigma_j^2},
\end{equation}
where
\begin{equation}
    \sigma_j^2= \frac{\sigma_z^2}{
2}\sqrt{\frac{d}{||\mathbf{y}'_j||^2}},\end{equation}
and use these LLRs for the decoding. 

To compare the noise distribution of high-dimensional reconciliation to multi-dimensional reconciliation with $d \leq 8$, we define $\mathbf{n}$ for high-dimensional reconciliation as $\mathbf{n} = [\mathbf{n}'_1/(||\mathbf{y}'_1||^2/d), \mathbf{n}'_1/(||\mathbf{y}'_2||^2/d), \cdots, \mathbf{n}'_{N/d}/(||\mathbf{y}'_{N/d}||^2/d)]$, i.e., we normalise the noise such that the signal power for each $j$ is equal 1. The virtual channel approaches a BI-AWGN with noise variance $\sigma_z^2/2$, which is shown in Fig. \ref{fig:pdf} where the histogram of $n$ when $d = 128$ follows a Gaussian probability density function almost exactly. We can once again use the law of large numbers \cite{revesz2014laws} to show why increasing $d$ causes to virtual channel to approach the BI-AWGN channel. Let $||\mathbf{y}'_j||^2 = \sum_{i = 1}^d ||y_{\cdot j - d + i}||^2$, where $||y_{\cdot j - d + i}||^2$ is i.i.d. with mean $\mu_{||y||^2}$ and variance $\sigma^2_{||y||^2}$, which are also both finite. Because $||\mathbf{y}'_j||^2$ is a sum of i.i.d. random variables with a finite mean and variance, $\text{Var}(||\mathbf{y}'||^2) \to 0$ as $d \to \infty$, hence the signal power $\frac{||\mathbf{y}'_j||^2}{d}$ is the same for each $j$.
\\\indent To analyse the effect of the multi-dimensional reconciliation on the performance of the error correction, we have simulated the FER curves for three different rate type-based protograph (TBP) LDPC codes ($R = \frac{1}{5}$, $R = \frac{1}{10}$, $R = \frac{1}{50}$) \cite{gumucs2021low} corresponding to short-, mid-, and long-range CV-QKD links for a large range of values of $d$. The quasi-cyclic parity check matrices for these codes were generated using progressive edge-growth with the approximate cycle extrinsic message degree constraint \cite{hu2005regular}, with cyclant sizes of 200, 200, and 500, and $N$ of $10^5$, $10^5$, and $10^6$ respectively. Gaussian modulation was used to generate the quantum states. The maximum amount of decoding iterations was set to 500, and for each data point, we simulated until at least 100 frame errors occurred or a maximum of $10^5$ frames had been simulated with the all-zero codeword assumption. 
\setcounter{figure}{7}
\begin{figure*}[!b]
    \resizebox{\linewidth}{!}{\begin{tikzpicture}
    \pgfmathsetmacro{\ROUNDEDCORNERS}{.0mm}

    \tikzstyle{arrowstyle} = [>={Triangle[scale=0.8]}, linestyle, rounded corners=1mm,]
    \def\loopstackgap{3.25cm}

    \tikzstyle{gradientc0} = [bottom color=C0!5, top color=C0!15]
    \tikzstyle{gradientc1} = [bottom color=C1!5, top color=C1!15]
    \tikzstyle{gradientc2} = [bottom color=C2!5, top color=C2!15]
    \tikzstyle{gradientc3} = [bottom color=C3!5, top color=C3!15]
    \tikzstyle{gradientc4} = [bottom color=C4!5, top color=C4!15]
    \tikzstyle{gradientc5} = [bottom color=C5!5, top color=C5!15]
    \tikzstyle{gradientc6} = [bottom color=C6!5, top color=C6!15]
    \tikzstyle{gradientc7} = [bottom color=C7!5, top color=C7!15]
    \tikzstyle{gradientO} = [bottom color=C0!5, top color=C0!15]
    \tikzstyle{gradientEO} = [bottom color=C4!5, top color=C4!15]

    \draw[gradientc1,color = C1,very thick] (-5.5,-3.7) rectangle(1,-0.2);
    \draw[gradientc7,color = C7,very thick] (1.03,-3.7) rectangle(8.7,-0.2);
    \draw[gradientc4,color = C4,very thick] (8.73,-3.7) rectangle(14.3,-0.2);
    \node at (-2.25,-3.4){\textcolor{C1}{Alice}};
    \node at (4.86,-3.5){\textcolor{C7}{Quantum FSO channel}};
    \node at (11.61,-3.5){\textcolor{C4}{Bob}};

    \node[box={color=O, width=0.5,height=0.5},yshift=-22.2mm,xshift=-12.2mm,gradientc1] at (0,0) (filterin2){VOA};

    \node[box={color=C4, linestyle={linestyle, gradientc4}}, left=4.5mm of filterin2,minimum height=5mm, inner sep=1.5mm,] (omft2){IQM};

    \node[coupler, right=6mm of filterin2, xshift=2mm] (coupler2) {};

    \node[box={direction=e, height=0.35, width=0.7}, left= 3mm of omft2, gradientc1,inner sep=1.5mm,align = center] (eclq1){ECL \\ 1550nm};

    \node[box={color=E, direction=s, width=0.35, height=1, nin=2, nout=2}, above=of omft2,yshift=-2.4mm, gradientc2,minimum height=5mm,inner sep=1.5mm,] (dac2){DAC};

    \node[box={color=O, width=0.5}, right=of dac2, gradientc1, inner sep=1.5mm, minimum height=3mm, align=center] (PM1){Power meter};

    \draw[-->, O] (eclq1) to (omft2){};

    \foreach \n in {0,1}{
            \draw[E, -AA-] (dac2.out\n) to (dac2.out\n |- omft2.n){};
        }

    \draw[-->, O] (omft2.out) to [out=0, in=180] (filterin2.in){};
     \draw[---, O] (filterin2.out) to [out=0, in=180] (coupler2.in){};
    
    \draw[-->, O] (coupler2.out) to [out=-30, in=-90] (PM1.s){};
    \draw[-PC-, O] (filterin2.out) --++ (7mm,0){};
  

        
    \node[mux={color=O, width=0.2,height=0.45, direction=w, nports=2, fillgradient=gradientc1}, right= \QNODESIZE of coupler2, xshift = 20mm] (colin){};
    \node[coupler, left=3mm of colin] (coupler3) {};
        \node[box={direction=e, height=0.35, width=0.7}, above= 7.3mm of coupler3, gradientc1,inner sep=1.5mm,xshift = -3mm, align = center] (eclq2){ECL\\1528nm};
    \draw[-->, O] (coupler2.out) to [out=0, in=180] (colin.out){};
\draw[---, O] (eclq2.s) to [out=-90, in=180] (coupler3.in){};
    \node[box={height=1.3, width=1.6, color=gray}, align=center, right= \QNODESIZE of colin, anchor=w] (otg){};

    \node[mux={color=O, width=0.2,height=0.45, nports=2, fillgradient=gradientO}, right= \QNODESIZE of otg] (cl){};

    \begin{scope}
        \tikzset{/tikz/linestyle/.append style = {C3, thin, rounded corners=0.1mm}}
        \path[fill=C4, fill opacity=0.2] (colin.in0) to (colin.in1) to (cl.in1) to (cl.in0) -- cycle;
    \end{scope}

    \def\eddieop{0.15}
    \def\eddiecol{gray}
    \node[ellipse,fill=\eddiecol, opacity=\eddieop,minimum width=20, minimum height=18] (T) at ([xshift=14, yshift=10]otg) {};
    \node[ellipse,fill=\eddiecol, opacity=\eddieop,minimum width=15, minimum height=17] (T) at ([xshift=-20, yshift=-5]otg) {};
    \node[ellipse,fill=\eddiecol, opacity=\eddieop,minimum size=12] (T) at ([xshift=-21, yshift=-20]otg) {};
    \node[ellipse,fill=\eddiecol, opacity=\eddieop,minimum size=5] (T) at ([xshift=-12, yshift=-24]otg) {};
    \node[ellipse,fill=\eddiecol, opacity=\eddieop,minimum size=6] (T) at ([xshift=-28, yshift=-13]otg) {};
    \node[ellipse,fill=\eddiecol, opacity=\eddieop,minimum size=8] (T) at ([xshift=-27, yshift=6]otg) {};
    \node[ellipse,fill=\eddiecol, opacity=\eddieop,minimum size=17] (T) at ([xshift=-21, yshift=18]otg) {};
    \node[ellipse,fill=\eddiecol, opacity=\eddieop,minimum size=10] (T) at ([xshift=-7, yshift=0]otg) {};
    \node[ellipse,fill=\eddiecol, opacity=\eddieop,minimum size=14] (T) at ([xshift=-5, yshift=13]otg) {};
    \node[ellipse,fill=\eddiecol, opacity=\eddieop,minimum size=7] (T) at ([xshift=-9, yshift=23]otg) {};
    \node[ellipse,fill=\eddiecol, opacity=\eddieop,minimum size=6] (T) at ([xshift=-15, yshift=7]otg) {};
    \node[ellipse,fill=\eddiecol, opacity=\eddieop,minimum size=10] (T) at ([xshift=-9, yshift=-16]otg) {};
    \node[ellipse,fill=\eddiecol, opacity=\eddieop,minimum width=7, minimum height=6] (T) at ([xshift=-3, yshift=-23]otg) {};
    \node[ellipse,fill=\eddiecol, opacity=\eddieop,minimum size=18] (T) at ([xshift=4, yshift=-10]otg) {};
    \node[ellipse,fill=\eddiecol, opacity=\eddieop,minimum size=9] (T) at ([xshift=4, yshift=22]otg) {};
    \node[ellipse,fill=\eddiecol, opacity=\eddieop,minimum width=7, minimum height=5] (T) at ([xshift=2, yshift=2]otg) {};
    \node[ellipse,fill=\eddiecol, opacity=\eddieop,minimum size=11] (T) at ([xshift=25, yshift=22]otg) {};
    \node[ellipse,fill=\eddiecol, opacity=\eddieop,minimum size=6] (T) at ([xshift=15, yshift=24]otg) {};
    \node[ellipse,fill=\eddiecol, opacity=\eddieop,minimum size=12] (T) at ([xshift=20, yshift=-6]otg) {};
    \node[ellipse,fill=\eddiecol, opacity=\eddieop,minimum size=15] (T) at ([xshift=20, yshift=-20]otg) {};
    \node[ellipse,fill=\eddiecol, opacity=\eddieop,minimum size=8] (T) at ([xshift=8, yshift=-23]otg) {};
    \node[ellipse,fill=\eddiecol, opacity=\eddieop,minimum size=8] (T) at ([xshift=27, yshift=4]otg) {};
    \node[ellipse,fill=\eddiecol, opacity=\eddieop,minimum size=5] (T) at ([xshift=28, yshift=12]otg) {};
    \node[ellipse,fill=\eddiecol, opacity=\eddieop,minimum size=6] (T) at ([xshift=28, yshift=-12]otg) {};

    \node[coupler, right=2mm of cl, xshift = 2mm] (coupler4) {};
    \node[box={color=O, width=0.5}, right=of cl, yshift = 11.17mm, xshift = -6mm, gradientc1, inner sep=1.5mm, minimum height=3mm, align=center] (PM2){High-speed \\ power meter};
    \node[aom={size=0.5,fillgradient=gradientO,},right=\QNODESIZE of cl,xshift=26mm] (switch) {};
    \node[box={nout=2,direction=n}, O, gradientO, minimum height=6mm, minimum width=20mm, right=15mm of switch.s, anchor=east, rotate=-90,xshift=0mm, inner sep = 1.5mm] (hybrid) {$\mathrm{90^{\circ}}$ hybrid};
    \node[box={color=O, width=0.45,height=0.45},gradientO, anchor=n, inner sep = 1.5mm] at (switch |- hybrid.w) (lo){Local LO};
    \node[box={width=0.7, height=3, nin=15, direction=n}, minimum height=6mm, minimum width=5mm, C2, align=center, gradientc2, xshift=17mm, right=of hybrid.n,  anchor=north, rotate=-90, inner sep = 1.5mm] (adc){GPU-based\\receiver};

    \foreach \n [evaluate=\n as \npd using int(\n+1)] in {0,1}{
            \node[pd={fillgradient=gradientEO}, minimum height=5mm, minimum width=5mm,anchor=ne, right=3mm of hybrid.out\n] (pd\n){};
            \draw[-->, O] ([yshift=1mm]hybrid.out\n) -- ([yshift=1mm]hybrid.out\n -| pd\n.in){};
            \draw[-->, O] ([yshift=-1mm]hybrid.out\n) -- ([yshift=-1mm]hybrid.out\n -| pd\n.in){};
            \draw[-->, E] (pd\n.out) -- (pd\n.out -| adc.s){};
        }

    \draw[---, O] (cl.out) to [out=0, in=180] (coupler4){};
    \draw[-->, O] (coupler4.out) to [out=0, in=-90] (PM2.s){};
    \draw[-->, O] (coupler4.out) to [out=0, in=180] (switch){};
    \draw[-->, O] (switch.out) -- (switch.out -| hybrid.s){};
    \draw[-->, O] (lo.out) to (lo.out -| hybrid.in);

    \node[label, yshift=6.9mm,xshift=0mm, O, below=of coupler2] {\small 90/10};
    \node[label, yshift=6.7mm,xshift=0.5mm,O, below=of coupler4] {\small 1/99};
    \node[label, yshift=6.7mm,xshift=0.5mm,O, below=of coupler3] {\small 50/50};
    \node[label, yshift=-13mm,xshift=-0.3mm ,above=of otg] {\textcolor{gray}{OTG}};

\end{tikzpicture}}
    \caption{The CV-QKD setup for transmission over an FSO channel with an optical turbulence generator.}
    \label{fig:setup}
\end{figure*}
\setcounter{figure}{5}
\\\indent The results of these simulations are shown in Fig. \ref{FER_MD}. The performance of the reconciliation depends on both $d$ and $R$. For $R = \frac{1}{5}$, there is a significant gap in performance between $d = 8$ and $d = 128$, namely 2.7\% $\beta$ for an FER of 10\%. When the rate of the code decreases, this gap diminishes, with a 1.6\% difference for $R = \frac{1}{10}$, and 0.5\% for $R = \frac{1}{50}$. We conjecture that this is because in the lower SNR regime the decrease in channel capacity of the virtual channel compared to the quantum channel, caused by using a finite $d$ in multi-dimensional reconciliation, is smaller. It is still worth considering using high-dimensional reconciliation for long-distance CV-QKD links. Even though the complexity increases by more than an order of magnitude when using $d = 128$ compared to $d = 8$, this remains negligible compared to the complexity of the decoding while offering an increase in $\beta$. However, it is mostly useful for short to mid-distance links, as this increase in $\beta$ can lead to significant increases in SKR, as will be shown later.
\\\indent We also simulated how different modulation formats impact the multi-dimensional reconciliation. The modulation formats are PS-QAM constellations, designed according to \cite{Denys_2021}. As shown in Fig. \ref{FER_Mod}, the modulation format has a slight impact on the error correction performance. QPSK modulation performs the best because all symbols have equal power, and therefore, the virtual channel created during reconciliation corresponds to a BI-AWGN channel regardless of what $d$ is. When increasing the cardinality of the constellation, the performance of the error correction code worsens when $d=8$, as the amount of different power levels in the constellation increases, hence the noise variance of each channel use varies more. This is especially clear for $R = \frac{1}{5}$, but for $R = \frac{1}{10}$ and $R = \frac{1}{50}$ the modulation formats are closer in performance. When we choose a high $d$, the error correction curves almost completely overlap for all of the codes for the different modulation formats. 

\section{Rate Adaptivity}
\label{Rate Adaptivity}
As mentioned before, rate adaptivity is an important part of reconciliation, as fluctuations in the quantum channel can cause the values of $I_{AB}$ and $\chi_{BE}$, which are estimated during the parameter estimation, to change. To operate at a steady $\beta$, the rate of the code needs to be changed to match $\beta I_{AB}$. We achieve this rate adaptivity using puncturing as described in the \textit{sp}-protocol proposed in \cite{wang2017efficient}. Using this protocol, it is possible to obtain a wide range of code rates without needing multiple LDPC codes designs. 

At the start of reconciliation, Bob knows both $I_{AB}$ from the parameter estimation, and the target $\beta$, and has an LDPC code with base rate $R_b$. Bob calculates the target code rate $R_t$, which should be larger than $R_b$ as with puncturing, it is only possible to increase the rate of the code. Bob determines the amount of bits that need to be punctured $p$:

 \begin{figure}[t!]
    \centering
      \begin{tikzpicture}
\begin{axis}[
every axis/.append style={font=\small},
tick label style={font=\footnotesize},
xlabel = $R_{t}/R_{b}$,
ylabel = $\beta$,
xmin = 1, xmax =2,
ymin = 0.91, ymax = 0.95,
x tick label style={yshift= -1mm},
y tick label style={xshift= -1mm},
ylabel shift = 1mm,
xlabel shift = 1mm,
width=0.49\textwidth,
height=7.5cm,
set layers, mark layer=axis tick labels,
grid = major,
 xlabel near ticks,  
 ylabel near ticks,  
 xticklabel style={/pgf/number format/fixed},
every axis plot/.append style={thick},legend style={at={(0.40,0.25)},anchor=west, font = \scriptsize,row sep=-0.75ex,inner sep=0.2ex},
legend cell align={left},
cycle list name = foo
]

\pgfplotstableread{Figures/Puncturing.txt}
\datatable
\foreach \x/\y in {0/1,2/3,4/5}{
\addplot+
         table
         [
          x expr=\thisrowno{\x}, 
          y expr=\thisrowno{\y} 
         ] {\datatable};
}
\addlegendentry{$R_{b} = \frac{1}{5}$}
\addlegendentry{$R_{b} = \frac{1}{10}$}
\addlegendentry{$R_{b} = \frac{1}{50}$}

\end{axis}
\end{tikzpicture}
    \caption{Simulated results. The $\beta$ at which an FER of 10\% is achieved vs. the relative rate of the punctured code for TBP-LDPC codes.}
    \label{fig:Puncturing}
\end{figure}

\begin{equation}
    p = \Bigg\lfloor N - \frac{R_bN}{R_t}\Bigg\rfloor,
\end{equation}
where we use the floor operation as $p$ has to be an integer. Bob then randomly generates a vector $\mathbf{w} = [w_1, w_2, \cdots w_p]$, where each value $w_i$ is a unique integer with a value between 1 and $N$, indicating the position of the punctured bits in the codeword. When calculating $\mathbf{m}$, Bob does not calculate a message for these specific bit positions, i.e., $\mathbf{m}$ will have a length of $N-p$. Bob transmits $\mathbf{w}$ alongisde $\mathbf{m}$. During decoding, Alice will set the LLRs of the punctured bits to 0.

We have simulated the performance of the puncturing for three different codes, the same $R_b = \frac{1}{5}$, $R_b = \frac{1}{10}$, and $R_b = \frac{1}{50}$ TBP-LDPC codes used in Fig. \ref{FER_MD} and \ref{FER_Mod}. We have randomly punctured the codes for a wide range of code rates, up until twice the base code rate. Figure~\ref{fig:Puncturing} shows at which $\beta$ an FER of 10\% is achieved for the different codes and rates, which we have determined from the $\beta$ vs. FER curves obtained during the simulation. For $R_b = \frac{1}{5}$, the error correction performance drops over a wide range of code rates, when doubling the rate there is a 2.9\% penalty in $\beta$. When $R_b$ decreases, we can see that the codes become more resilient to rate changes, as for the $R_b = \frac{1}{50}$ code the error correction performance stays almost constant when increasing the rate. 

\begin{figure}[t!]
    \centering
      \begin{tikzpicture}
\begin{axis}[
every axis/.append style={font=\small},
tick label style={font=\footnotesize},
xlabel = $R_{t}/R_{b}$,
ylabel = $\beta$,
xmin = 1, xmax =2,
ymin = 0.91, ymax = 0.95,
x tick label style={yshift= -1mm},
y tick label style={xshift= -1mm},
ylabel shift = 1mm,
xlabel shift = 1mm,
width=0.49\textwidth,
height=7.5cm,
set layers, mark layer=axis tick labels,
grid = major,
 xlabel near ticks,  
 ylabel near ticks,  
 xticklabel style={/pgf/number format/fixed},
every axis plot/.append style={thick},legend style={at={(0.20,0.15)},anchor=west, font = \scriptsize,row sep=-0.75ex,inner sep=0.2ex},
legend cell align={left},
cycle list name = foo
]

\addlegendimage{no markers}
\addlegendimage{no markers, dashed}

\pgfplotstableread{Figures/Puncturing2.txt}
\datatable
\addplot+
         table
         [
          x expr=\thisrowno{0}, 
          y expr=\thisrowno{1} 
         ] {\datatable};
\foreach \x in {2,3,4,5,6}{
\addplot+[dashed]
         table
         [
          x expr=\thisrowno{0}, 
          y expr=\thisrowno{\x} 
         ] {\datatable};
}
\addlegendentry{Random puncturing}
\addlegendentry{Set puncturing pattern}
\end{axis}
\end{tikzpicture}
    \caption{Simulated results. The $\beta$ at which an FER of 10\% is achieved vs the relative rate of the punctured code for a $R = \frac{1}{5}$ TBP-LDPC code, comparing random puncturing with a set puncturing pattern. For the simulations, 5 different set puncturing patterns were simulated.}
    \label{fig:Puncturing2}
\end{figure}
\setcounter{figure}{8}

\begin{figure*}[t!]
\begin{subfigure}{0.49\linewidth}
      \begin{tikzpicture}[>=latex]
\begin{axis}[
every axis/.append style={font=\small},
tick label style={font=\footnotesize},
xlabel = $\beta$(\%),
ylabel = FER,
xmin = 89, xmax =98,
ymin = 0.001, ymax = 1,
ymode = log,
xtick distance=1,
x tick label style={yshift= -1mm},
y tick label style={xshift= -1mm},
ylabel shift = 1mm,
xlabel shift = 1mm,
set layers, mark layer=axis tick labels,
grid = major,
 xlabel near ticks,  
 ylabel near ticks,  
 xticklabel style={/pgf/number format/fixed},
every axis plot/.append style={thick},legend style={at={(0.40,0.15)},anchor=west, font = \scriptsize,row sep=-0.75ex,inner sep=0.2ex},
legend cell align={left},
cycle list name = foo
]

\pgfplotstableread{Figures/FER_Experiment.txt}
\datatable
\pgfplotsinvokeforeach {3,2,1}{
\addplot+
         table
         [
          x expr=\thisrowno{0}, 
          y expr=\thisrowno{#1} 
         ] {\datatable};
}
\addlegendentry{BI-AWGN}
\addlegendentry{$d = 8$}
\addlegendentry{$d = 128$}
\draw[<->,thick](axis cs: 90.2,0.1) -- (axis cs:93.6,0.1);

\node at (axis cs:91.9,0.073){\scriptsize3.4\%};
\end{axis}
\end{tikzpicture}
\end{subfigure}\hfill
\begin{subfigure}{0.49\linewidth}
      \begin{tikzpicture}[>=latex]
\begin{axis}[
every axis/.append style={font=\small},
tick label style={font=\footnotesize},
xlabel = $\beta$(\%),
ylabel = SKR (bits/symbol),
xmin = 89, xmax =98,
ymin = 0, ymax = 0.021,
xtick distance=1,
grid = major,
set layers, mark layer=axis tick labels,
 xlabel near ticks,  
 ylabel near ticks,  
 ylabel shift = 1mm,
xlabel shift = 1mm,
 x tick label style={yshift= -1mm},
y tick label style={xshift= -1mm},
   scaled y ticks = false, 
   ytick = {0,0.003,0.006,0.009,0.012,0.015,0.018,0.021},
   yticklabels={0,0.003,0.006,0.009,0.012,0.015,0.018,0.021},
 xticklabel style={/pgf/number format/fixed},
every axis plot/.append style={thick},legend style={at={(0.45,0.23)},anchor=west, font = \scriptsize,row sep=-0.75ex,inner sep=0.2ex},
legend cell align={left},
cycle list name = foo
]

\pgfplotstableread{Figures/SKR_Experiment.txt}
\datatable
\pgfplotsinvokeforeach {3,2,1}{
\addplot+
         table
         [
          x expr=\thisrowno{0}, 
          y expr=\thisrowno{#1} 
         ] {\datatable};
}
\addlegendentry{BI-AWGN}
\addlegendentry{$d = 8$}
\addlegendentry{$d = 128$}
\draw[dashed,thick](axis cs:90,0.00735) -- (axis cs:94,0.00735);
\draw[<->,thick](axis cs: 94,0.01945) -- (axis cs:94,0.00735);
\node at (axis cs: 94.5,0.0134){\scriptsize 165\%};

\end{axis}
\end{tikzpicture}
    \end{subfigure}
    \caption{Experimental results. Left: The FER of the $R=\frac{1}{5}$ expanded TBP-LDPC code when punctured to $R = 0.3$ for different $d$ compared to the BI-AWGN channel. Right: The SKR of the $R=\frac{1}{5}$ expanded TBP-LDPC code when punctured to $R = 0.3$ for different $d$ compared to the BI-AWGN channel for the FSO channel with $\sigma_I$~=~0.001 and $\beta_{jitter}$~=~123.8.}
    \label{experimental results}
\end{figure*}

\indent Instead of randomly generating and transmitting $\mathbf{w}$ every single time, it is also possible to generate a set puncturing pattern $\mathbf{v} = [v_1,v_2,\cdots,v_N]$, which is a random scrambling of all integers from 1 to N. This puncturing pattern is known to all parties. Bob then only needs to transmit $p$ over the channel, indicating that bits $v_1$ to $v_p$ are punctured. This simplifies the puncturing process, as Bob does not need to generate and transmit $\mathbf{w}$ every single frame.

\indent We have simulated the puncturing for five different $\mathbf{v}$, which were all randomly generated, and compared it to random puncturing for the $R = \frac{1}{5}$ TBP-LDPC code with $N = 10^5$. Figure \ref{fig:Puncturing2} shows that using a set puncturing pattern has little influence on the error correction performance compared to generating a random puncturing pattern every frame. It should be noted that it is possible to optimise $\mathbf{v}$ for a specific code to improve the puncturing performance \cite{Beerman2014}, however, that is outside of the scope of this paper. 
\\\indent Until now, we have assumed that $\beta$ has a constant value during reconciliation. When there are fluctuations in the quantum channel, the rate of the code is changed to obtain this $\beta$. This $\beta$ is chosen to optimise the SKR of the system, assuming that the parameters of the system, such as the excess noise, are constant. This approach is valid for CV-QKD transmission over fibre, as the fibre channel is stable. However, factors such as atmospheric turbulence cause additional time-dependent instabilities in FSO channels. Hence, the optimal $\beta$ changes over time. During the parameters estimation, Bob continuously determines $I_{AB}$ and $\chi_{BE}$ of the measured quantum symbols during the parameter estimation. Bob does this on a per-block basis, where Bob measures $N_p$ quantum symbols and does the parameter estimation over this entire CV-QKD block. For each CV-QKD block, $\beta$ can be optimised. This can be simply implemented by using a $\beta$-FER look-up table, and calculating which $\beta$-FER pair optimises the SKR for a specific block. This adds a negligible amount of increase in the complexity of the parameter estimation. In the next section, we show how $\beta$-optimisation can increase the SKR.

\section{Experimental Setup}
\label{Experimental Results}

\indent Fig. \ref{fig:setup} shows the experimental setup employed for CV-QKD transmission validation over a free-space optical (FSO) channel. On Alice's side, we deploy a \SI{<100}{kHz} linewidth external cavity laser (ECL) tuned to 1550~nm (this is similar to the linewidth of lasers in commercial transceivers), a digital-to-analogue (DAC) converter, and an optical IQ-modulator (IQM) to transmit a PS-256QAM constellation with a symbol rate of 250~MBaud. A variable optical attenuator (VOA) and a power meter are used to attenuate the signal to an average power of 7.44 shot noise units (SNU) ($-69.2$~dBm). We combine a second tone produced by a second ECL for turbulence characterisation placed at 1528~nm with the attenuated 1550~nm quantum signal and coupled to free space using a collimator. The light then traverses an optical turbulence generator (OTG) \cite{OTGthesis, OTG}, which can mimic FSO channels with various turbulence strengths. We split off 1\% of the light to a high-speed power meter for FSO channel characterisation. For our setup, we have measured four different turbulence strengths generated by the OTG with scintillation indices $\sigma_I$ = 0.001, 0.009, 0.01, 0.013 and pointing jitter $\beta_{jitter}$ = 123.8, 8.6, 3.0, 1.6 respectively, which would all be classified as weak fluctuations, corresponding to calm weather conditions \cite{1994OptEn..33.3748K}. 
\begin{figure}[b!]
\centering
\begin{subfigure}{1.05\linewidth}
          \begin{tikzpicture}[>=latex]
\begin{axis}[
every axis/.append style={font=\small},
tick label style={font=\footnotesize},
xlabel = CV-QKD block,
ylabel = Bits/symbol,
xmin = 0, xmax =120,
ymin = 0.45, ymax = 0.65,
y tick label style={xshift= -1mm},
x tick label style={yshift= -1mm},
yticklabels = {0,0.45,0.50,0.55,0.60,0.65},
ylabel shift = 1mm,
xlabel shift = 1mm,
width =\linewidth,
height=4cm,
set layers, mark layer=axis tick labels,
grid = major,
 xlabel near ticks,  
 ylabel near ticks,  
 xticklabel style={/pgf/number format/fixed},
every axis plot/.append style={thick},legend style={at={(0.05,0.15)},anchor=west, font = \scriptsize,row sep=-0.75ex,inner sep=0.2ex},
legend cell align={left},
cycle list name = foo
]

\pgfplotstableread{Figures/MI_chi.txt}
\datatable
\pgfplotsinvokeforeach {1,2}{
\addplot+[only marks]
         table
         [
          x expr=\thisrowno{0}, 
          y expr=\thisrowno{#1} 
         ] {\datatable};
}
\addlegendentry{$I_{AB}$}
\addlegendentry{$\chi_{BE}$}
\end{axis}
\end{tikzpicture}
\end{subfigure}
\begin{subfigure}{1.05\linewidth}
  \begin{tikzpicture}[>=latex]
\begin{axis}[
every axis/.append style={font=\small},
tick label style={font=\footnotesize},
xlabel = CV-QKD block,
ylabel = $\beta$,
xmin = 0, xmax =120,
ymin = 0.92, ymax = 0.98,
  scaled y ticks = false, 
y tick label style={xshift= -1mm},
x tick label style={yshift= -1mm},
ylabel shift = 1mm,
xlabel shift = 1mm,
height=4cm,
width = \linewidth,
set layers, mark layer=axis tick labels,
grid = major,
 xlabel near ticks,  
 ylabel near ticks,  
 xticklabel style={/pgf/number format/fixed},
every axis plot/.append style={thick},legend style={at={(0.80,0.8)},anchor=west, font = \scriptsize,row sep=-0.75ex,inner sep=0.2ex},
legend cell align={left},
cycle list name = foo
]

\pgfplotstableread{Figures/MI_chi.txt}
\datatable

\addplot+[only marks,C5]
         table
         [
          x expr=\thisrowno{0}, 
          y expr=\thisrowno{4} 
         ] {\datatable};

\addplot[only marks, mark = x, C4]
coordinates{(18,0.92)(19,0.92)(65,0.92)(72,0.92)(89,0.92)(116,0.92)(118,0.92)};
\end{axis}
\end{tikzpicture}
\end{subfigure}
    \caption{Experimental results. Top: The $I_{AB}$ and $\chi_{BE}$ for 120 CV-QKD blocks for the FSO channel with $\sigma_I = 0.009$ and $\beta_{jitter}$ = 8.6. Bottom: The optimal $\beta$ for each CV-QKD block, using a $R = \frac{1}{5}$ TBP-LDPC code punctured to the $R = \beta I_{AB}$ for each block. The red crosses correspond to the cases where reconciliation is not possible because the SKR is negative regardless of which $\beta$ is chosen.}
    \label{fig:MI_chil}
\end{figure}
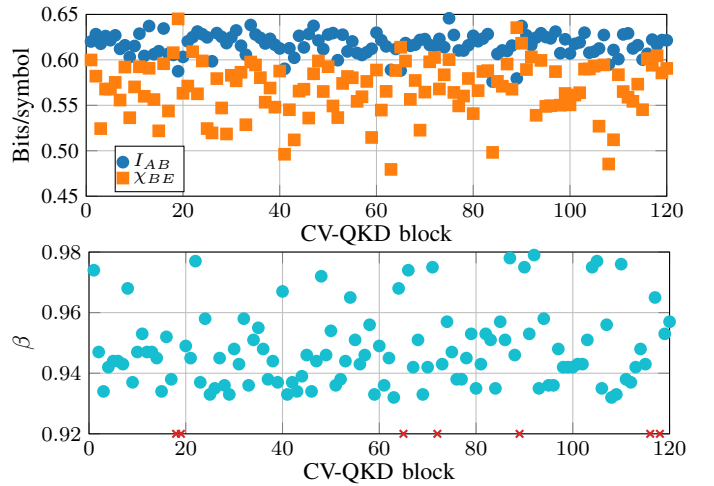
\\\indent The remaining 99\% of light is directed to Bob's side, where for the 90$^{\circ}$ optical hybrid, a local ECL is used as a local local oscillator (LLO), after which the outputs are digitised. Note that a single polarisation optical set-up was used for this experiment. Polarisation is aligned beforehand using a polarisation controller and is stable during measurements. Calibration and recovery of the quantum signal is done using digital signal processing \cite{Sjoerd2023}. During the parameter estimation, $I_{AB}$, the excess noise $\xi_{Bob}$, and $\chi_{BE}$ are estimated, taking into account finite-size effects \cite{Jouguet_2012_fs}. Relevant parameters for the system are a quantum efficiency of 40\%, $N_{privacy} = 6.8\cdot10^6$, a clearance of 10~dB, an average $\xi_{Bob}$ of 0.0045 SNU, and an average transmittance $T$ of 0.41. 
\begin{figure}[t!]
    \centering
      \begin{tikzpicture}[>=latex]
\begin{axis}[
every axis/.append style={font=\small},
tick label style={font=\footnotesize},
xlabel = $\beta$(\%),
ylabel = SKR (bits/symbol),
xmin = 89, xmax = 98,
ymin = 0, ymax = 0.021,
width=0.49\textwidth,
xtick distance=1,
ytick distance = 0.003,
yticklabels={0,0,0.003,0.006,0.009,0.012,0.015,0.018,0.021,0.024,0.027,0.03},
height=7.5cm,
set layers, mark layer=axis tick labels,
grid = major,
  scaled y ticks = false, 
  x tick label style={yshift= -1mm},
y tick label style={xshift= -1mm},
ylabel shift = 1mm,
xlabel shift = 1mm,
 xlabel near ticks,  
 ylabel near ticks,  
 xticklabel style={/pgf/number format/fixed},
every axis plot/.append style={thick},legend style={at={(0.15,0.15)},anchor=west, font = \scriptsize,row sep=-0.75ex,inner sep=0.2ex},
legend cell align={left},
cycle list name = foo
]

\pgfplotstableread{Figures/Beta_SKR_RA.txt}
\datatable
\pgfplotsinvokeforeach {1,2,3,4}{
\addplot+[thick]
         table
         [
          x expr=\thisrowno{0}, 
          y expr=\thisrowno{#1} 
         ] {\datatable};
}
\addlegendentry{$\sigma_I$ = 0.001, $\beta_{jitter}$ = 123.8}
\addlegendentry{$\sigma_I$ = 0.009, $\beta_{jitter}$ = 8.6}
\addlegendentry{$\sigma_I$ = 0.010, $\beta_{jitter}$ = 3.0}
\addlegendentry{$\sigma_I$ = 0.013, $\beta_{jitter}$ = 1.6}

\addlegendimage{dashed};
\addlegendentry{$\beta$-optimisation};
\draw[thick,dashed, color = C1] (axis cs: 89,0.0208) -- (axis cs: 98,0.0208);
\draw[thick,dashed, color = C2] (axis cs: 89,0.0155) -- (axis cs: 98,0.0155);
\draw[thick,dashed, color = C3] (axis cs: 89,0.0153) -- (axis cs: 98,0.0153);
\draw[thick,dashed, color = C4] (axis cs: 89,0.0151) -- (axis cs: 98,0.0151);
\draw[thick,<->](axis cs:94,0.0143) to[in = -120,out = -60] (axis cs:97.5,0.0155) ;
\node at (95.7,0.013) {\scriptsize 7.6\%};

\end{axis}
\end{tikzpicture}
    \caption{Experimental Results. $\beta$ vs. SKR for FSO channels with different turbulence settings. A $R = \frac{1}{5}$ TBP-LDPC code is used, punctured to the appropriate rate. The dashed lines correspond to the SKR when $\beta$-optimisation is used.}
    \label{fig:SKR_RA}
\end{figure}
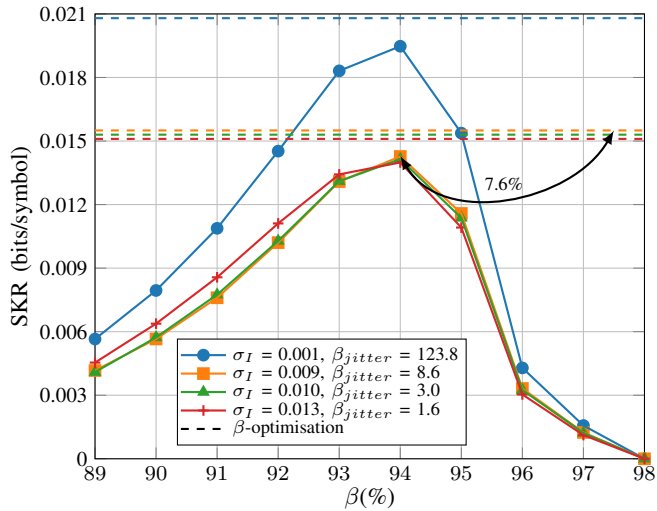
\\\indent We use 128-dimensional reconciliation. Furthermore, we use the $R = 0.2$ expanded TBP-LDPC code punctured to $R \approx 0.3$, around the average $I_{AB}$ of the system, for error correction. We choose a block length $N = 1.024\cdot10^5$ with a maximum of 500 decoding iterations. We have also implemented the $\beta$-optimisation as well.
\\\indent Figure \ref{experimental results} shows both the FER (left) and the SKR (right) of the experimental results for the case when $\sigma_I = 0.001$ and $\beta_{jitter} = 123.8$ assuming a constant $\beta$ for all CV-QKD blocks. As expected, the FER of the 128-dimensional reconciliation is close in performance to the BI-AWGN, and at an FER of 10\% there is an increase in $\beta$ of approximately 3.4\% compared to the 8-dimensional case. This is slightly more gain compared to the results in Fig. \ref{FER_Mod} because of the higher code rate. As a result, the SKR sees an increase of 165\%, where the optimal SKR is achieved at $\beta = 94\%$.  
\\\indent In Fig. \ref{fig:MI_chil} we show how $I_{AB}$ and $\chi_{BE}$ and the optimal $\beta$ changes over time for the FSO transmission with $\sigma_I = 0.009$, $\beta_{jitter}$ = 8.6, and $N_p = 10^7$. We have transmitted a total of 120 CV-QKD blocks. We can see from the figure that $I_{AB}$ fluctuates slightly over time, while for $\chi_{BE}$, the fluctuations are more significant. This is because the fluctuations are mainly caused by fluctuations in $\xi_{bob}$, which has a larger influence on $\chi_{BE}$ than on $I_{AB}$ \cite{Laudenbach_2018}. For the reconciliation, the optimal $\beta$ for most CV-QKD blocks tends to be close to the previously mentioned 94\%. For some CV-QKD block the optimal $\beta$ is much higher, because the gap between $I_{AB}$ and $\chi_{BE}$ is much smaller, hence a higher $\beta$ is required to get a positive SKR. There are also some CV-QKD blocks where reconciliation is not possible, as indicated by the red crosses. In these cases, $\chi_{BE}$ is either very close to or even larger than $I_{AB}$, and for positive SKRs, the required $\beta$ is not achievable with our error correction code. 
\\\indent The effects of both the $\beta$-optimisation and the different turbulence settings on the overall SKR of the system are shown in Fig. \ref{fig:SKR_RA}. For all turbulence settings, the optimal SKR without $\beta$-optimisation, i.e., when $\beta$ is chosen to be the same for each CV-QKD block, is achieved at $\beta = 94\%$. The overall SKR of the system decreases when the turbulence increases, however, three of the turbulence settings are very close in performance to each other, as for these settings the turbulence strength is similar. When applying $\beta$-optimisation, gains of up to 7.6\% are reported. These gains are dependent on both the error correction code used and the stability of the system. The SKR after $\beta$-optimisation will always be at least as high as the SKR without $\beta$-optimisation, and it is therefore always advantageous to implement it.

\section{Conclusion}
\label{Conclusion}
In this paper, we have investigated the use of high-dimensional reconciliation for CV-QKD. We have analysed how the different modulation formats impact the reconciliation, and using experimental results, we show SKR gains by up to 165\%. Furthermore, we have discussed rate-adaptivity for reconciliation in CV-QKD, and have shown that with $\beta$-optimisation it is possible to get an additional 7.6\% gain in SKR. Future works could focus on efficient implementations of high-dimensional reconciliation in hardware, in addition to a more extensive study on reconciliation in different experimental settings.
\ifCLASSOPTIONcaptionsoff
  \newpage
\fi

\IEEEtriggeratref{40}

\printbibliography[notcategory=ignore]

@article{gumucs2021low,
  title={Low Rate Protograph-Based {LDPC} Codes for Continuous Variable Quantum Key Distribution},
  author={G{\"u}m{\"u}{\c{s}}, Kadir and Schmalen, Laurent},
  journal={Proc. ISWCS 2021},
  year={2021},
  organization={IEEE}
}

@article{wang2017efficient,
  title={Efficient rate-adaptive reconciliation for continuous-variable quantum key distribution},
  author={Wang, Xiangyu and Zhang, Yi-Chen and Li, Zhengyu and Xu, Bingjie and Yu, Song and Guo, Hong},
  journal={arXiv preprint arXiv:1703.04916},
  year={2017}
}

@article{Jouguet_2011,
	year = 2011,

	publisher = {American Physical Society ({APS})},
	volume = {84},
	number = {6},
	author = {Paul Jouguet and S\'{e}bastien Kunz-Jacques and Anthony Leverrier},
	title = {Long-distance continuous-variable quantum key distribution with a {G}aussian modulation},
	journal = {Physical Review A}
}

@article{Leverrier_2008,


	year = 2008,

	publisher = {American Physical Society ({APS})},
	volume = {77},
	number = {4},
	author = {Anthony Leverrier and Romain All{\'{e}
}aume and Joseph Boutros and Gilles Z{\'{e}}mor and Philippe Grangier},
	title = {Multidimensional reconciliation for a continuous-variable quantum key distribution},
	journal = {Physical Review A}
}

@article{Milicevic_2018,
	year = 2018,

	publisher = {Springer Science and Business Media {LLC}
},
	volume = {4},
	number = {1},
	author = {Mario Milicevic and Chen Feng and Lei M. Zhang and P. Glenn Gulak},
	title = {Quasi-cyclic multi-edge {LDPC} codes for long-distance quantum cryptography},
	journal = {npj Quantum Information}
}

@article{gyongyosi2019survey,
  title={A survey on quantum computing technology},
  author={Gyongyosi, Laszlo and Imre, Sandor},
  journal={Computer Science Review},
  volume={31},
  year={2019},
  publisher={Elsevier}
}

@article{Wang:21,
author = {Shiyu Wang and Peng Huang and Tao Wang and Guihua Zeng},
journal = {Opt. Lett.},
number = {23},
publisher = {Optica Publishing Group},
title = {Feasibility of continuous-variable quantum key distribution through fog},
volume = {46},

year = {2021},
}

@article{Shen2019,
  title = {Free-space continuous-variable quantum key distribution of unidimensional {G}aussian modulation using polarized coherent states in an urban environment},
  author = {Shen, Shi-Yang and Dai, Ming-Wei and Zheng, Xue-Tao and Sun, Qi-Yao and Guo, Guang-Can and Han, Zheng-Fu},
  journal = {Phys. Rev. A},
  volume = {100},
  issue = {1},
  numpages = {8},
  year = {2019},

  publisher = {American Physical Society},
}

@article{Laudenbach_2018,
  
	year = 2018,

  
	publisher = {Wiley},
  
	volume = {1},
  
	number = {1},
  
	author = {Fabian Laudenbach and Christoph Pacher and Chi-Hang Fred Fung and Andreas Poppe and Momtchil Peev and Bernhard Schrenk and Michael Hentschel and Philip Walther and Hannes Hübel},
  
	title = {Continuous-Variable Quantum Key Distribution with {G}aussian Modulation{\textemdash}The Theory of Practical Implementations},
  
	journal = {Advanced Quantum Technologies}
}

@article{Denys_2021,
	year = 2021,

  
	publisher = {Verein zur Forderung des Open Access Publizierens in den Quantenwissenschaften},
  
	volume = {5},
  
  
	author = {Aur{\'{e}
}lie Denys and Peter Brown and Anthony Leverrier},
  
	title = {Explicit asymptotic secret key rate of continuous-variable quantum key distribution with an arbitrary modulation},
  
	journal = {Quantum}
}

@article{Sjoerd2023,

  author={van der Heide, Sjoerd and Frazão, João and Albores-Mejía, Aaron and Okonkwo, Chigo},

  journal = {Proc. OFC 2023}, 

  title={Receiver Noise Stability Calibration for {CV-QKD}}, 

  year={2023},

  volume={},

  number={},
}

@article{Jouguet_2012_fs,
  
	year = 2012,

  
	publisher = {American Physical Society ({APS})},
  
	volume = {86},
  
	number = {3},
  
	author = {Paul Jouguet and S{\'{e}
}bastien Kunz-Jacques and Eleni Diamanti and Anthony Leverrier},
  
	title = {Analysis of imperfections in practical continuous-variable quantum key distribution},
  
	journal = {Physical Review A}
}

@mastersthesis{OTGthesis,
    title  = {{Optical Turbulence Generator for Lab-based Experimental Studies of Atmospheric Turbulence in Vertical Optical Communication Links}},
    year   = {2022},
    author = {van Vliet, Vincent},
    school = {TU/e},
}

@ARTICLE{1994OptEn..33.3748K,
       author = {{Kiasaleh}, Kamran},
        title = "{On the probability density function of signal intensity in free-space optical communications systems impaired by pointing jitter and turbulence}",
      journal = {Optical Engineering},
         year = 1994,

       volume = {33},
}

@article{gumucs2023adaptive,
  title={{Adaptive Reconciliation for Experimental Continuous-Variable Quantum Key Distribution Over a Turbulent Free-Space Optical Channel}},
  author={G{\"u}m{\"u}{\c{s}}, Kadir and Fraz{\~a}o, Jo{\~a}o dos Reis and van Vliet, Vincent and van der Heide, Sjoerd and Hout, Menno van den and Albores-Mejia, Aaron and Bradley, Thomas and Okonkwo, Chigo},
  journal={OFC 2024},
  year={2024}
}

@article{zhou2018practical,
  title={Practical security of continuous-variable quantum key distribution under finite-dimensional effect of multi-dimensional reconciliation},
  author={Zhou, Yingming and Jiang, Xue-Qin and Liu, Weiqi and Wang, Tao and Huang, Peng and Zeng, Guihua},
  journal={Chinese Physics B},
  volume={27},
  number={5},
  pages={050301},
  year={2018},
  publisher={IOP Publishing}
}

@article{shor1999polynomial,
  title={Polynomial-time algorithms for prime factorization and discrete logarithms on a quantum computer},
  author={Shor, Peter W},
  journal={SIAM review},
  volume={41},
  number={2},
  pages={303--332},
  year={1999},
  publisher={SIAM}
}

@article{yang2023information,
  title={Information reconciliation of continuous-variables quantum key distribution: principles, implementations and applications},
  author={Yang, Shenshen and Yan, Zhilei and Yang, Hongzhao and Lu, Qing and Lu, Zhenguo and Cheng, Liuyong and Miao, Xiangyang and Li, Yongmin},
  journal={EPJ Quantum Technology},
  volume={10},
  number={1},
  pages={40},
  year={2023},
  publisher={Springer Berlin Heidelberg}
}

@ARTICLE{Assche2004,

  author={Van Assche, G. and Cardinal, J. and Cerf, N.J.},

  journal={IEEE Transactions on Information Theory}, 

  title={Reconciliation of a quantum-distributed Gaussian key}, 

  year={2004},

  volume={50},

  number={2},

  pages={394-400},
}

@misc{Joe2024,
      title={{Co-propagation of Classical and Continuous-variable QKD Signals over a Turbulent Optical Channel with a Real-time QKD Receiver}}, 
      author={João dos Reis Frazão and Vincent van Vliet and Sjoerd van der Heide and Menno van den Hout and Kadir Gümüş and Aaron Albores-Mejía and Boris Škorić and Chigo Okonkwo},
      year={2024},
      eprint={2401.10581},
      archivePrefix={arXiv},
      primaryClass={quant-ph},
      note={Accepted for OFC 2024}
}

@article{eriksson2019wavelength,
  title={Wavelength division multiplexing of continuous variable quantum key distribution and 18.3 Tbit/s data channels},
  author={Eriksson, Tobias A and Hirano, Takuya and Puttnam, Benjamin J and Rademacher, Georg and Lu{\'\i}s, Ruben S and Fujiwara, Mikio and Namiki, Ryo and Awaji, Yoshinari and Takeoka, Masahiro and Wada, Naoya and others},
  journal={Communications Physics},
  volume={2},
  number={1},
  pages={9},
  year={2019},
  publisher={Nature Publishing Group UK London}
}

@article{hajomer2022modulation,
  title={Modulation leakage-free continuous-variable quantum key distribution},
  author={Hajomer, Adnan AE and Jain, Nitin and Mani, Hossein and Chin, Hou-Man and Andersen, Ulrik L and Gehring, Tobias},
  journal={npj Quantum Information},
  volume={8},
  number={1},
  pages={136},
  year={2022},
  publisher={Nature Publishing Group UK London}
}

@article{Zhang_2020,
   title={Long-Distance Continuous-Variable Quantum Key Distribution over 202.81 km of Fiber},
   volume={125},
   ISSN={1079-7114},
   number={1},
   journal={Physical Review Letters},
   publisher={American Physical Society (APS)},
   author={Zhang, Yichen and Chen, Ziyang and Pirandola, Stefano and Wang, Xiangyu and Zhou, Chao and Chu, Binjie and Zhao, Yijia and Xu, Bingjie and Yu, Song and Guo, Hong},
   year={2020},
}

@article{GG02,
  title = {Continuous Variable Quantum Cryptography Using Coherent States},
  author = {Grosshans, Fr\'ed\'eric and Grangier, Philippe},
  journal = {Phys. Rev. Lett.},
  volume = {88},
  issue = {5},
  pages = {057902},
  numpages = {4},
  year = {2002},

  publisher = {American Physical Society},
}

@article{dickson1919quaternions,
  title={On quaternions and their generalization and the history of the eight square theorem},
  author={Dickson, Leonard E},
  journal={Annals of Mathematics},
  pages={155--171},
  year={1919},
  publisher={JSTOR}
}

@article{Kaur2021,
  title = {Asymptotic security of discrete-modulation protocols for continuous-variable quantum key distribution},
  author = {Kaur, Eneet and Guha, Saikat and Wilde, Mark M.},
  journal = {Phys. Rev. A},
  volume = {103},
  issue = {1},
  pages = {012412},
  numpages = {20},
  year = {2021},

  publisher = {American Physical Society},
}

@article{householder1958unitary,
  title={Unitary triangularization of a nonsymmetric matrix},
  author={Householder, Alston S},
  journal={Journal of the ACM (JACM)},
  volume={5},
  number={4},
  pages={339--342},
  year={1958},
  publisher={ACM New York, NY, USA}
}

@article{hu2005regular,
  title={Regular and irregular progressive edge-growth tanner graphs},
  author={Hu, Xiao-Yu and Eleftheriou, Evangelos and Arnold, Dieter-Michael},
  journal={IEEE transactions on information theory},
  volume={51},
  number={1},
  pages={386--398},
  year={2005},
  publisher={IEEE}
}

@article{Notarnicola_2024,
   title={Probabilistic Amplitude Shaping for Continuous-Variable Quantum Key Distribution With Discrete Modulation Over a Wiretap Channel},
   volume={72},
   ISSN={1558-0857},
   number={1},
   journal={IEEE Transactions on Communications},
   publisher={Institute of Electrical and Electronics Engineers (IEEE)},
   author={Notarnicola, Michele N. and Olivares, Stefano and Forestieri, Enrico and Parente, Emanuele and Potì, Luca and Secondini, Marco},
   year={2024},
 pages={375–386} }

@article{cho2019probabilistic,
  title={Probabilistic constellation shaping for optical fiber communications},
  author={Cho, Junho and Winzer, Peter J},
  journal={Journal of Lightwave Technology},
  volume={37},
  number={6},
  pages={1590--1607},
  year={2019},
  publisher={IEEE}
}

@article{Bennett_2014,
   title={Quantum cryptography: Public key distribution and coin tossing},
   volume={560},
   ISSN={0304-3975},
   journal={Theoretical Computer Science},
   publisher={Elsevier BV},
   author={Bennett, Charles H. and Brassard, Gilles},
   year={2014},
 pages={7–11} }

@INPROCEEDINGS{Li2005,

  author={Li, J. and Bose, A. and Zhao, Y.Q.},

  booktitle={3rd Annual Communication Networks and Services Research Conference (CNSR'05)}, 

  title={Rayleigh flat fading channels' capacity}, 

  year={2005},

  volume={},

  number={},

  pages={214-217},
}

@article{collins2006integration,
  title={Integration with respect to the Haar measure on unitary, orthogonal and symplectic group},
  author={Collins, Beno{\^\i}t and {\'S}niady, Piotr},
  journal={Communications in Mathematical Physics},
  volume={264},
  number={3},
  pages={773--795},
  year={2006},
  publisher={Springer}
}

@inproceedings{OTG,
title = {{Design, Characterisation, and Demonstration of a Hot-Air-Based Optical Turbulence Generator}},
author = "{van Vliet}, V. and {van den Hout}, M. and {van der Heide}, S. and C.M. Okonkwo",
year = "2022",
booktitle = {{Proceedings of the 26th Annual Symposium of the IEEE Photonics Benelux Chapter}},
note = {{26th Annual Symposium of the IEEE Photonics Benelux Chapter}},
}

@article{grosshans2003quantum,
  title={Quantum key distribution using gaussian-modulated coherent states},
  author={Grosshans, Fr{\'e}d{\'e}ric and Van Assche, Gilles and Wenger, J{\'e}r{\^o}me and Brouri, Rosa and Cerf, Nicolas J and Grangier, Philippe},
  journal={Nature},
  volume={421},
  number={6920},
  pages={238--241},
  year={2003},
  publisher={Nature Publishing Group UK London}
}

@article{ghorai2019asymptotic,
  title={Asymptotic security of continuous-variable quantum key distribution with a discrete modulation},
  author={Ghorai, Shouvik and Grangier, Philippe and Diamanti, Eleni and Leverrier, Anthony},
  journal={Physical Review X},
  volume={9},
  number={2},
  pages={021059},
  year={2019},
  publisher={APS}
}

@inproceedings{Zhou2023,
author = {Chuang Zhou and Yang Li and Li Ma and Yujie Luo and Jie Yang and Mei Wu and Shuai Zhang and Wei Huang and Bingjie Xu},
title = {{A high-throughput and FPGA-based LDPC decoder for continuous-variable quantum key distribution system}},
volume = {12775},
booktitle = {Quantum and Nonlinear Optics X},
editor = {Qiongyi He and Dai-Sik Kim and Chuan-Feng Li},
organization = {International Society for Optics and Photonics},
publisher = {SPIE},
pages = {1277517},
year = {2023},
}

@book{revesz2014laws,
  title={The laws of large numbers},
  author={R{\'e}v{\'e}sz, P{\'a}l},
  volume={4},
  year={2014},
  publisher={Academic Press}
}

@INPROCEEDINGS{Beerman2014,

  author={Beermann, Moritz and Vary, Peter},

  booktitle={2014 IEEE Wireless Communications and Networking Conference (WCNC)}, 

  title={Joint optimization of multi-rate LDPC code ensembles for the AWGN channel based on shortening and puncturing}, 

  year={2014},

  volume={},

  number={},

  pages={200-205},
}

@inproceedings{Cil24QCrypt,
	author = {E. E. Cil and L. Schmalen},
	title = "An open-source library for information reconciliation in continuous-variable {QKD}",
	booktitle = {Proc. International Conference on Quantum Cryptography (QCRYPT)},
	address = {Vigo, Spain},
	month = Sep,
	year = {2024},}

@inproceedings{Eriksson:20,
author = {Tobias A. Eriksson and Ruben S. Lu\'{i}s and Kadir G\"{u}m\"{u}\c{s} and Georg Rademacher and Benjamin J. Puttnam and Hideaki Furukawa and Naoya Wada and Yoshinari Awaji and Alex Alvarado and Masahide Sasaki and Masahiro Takeoka},
booktitle = {Optical Fiber Communication Conference (OFC) 2020},
journal = {Optical Fiber Communication Conference (OFC) 2020},
pages = {T3D.5},
publisher = {Optica Publishing Group},
title = {Digital Self-Coherent Continuous Variable Quantum Key Distribution System},
year = {2020},
url = {https://opg.optica.org/abstract.cfm?URI=OFC-2020-T3D.5},
doi = {10.1364/OFC.2020.T3D.5},
}

@article{leverrier2010finite,
  title={Finite-size analysis of a continuous-variable quantum key distribution},
  author={Leverrier, Anthony and Grosshans, Fr{\'e}d{\'e}ric and Grangier, Philippe},
  journal={Physical Review A—Atomic, Molecular, and Optical Physics},
  volume={81},
  number={6},
  pages={062343},
  year={2010},
  publisher={APS}
}

\end{document}